\documentstyle[12pt,hyperref,graphicx,amssymb,amsmath,tensind]{article}

\tensordelimiter{?}

\def\mathfrak{\bf}

\def\be{\begin{equation}}
\def\ee{\end{equation}}
\def\bea{\begin{eqnarray}}
\def\eea{\end{eqnarray}}

\def\dt#1{\on{\hbox{\bf .}}{#1}}                
\def\Dot#1{\dt{#1}}
\def\IR{\relax{\rm I\kern-.18em R}}
\def\binomial#1#2{\left(\,{\buildrel
{\raise4pt\hbox{$\displaystyle{#1}$}}\over
{\raise-6pt\hbox{$\displaystyle{#2}$}}}\,\right)}

\def\[{\lfloor{\hskip 0.35pt}\!\!\!\lceil}
\def\]{\rfloor{\hskip 0.35pt}\!\!\!\rceil}

\catcode`@=11
\def\un#1{\relax\ifmmode\@@underline#1\else
        $\@@underline{\hbox{#1}}$\relax\fi}
\catcode`@=12

\def\fracm#1#2{\hbox{\large{${\frac{{#1}}{{#2}}}$}}}

\def\ad{{\kern0.5pt
                   \alpha \kern-5.05pt
\raise5.8pt\hbox{$\textstyle.$}\kern
0.5pt}}

\def\Dot#1{{\kern0.5pt
     {#1} \kern-5.05pt \raise5.8pt\hbox{$\textstyle.$}\kern
0.5pt}}

\def\d{\delta}

\def\i{\iota}

\def\l{\lambda}

\def\bo{{\raise.15ex\hbox{\large$\Box$}}}               
\def\TH{{\raise.2ex\hbox{$\displaystyle \bigodot$}\mskip-4.7mu \llap H
\;}}
\def\face{{\raise.2ex\hbox{$\displaystyle \bigodot$}\mskip-2.2mu \llap
{$\ddot
        \smile$}}}                                      

\def\leftrightarrowfill{$\mathsurround=0pt \mathord\leftarrow \mkern-6mu
        \cleaders\hbox{$\mkern-2mu \mathord- \mkern-2mu$}\hfill
        \mkern-6mu \mathord\rightarrow$}
\def\dvec#1{\vbox{\ialign{##\crcr
        \leftrightarrowfill\crcr\noalign{\kern-1pt\nointerlineskip}
        $\hfil\displaystyle{#1}\hfil$\crcr}}}           
\def\dt#1{{\buildrel {\hbox{\LARGE .}} \over {#1}}}     

\def\fracm#1#2{\hbox{\large{${\frac{{#1}}{{#2}}}$}}}
\def\frac#1#2{{\textstyle{#1\over\vphantom2\smash{\raise.20ex
        \hbox{$\scriptstyle{#2}$}}}}}                   
\def\sfrac#1#2{{\vphantom1\smash{\lower.5ex\hbox{\small$#1$}}\over
        \vphantom1\smash{\raise.4ex\hbox{\small$#2$}}}} 
\def\bfrac#1#2{{\vphantom1\smash{\lower.5ex\hbox{$#1$}}\over
        \vphantom1\smash{\raise.3ex\hbox{$#2$}}}}       
\def\afrac#1#2{{\vphantom1\smash{\lower.5ex\hbox{$#1$}}\over#2}}    
\def\on#1#2{\mathop{\null#2}\limits^{#1}}               

\newskip\humongous \humongous=0pt plus 1000pt minus 1000pt
\def\caja{\mathsurround=0pt}
\def\eqalign#1{\,\vcenter{\openup2\jot \caja
        \ialign{\strut \hfil$\displaystyle{##}$&$
        \displaystyle{{}##}$\hfil\crcr#1\crcr}}\,}
\newif\ifdtup

  \def\pp{{\mathchoice
          {
              \kern 1pt%
              \raise 1pt
              \vbox{\hrule width5pt height0.4pt depth0pt
                    \kern -2pt
                    \hbox{\kern 2.3pt
                          \vrule width0.4pt height6pt depth0pt
                          }
                    \kern -2pt
                    \hrule width5pt height0.4pt depth0pt}%
                    \kern 1pt
           }
            {
              \kern 1pt%
              \raise 1pt
              \vbox{\hrule width4.3pt height0.4pt depth0pt
                    \kern -1.8pt
                    \hbox{\kern 1.95pt
                          \vrule width0.4pt height5.4pt depth0pt
                          }
                    \kern -1.8pt
                    \hrule width4.3pt height0.4pt depth0pt}%
                    \kern 1pt
            }
            {
              \kern 0.5pt%
              \raise 1pt
              \vbox{\hrule width4.0pt height0.3pt depth0pt
                    \kern -1.9pt  
                    \hbox{\kern 1.85pt
                          \vrule width0.3pt height5.7pt depth0pt
                          }
                    \kern -1.9pt
                    \hrule width4.0pt height0.3pt depth0pt}%
                    \kern 0.5pt
            }
            {
              \kern 0.5pt%
              \raise 1pt
              \vbox{\hrule width3.6pt height0.3pt depth0pt
                    \kern -1.5pt
                    \hbox{\kern 1.65pt
                          \vrule width0.3pt height4.5pt depth0pt
                          }
                    \kern -1.5pt
                    \hrule width3.6pt height0.3pt depth0pt}%
                    \kern 0.5pt
            }
        }}

  \def\mm{{\mathchoice
                       {
                             \kern 1pt
               \raise 1pt    \vbox{\hrule width5pt height0.4pt depth0pt
                                  \kern 2pt
                                  \hrule width5pt height0.4pt depth0pt}
                             \kern 1pt}
                       {
                            \kern 1pt
               \raise 1pt \vbox{\hrule width4.3pt height0.4pt depth0pt
                                  \kern 1.8pt
                                  \hrule width4.3pt height0.4pt depth0pt}
                             \kern 1pt}
                       {
                            \kern 0.5pt
               \raise 1pt
                            \vbox{\hrule width4.0pt height0.3pt depth0pt
                                  \kern 1.9pt
                                  \hrule width4.0pt height0.3pt depth0pt}
                            \kern 1pt}
                       {
                           \kern 0.5pt
             \raise 1pt  \vbox{\hrule width3.6pt height0.3pt depth0pt
                                  \kern 1.5pt
                                  \hrule width3.6pt height0.3pt depth0pt}
                           \kern 0.5pt}
                       }}

\def\pd{{\kern0.5pt
                   + \kern-5.05pt \raise5.8pt\hbox{$\textstyle.$}\kern
0.5pt}}

\def\pmd{{\kern0.5pt
                  \pm \kern-5.05pt \raise6.3pt\hbox{$\textstyle.$}\kern1.5pt}}

\def\md{{\mathchoice
   {
      {{\kern 1pt - \kern-6.2pt \raise5pt\hbox{$\textstyle.$}\kern 1pt}}}
    {
      {{\kern 1pt - \kern-6.2pt \raise5pt\hbox{$\textstyle.$}\kern 1pt}}}
    {
      {\kern0.5pt - \kern-5.05pt \raise3.4pt\hbox{$\textstyle.$}\kern0.5pt}}
    {
      {\kern0.5pt - \kern-5.05pt \raise3.4pt\hbox{$\textstyle.$}\kern0.5pt}}}}

\def\ad{{\dot{\alpha}}}

\def\pp{{\mathchoice
          {
              \kern 1pt%
              \raise 1pt
              \vbox{\hrule width5pt height0.4pt depth0pt
                    \kern -2pt
                    \hbox{\kern 2.3pt
                          \vrule width0.4pt height6pt depth0pt
                          }
                    \kern -2pt
                    \hrule width5pt height0.4pt depth0pt}%
                    \kern 1pt
           }
            {
              \kern 1pt%
              \raise 1pt
              \vbox{\hrule width4.3pt height0.4pt depth0pt
                    \kern -1.8pt
                    \hbox{\kern 1.95pt
                          \vrule width0.4pt height5.4pt depth0pt
                          }
                    \kern -1.8pt
                    \hrule width4.3pt height0.4pt depth0pt}%
                    \kern 1pt
            }
            {
              \kern 0.5pt%
              \raise 1pt
              \vbox{\hrule width4.0pt height0.3pt depth0pt
                    \kern -1.9pt  
                    \hbox{\kern 1.85pt
                          \vrule width0.3pt height5.7pt depth0pt
                          }
                    \kern -1.9pt
                    \hrule width4.0pt height0.3pt depth0pt}%
                    \kern 0.5pt
            }
            {
              \kern 0.5pt%
              \raise 1pt
              \vbox{\hrule width3.6pt height0.3pt depth0pt
                    \kern -1.5pt
                    \hbox{\kern 1.65pt
                          \vrule width0.3pt height4.5pt depth0pt
                          }
                    \kern -1.5pt
                    \hrule width3.6pt height0.3pt depth0pt}%
                    \kern 0.5pt
            }
        }}

  \def\mm{{\mathchoice
                       {
                             \kern 1pt
               \raise 1pt    \vbox{\hrule width5pt height0.4pt depth0pt
                                  \kern 2pt
                                  \hrule width5pt height0.4pt depth0pt}
                             \kern 1pt}
                       {
                            \kern 1pt
               \raise 1pt \vbox{\hrule width4.3pt height0.4pt depth0pt
                                  \kern 1.8pt
                                  \hrule width4.3pt height0.4pt depth0pt}
                             \kern 1pt}
                       {
                            \kern 0.5pt
               \raise 1pt
                            \vbox{\hrule width4.0pt height0.3pt depth0pt
                                  \kern 1.9pt
                                  \hrule width4.0pt height0.3pt depth0pt}
                            \kern 1pt}
                       {
                           \kern 0.5pt
             \raise 1pt  \vbox{\hrule width3.6pt height0.3pt depth0pt
                                  \kern 1.5pt
                                  \hrule width3.6pt height0.3pt depth0pt}
                           \kern 0.5pt}
                       }}

\def\pd{{\kern0.5pt
                   + \kern-5.05pt \raise5.8pt\hbox{$\textstyle.$}\kern
0.5pt}}

\def\pmd{{\kern0.5pt
                  \pm \kern-5.05pt \raise6.3pt\hbox{$\textstyle.$}\kern1.5pt}}

\def\md{{\mathchoice
   {
      {{\kern 1pt - \kern-6.2pt \raise5pt\hbox{$\textstyle.$}\kern 1pt}}}
    {
      {{\kern 1pt - \kern-6.2pt \raise5pt\hbox{$\textstyle.$}\kern 1pt}}}
    {
      {\kern0.5pt - \kern-5.05pt \raise3.4pt\hbox{$\textstyle.$}\kern0.5pt}}
    {
      {\kern0.5pt - \kern-5.05pt \raise3.4pt\hbox{$\textstyle.$}\kern0.5pt}}}}

\def\dslash{\not{\hbox{\kern-2pt $\partial$}}}
\def\Dslash{\not{\hbox{\kern-4pt $D$}}}
\def\pslash{\not{\hbox{\kern-2.3pt $p$}}}
 \newtoks\slashfraction
 \slashfraction={.13}
 \def\slash#1{\setbox0\hbox{$ #1 $}
 \setbox0\hbox to \the\slashfraction\wd0{\hss \box0}/\box0 }

\font\ro=cmsy10                          
\def\kcr{{\hbox{\ro \char'170}}}                
\def\ktl{{\hbox{\ro \char'170}}}        
\def\ktr{{\hbox{\ro \char'170}}}        
\def\kbl{{\hbox{\ro \char'170}}}        
\def\kbr{{\hbox{\ro \char'170}}}        

\def\plpl{\raise-2pt\hbox{$\raise3pt\hbox{$_+$}\hskip-6.67pt\raise0.0pt
\hbox{$^+$}\hskip 0.01pt$}}
\def\mimi{\raise-2pt\hbox{$\raise3pt\hbox{$_-$}\hskip-6.67pt\raise0.0pt
\hbox{$^-$}\hskip 0.01pt$}}

\def\bo{{\raise.15ex\hbox{\large$\Box$}}}               
\def\TH{{\raise.2ex\hbox{$\displaystyle \bigodot$}\mskip-4.7mu \llap H \;}}
\def\face{{\raise.2ex\hbox{$\displaystyle \bigodot$}\mskip-2.2mu \llap {$\ddot
        \smile$}}}                                      

\def\leftrightarrowfill{$\mathsurround=0pt \mathord\leftarrow \mkern-6mu
        \cleaders\hbox{$\mkern-2mu \mathord- \mkern-2mu$}\hfill
        \mkern-6mu \mathord\rightarrow$}
\def\dvec#1{\vbox{\ialign{##\crcr
        \leftrightarrowfill\crcr\noalign{\kern-1pt\nointerlineskip}
        $\hfil\displaystyle{#1}\hfil$\crcr}}}           
\def\dt#1{{\buildrel {\hbox{\LARGE .}} \over {#1}}}     

\def\fracm#1#2{\hbox{\large{${\frac{{#1}}{{#2}}}$}}}
\def\frac#1#2{{\textstyle{#1\over\vphantom2\smash{\raise.20ex
        \hbox{$\scriptstyle{#2}$}}}}}                   
\def\sfrac#1#2{{\vphantom1\smash{\lower.5ex\hbox{\small$#1$}}\over
        \vphantom1\smash{\raise.4ex\hbox{\small$#2$}}}} 
\def\bfrac#1#2{{\vphantom1\smash{\lower.5ex\hbox{$#1$}}\over
        \vphantom1\smash{\raise.3ex\hbox{$#2$}}}}       
\def\afrac#1#2{{\vphantom1\smash{\lower.5ex\hbox{$#1$}}\over#2}}    
\def\on#1#2{\mathop{\null#2}\limits^{#1}}               

\parskip=\medskipamount                 
\thicklines                         

\def\oldheadpic{                                
        \setlength{\unitlength}{.4mm}
        \thinlines
        \par
        \begin{picture}(349,16)
        \put(325,16){\line(1,0){4}}
        \put(330,16){\line(1,0){4}}
        \put(340,16){\line(1,0){4}}
        \put(335,0){\line(1,0){4}}
        \put(340,0){\line(1,0){4}}
        \put(345,0){\line(1,0){4}}
        \put(329,0){\line(0,1){16}}
        \put(330,0){\line(0,1){16}}
        \put(339,0){\line(0,1){16}}
        \put(340,0){\line(0,1){16}}
        \put(344,0){\line(0,1){16}}
        \put(345,0){\line(0,1){16}}
        \put(329,16){\oval(8,32)[bl]}
        \put(330,16){\oval(8,32)[br]}
        \put(339,0){\oval(8,32)[tl]}
        \put(345,0){\oval(8,32)[tr]}
        \end{picture}
        \par
        \thicklines
        \vskip.2in}
\def\oldtitle#1#2#3#4{\oldheadpic\begin{center}\vglue.5in{\large\bf #1}\\[.6in]
        {#2}\\[.1in] {\it Department of Physics and Astronomy}\\
        {\it University of Maryland, College Park, MD 20742}\\[.6in]
        Physics Publication \#{#3}\\ {#4}\\[1.5in] {\bf ABSTRACT}\\[.1in]
        \end{center} \begin{quotation}}                 
\def\oldTitle#1#2#3#4#5#6#7{\oldheadpic\begin{center} \vglue .4in
        {\large\bf #1}\\[.4in]
        {#2}\\[.1in] {\it Department of Physics and Astronomy}\\
        {\it University of Maryland, College Park, MD 20742}\\[.1in]
        {#3}\\[.1in] {\it {#4}}\\ {\it {#5}}\\[.4in]
        Physics Publication \#{#6}\\ {#7}\\[.5in] {\bf ABSTRACT}\\[.1in]
        \end{center} \begin{quotation}}                 
\def\border{                                            
        \setlength{\unitlength}{1mm}
        \newcount\xco
        \newcount\yco
        \xco=-21
        \yco=12
        \begin{picture}(140,0)
        \put(\xco,\yco){$\ktl$}
        \advance\yco by-1
        {\loop
        \put(\xco,\yco){$\kcr$}
        \advance\yco by-2
        \ifnum\yco>-240
        \repeat
        \put(\xco,\yco){$\kbl$}}
        \xco=158
        \yco=12
        \put(\xco,\yco){$\ktr$}
        \advance\yco by-1
        {\loop
        \put(\xco,\yco){$\kcr$}
        \advance\yco by-2
        \ifnum\yco>-240
        \repeat
        \put(\xco,\yco){$\kbr$}}
        \put(-20,13){\tiny ~The University of Iowa Particle Theory Group ** The University of Maryland CSPT and Department of Physics ** Grinnell College Department of Physics}
        \put(-20,-241.5){\tiny ~The University of Iowa Particle Theory Group ** The University of Maryland CSPT and Department of Physics ** Grinnell College Department of Physics}
        \end{picture}
        \par\vskip-8mm}
\def\bordero{                                           
        \setlength{\unitlength}{1mm}
        \newcount\xco
        \newcount\yco
        \xco=-31
        \yco=12
        \begin{picture}(140,0)
        \put(\xco,\yco){$\ktl$}
        \advance\yco by-1
        {\loop
        \put(\xco,\yco){$\kclr$}
        \advance\yco by-2
        \ifnum\yco>-240
        \repeat
        \put(\xco,\yco){$\kbl$}}
        \xco=151
        \yco=12
        \put(\xco,\yco){$\ktr$}
        \advance\yco by-1
        {\loop
        \put(\xco,\yco){$\kcr$}
        \advance\yco by-2
        \ifnum\yco>-240
        \repeat
        \put(\xco,\yco){$\kbr$}}
        \put(-20,12){\ooo bacdefghidfghghdhededbihdgdfdfhhdheidhdhebaaahjhhdahba

hgdedge
   hgfdiehhgdigicba}
        \put(-20,-241.5){\ooo ababaighefdbfghgeahgdfgafagihdidihiidhiagfedhadbfd

ecdcdfa
   gdcbhaddhbgfchbgfdacfediacbabab}
        \end{picture}
        \par\vskip-8mm}
\def\headpic{                                           
        \indent
        \setlength{\unitlength}{.4mm}
        \thinlines
        \par
        \begin{picture}(29,16)
        \put(165,16){\line(1,0){4}}
        \put(170,16){\line(1,0){4}}
        \put(180,16){\line(1,0){4}}
        \put(175,0){\line(1,0){4}}
        \put(180,0){\line(1,0){4}}
        \put(185,0){\line(1,0){4}}
        \put(169,0){\line(0,1){16}}
        \put(170,0){\line(0,1){16}}
        \put(179,0){\line(0,1){16}}
        \put(180,0){\line(0,1){16}}
        \put(184,0){\line(0,1){16}}
        \put(185,0){\line(0,1){16}}
        \put(169,16){\oval(8,32)[bl]}
        \put(170,16){\oval(8,32)[br]}
        \put(179,0){\oval(8,32)[tl]}
        \put(185,0){\oval(8,32)[tr]}
        \end{picture}
        \par\vskip-6.5mm
        \thicklines}
\def\title#1#2#3#4{\border\headpic {\hbox to\hsize{#4 \hfill UMDEPP #3}}\par
        \begin{center} \vglue .5in {\large\bf #1}\\[.6in]
        {#2}\\[.1in] {\it Department of Physics and Astronomy}\\
        {\it University of Maryland, College Park, MD 20742}\\[1.5in]
        {\bf ABSTRACT}\\[.1in] \end{center} \begin{quotation}}  
\def\Title#1#2#3#4#5#6#7{\border\headpic
        {\hbox to\hsize{#7 \hfill UMDEPP #6}}\par
        \begin{center} \vglue .4in {\large\bf #1}\\[.4in]
        {#2}\\[.1in] {\it Department of Physics and Astronomy}\\
        {\it University of Maryland, College Park, MD 20742}\\[.1in]
        {#3}\\[.1in] {\it {#4}}\\ {\it {#5}}\\[.5in] {\bf ABSTRACT}\\[.1in]
        \end{center} \begin{quotation}}                 
\def\endtitle{\end{quotation}\newpage}                  

\numberwithin{equation}{section}
\def\qd{{\kern0.5pt
                   q \kern-5.05pt \raise5.8pt\hbox{$\textstyle.$}\kern
0.5pt}}

\begin{document}

\def\dt#1{\on{\hbox{\bf .}}{#1}}                
\def\Dot#1{\dt{#1}}

\def\gfrac#1#2{\frac {\scriptstyle{#1}}
        {\mbox{\raisebox{-.6ex}{$\scriptstyle{#2}$}}}}
\def\gg{{\hbox{\sc g}}}
\border\headpic {\hbox to\hsize{\today \hfill
{UMDEPP-11-009}}}
\par
{$~$ \hfill
~{1106.5475 [hep-th]}}
\par

\setlength{\oddsidemargin}{0.3in}
\setlength{\evensidemargin}{-0.3in}
\begin{center}
\vglue .10in
{\large\bf A Detailed Investigation of \\ First and Second Order Supersymmetries for \\ Off-Shell  ${\mathcal N}=2$ and ${\mathcal N} = 4$ Supermultiplets}\\[.5in]

S.\, James Gates, Jr.\footnote{gatess@wam.umd.edu}${}^{\dagger}$,
James Parker\footnote{jp@jamesparker.me}${}^{\dagger}$,~Vincent G. J. Rodgers\footnote{vrodgers@newton.physics.uiowa.edu}${}^*$, \\~Leo Rodriguez\footnote{rodrigul@grinnell.edu}${}^{**}$,~and Kory Stiffler\footnote{corresponding author: kstiffle@gmail.com}${}^{\dagger}$
\\[0.3in]
${}^\dag${\small {\it Department of Physics, University of Maryland,
College Park, MD 20742-4111}}
\\[0.1in]
${}^*${\small {\it Department of Physics and Astronomy,
The University of Iowa, Iowa City, IA 52242}}
\\[.1in]
${}^{**}${\small {\it Department of Physics, Grinnell College, Grinnell, IA 50112-1690}}
\\[.3in]{\bf ABSTRACT}\\[.01in]
\end{center}
\begin{quotation}
{ {\small This paper investigates the $d=4$, ${\mathcal N} = 4$ Abelian, global Super-Yang Mills system (SUSY-YM).  It is shown how the ${\mathcal N} =2$ Fayet Hypermultiplet (FH) and ${\mathcal N} = 2$ vector multiplet (VM) are embedded within.  The central charges and internal symmetries provide a plethora of information as to further symmetries of the Lagrangian.  Several of these symmetries are calculated to second order.  It is hoped that investigations such as these may yield avenues to help solve the auxiliary field closure problem for $d=4$, ${\mathcal N} = 4$, SUSY-YM and the $d=4$, ${\mathcal N} =2$ Fayet-Hypermultiplet, without using an infinite number of auxiliary fields.  }}

\endtitle

\setlength{\oddsidemargin}{0.3in}
\setlength{\evensidemargin}{-0.3in}

\section{Introduction}

$~~~$ The ${\mathcal N} = 4$ Super-Yang Mills (SUSY-YM) system is a very active area of study, and has become even more so over the past decade with the emergence of the $AdS/CFT$ correspondence~\cite{Maldacena:1997re}.   One very powerful aspect of this correspondence is that it relates a perturbation theory to a strongly coupled system. As ${\mathcal N} = 4$ SUSY-YM is a conformal field theory, an important undertaking has been to find dualities between string theory and theories that are more \emph{QCD-like}.  Klebanov and Strassler took a step in this direction in~\cite{Klebanov:2000hb}, where they unveiled a background which breaks the supersymmetry to ${\mathcal N} =1$, while regulating the IR divergence behavior.  Following this work, several other supersymmetry breaking backgrounds were discovered~\cite{Maldacena:2000yy,Cvetic:2001ma,Maldacena:2001pb,Canoura:2005uz}.

In parallel to the unveiling of these duality backgrounds, specific calculations were done showing duality to confining gauge theory calculations.  Herzog and Klebanov showed duality in the tree level energy calculations between branes on the supergravity side and confining strings on the gauge theory side~\cite{Herzog:2000rz,Herzog:2002ss}.  In this newly emerging gauge/gravity picture, Regge trajectories were resurrected from the old dual resonance models and reinvestigated by Pando Zayas, Sonnenschein, and Vaman in~\cite{PandoZayas:2003yb}, including some one loop level calculations.  Most recently, one loop corrections to the $k$-string energy has been investigated, the so-called L\"uscher term.  This emerges on the string theory side through the bosonic part of the D-brane energy, although in addition different one loop information of the \emph{fermionic} part has also been unveiled~\cite{PandoZayas:2008hw,Doran:2009pp,Stiffler:2009ma,Stiffler:2010pz}.  So we see a nice picture developing showing dualities between objects on the string theory and gauge theory sides.

In this paper, we take a step back from this picture. Even though this is the best understood of the gauge/gravity dualities, the $d=4$, ${\mathcal N} = 4$ SUSY-YM theory part of the correspondence itself still has unknown attributes.  Most glaring is the auxiliary field closure problem: \emph{it is still unknown how to augment this theory with finite numbers of auxiliary fields such that the charges satisfy the following algebra}:
\begin{align}\label{eq:ClosedAlgebra}
    \{ Q_a^I, Q_b^J \} &=  2~i~ \d^{I  J}(\gamma^{\mu})_{ab} \partial_\mu 
\end{align}
This is a problem which has been well known for at least thirty years.  In 1981, Siegel and Rocek (SR) investigated a solution within the known framework that existed at the time and found a no-go theorem~\cite{Siegel:1981dx}.  This result has been interpreted
as the definitive statement on this issue.  

However, there are some loose ends that challenge this conventional wisdom about the SR no-go theorem.
The first of these is contained within the SR work itself.  In an often overlooked final commentary in the work, the authors
state a possible way to avoid the SR no-go theorem.  It is also often overlooked that the derivation of the SR no-go theorem
is based on a particular assumption of dynamics.  In particular, the authors assume the gauge field is subject to the
dynamics of the usual Yang-Mills action.   It is simple to consider a different starting point.  It is easy to
negate this assumption.  

Though mostly unknown, the action for the ABJM model 
\cite{ABJM} 
together with a discussion of 3D, $\cal N$ $=$ 6
superconformal invariance first appeared in works written in the period of 1991-1995 on the 
importance of Chern-Simons models~\cite{3DSUSY1,3DSUSY2,3DSUSY3,3DSUSY4}.
So instead of considering the fields of a vector multiplet in 4D 
hypermultiplet in 4D that realizes $\cal N$ $=$ 2 SUSY, one could attempt to construct respective
3D Chern-Simons models with $\cal N$ $=$ 8 SUSY or $\cal N$ $=$ 4 SUSY that are based on
the dimensional reduction of 4D multiplets.  The SR no-go theorem cannot be applied to such constructions!
Thus, the study of 3D Chern-Simons theories provides a new way to attack this very old problem.

The methods in harmonic~\cite{Galperin1984,Galperin1985} or projective~\cite{Gates:1984nk,Karlhede:1984vr} superspace absolutely offer solutions, however these add an infinite number of auxiliary fields.  In this paper we offer an in-depth analysis of the Lagrangian symmetries generated by the central charges and internal symmetries of the algebra as a possible window into algebraic closure with a finite number of auxiliary fields.  To the knowledge of the authors, these symmetries have never been discussed in this detail; almost certainly not in the 4-D Majorana component notation that is used in this paper.  In short, we are trying to push the bounds of understanding further as to precisely how the algebra fails to close with a finite number of auxiliary fields.  Furthermore, this paper analyzes the central charges and internal symmetries, or lack thereof, of other SUSY systems embedded into the overarching $d=4$ ${\mathcal N}=4$ SUSY-YM system.


This paper is structured as follows.  We begin by showing how the Abelian $d=4$, ${\mathcal N} = 4$ super 
Yang-Mills (SUSY-YM) system can be made to split into the ${\mathcal N} = 2$ vector multiplet (VM), which closes, and the ${\mathcal N} = 2$ Fayet Hypermultiplet (FH) systems, which doesn't~\cite{Fayet:1975yi}.  Then we quote the main result: the recovery of many first and second order supersymmetries from the central charges and internal symmetries of this algebra. 

Unless otherwise specified throughout the document, our notation convention is as follows.  Capital Latin indices are euclidean and go from one to three:  $I,J,K,M,\dots = 1,2,3$.  Lower case Latin indices $i,j,k,m,\dots = 1,2$ are also Euclidean.  This is not to be confused with the spinor indices, which are the other half of the lower case latin alphabet $a,b,c,d,\dots = 1,2,3,4$, ranging from one to four.  Greek indices are four dimensional Minkowski space-time indices and go from zero to three: $\mu,\nu,\alpha,\beta,\dots = 0,1,2,3$.  Symmetrization and antisymmetrization are defined without normalization:
\begin{align}
   \Lambda_{(\mu\nu)} &= \Lambda_{\mu\nu} + \Lambda_{\nu\mu} \\
   \Lambda_{[\mu\nu]} &= \Lambda_{\mu\nu} - \Lambda_{\nu\mu} 
\end{align}

\section{Reduction of \texorpdfstring{${\mathcal N} = 4$}{{\mathcal N} = 4} SUSY-YM to \texorpdfstring{${\mathcal N} = 2$}{{\mathcal N} = 2} FH and VM}\label{sec:CCCVReduction}
In this section, the algebra for ${\mathcal N} = 4$ is laid out in component notation.  The Lagrangian is presented which is globally invariant to these transformations.  Next, the algebra is uncovered, which of course does not close.  Finally, it is shown how this algebra splits into both the ${\mathcal N} = 2$ FH and ${\mathcal N} = 2$ VM multiplets, the latter of which closes, the former which does not.  It is commented on how after reduction to the FH system, certain central charges and internal symmetries are removed from the algebra.  Of course, all central charges and internal symmetries are removed from the algebra under reduction to the ${\mathcal N} = 2$ VM multiplet.

\subsection{\texorpdfstring{${\mathcal N} = 4$}{{\mathcal N} = 4} Transformation Laws}\label{CCCVReductionNequals4TLaws}
The Lagrangian for the Abelian $d=4$, ${\mathcal N} = 4$ SUSY-YM system
\be\label{eq:SUSYYMLagrangian}\eqalign{
{\mathcal{L}} = &-\frac{1}{2}(\partial_{\mu}A^{J})(\partial^{\mu}A^{J}) -\frac{1}{2}(\partial_{\mu}B^{J})(\partial^{\mu}B^{J})\cr
&+i\frac{1}{2}(\gamma^{\mu})^{ab}\psi_{a}^{J}\partial_{\mu}\psi_{b}^{J} +\frac{1}{2}(F^{J})^{2}+\frac{1}{2}(G^{J})^{2}\cr
&-\frac{1}{4}F_{\mu\nu}F^{\mu\nu} +\frac{1}{2}i(\gamma^{\mu})^{cd}\lambda_{c}\partial_{\mu}\lambda_{d}+\frac{1}{2}{\rm d}^2
}\ee
is invariant with respect to the global supersymmetric transformations
\be\eqalign{
{\rm D}_a A^J ~=~&~ \psi_a^J  ~~~, \cr
{\rm D}_a B^J ~=~ &~i \, (\gamma^5){}_a{}^b \, \psi_b^J  ~~~, \cr
{\rm D}_a \psi_b^J ~=~ &~i\, (\gamma^\mu){}_{a \,b}\,  \partial_\mu A^J 
~-~ ~ (\gamma^5\gamma^\mu){}_{a \,b} \, \partial_\mu B^J ~\cr
&-~ ~i \, C_{a\, b} 
\,F^J  ~+~ ~ (\gamma^5){}_{ a \, b} G^J  ~~, \cr
{\rm D}_a F^J ~=~&~ (\gamma^\mu){}_a{}^b \, \partial_\mu \, \psi_b^J   ~~~, \cr
{\rm D}_a G^J ~=~ &~i \,(\gamma^5\gamma^\mu){}_a{}^b \, \partial_\mu \,  
\psi_b^J  ~~~.
} \label{eq:chi0specific}
\ee
\\
\be \eqalign{
{\rm D}_a \, A{}_{\mu} ~=~ & (\gamma_\mu){}_a {}^b \,  \l_b  ~~~, \cr
{\rm D}_a \l_b ~=~ &  - \, \, \fracm 12 (\sigma^{\mu\nu})_a{}_b \, F_{\mu\nu}
~+~  (\gamma^5){}_{a \,b} \,    {\rm d} ~~,  \cr
{\rm D}_a \, {\rm d} ~=~ & i \, (\gamma^5\gamma^\mu){}_a {}^b \, 
\,  \partial_\mu \l_b  ~~~. \cr
} \label{eq:V0specific}
\ee
\\
\be \eqalign{
{\rm D}_a^{I} A^J ~=~& \delta^{IJ}~ \l_a - \epsilon^{IJ}_{~~K} \psi_a^K~~~, \cr
{\rm D}_a^{I} B^J ~=~& ~i \, (\gamma^5){}_a{}^b \, [~\delta^{IJ}~ \l_b +\epsilon^{IJ}_{~~K} \psi_b^K~]  ~~~, \cr
{\rm D}_a^{I} \psi_b^J ~=~& \delta^{IJ}[~\, \, \fracm 12 (\sigma^{\mu\nu})_{ab} \,F_{\mu\nu}
~+~  (\gamma^5){}_{a \,b} \,    {\rm d}~]\cr
&-~\epsilon^{IJ}_{~~K}[~
~-i\, (\gamma^\mu){}_{a \,b}\,  \partial_\mu A^K 
~-~ (\gamma^5\gamma^\mu){}_{a \,b} \, \partial_\mu B^K ~\cr
&+~ i \, C_{a\, b} 
\,F^K  ~+~ (\gamma^5){}_{ a \, b} G^K~]  ~~, \cr
{\rm D}_a^{I} F^J ~=~& ~ (\gamma^\mu){}_a{}^b \, \partial_\mu \, [~\delta^{IJ}~ \l_b - \epsilon^{IJ}_{~~K} \psi_b^K~]   ~~~, \cr
{\rm D}_a^{I} G^J ~=~& ~i \,(\gamma^5\gamma^\mu){}_a{}^b \, \partial_\mu \,  
[~-\delta^{IJ}~ \l_b + \epsilon^{IJ}_{~~K} \psi_b^K~]  ~~~.
}\label{eq:chiIspecific}
\ee
\\
\be \eqalign{
{\rm D}_a^{I} \, A{}_{\mu} ~=~ & -(\gamma_\mu){}_a {}^b \,  \psi_b^{I}  ~~~, \cr
{\rm D}_a^{I} \l_b ~=~ & ~i\, (\gamma^\mu){}_{a \,b}\,  \partial_\mu A^I 
~-~ ~ (\gamma^5\gamma^\mu){}_{a \,b} \, \partial_\mu B^I ~\cr
&-~i \, C_{a\, b} 
\,F^I  ~-~ (\gamma^5){}_{ a \, b} G^I  ~~, \cr
{\rm D}_a^{I} \, {\rm d} ~=~& i \, (\gamma^5\gamma^\mu){}_a {}^b \, 
\,  \partial_\mu \psi_b^{I}  ~~~. \cr
} \label{eq:VIspecific}
\ee
where
\be
\sigma^{\mu\nu} = \frac{i}{2}[\gamma^{\mu},\gamma^{\nu}],~~~F_{\mu\nu} = \partial_{\mu}A_{\nu} - \partial_{\nu}A_{\mu}.
\ee
and our conventions for the gamma matrices are as in Appendix A of~\cite{Gates:2009me}.

These transformations are known as \emph{zeroth} order symmetries of the Lagrangian.  The main result of this paper will be the first and second order symmetries of the Lagrangian, and how they can be recovered from the algebra.
\subsection{Algebra}\label{CCCVReductionNequals4Algebra}
In this section, we will discover the central charges and internal symmetries of this algebra which will lead us to the Lagrangian symmetries in section~\ref{sec:ExtraSymmetries}.  Using the shorthand
\be\eqalign{
\chi = ( A^I, B^I, F^I, G^I, {\rm d}, \psi^J_c, \lambda_c) ,
}\ee

\noindent the algebra can be written
\be\label{eq:00termsspecific1} \eqalign{
 \{{\rm D}_a, {\rm D}_b\} \chi &=   2 i (\gamma^{\mu})_{ab}\partial_\mu \chi,~~~\{{\rm D}_a, {\rm D}_b\}A_\nu = 2 i (\gamma^{\mu})_{ab}F_{\mu\nu}
}\ee

\noindent and
\be\label{eq:IJtermsspecific1} \eqalign{
\{{\rm D}_a^I, {\rm D}_b^J\} A^K = & 2 i \delta^{IJ} ( \gamma^\mu)_{ab}  \partial_\mu A^K - 2 \epsilon^{IJK} (\gamma^5)_{ab} {\rm d} + \cr
&-2 Z^{IJKM}[i C_{ab} F^M + (\gamma^5)_{ab}G^M],  \cr
\{{\rm D}_a^I, {\rm D}_b^J\} B^K = & 2 i \delta^{IJ} ( \gamma^\mu)_{ab}  \partial_\mu B^K + 2~i \epsilon^{IJK} C_{ab} {\rm d},  \cr
\{{\rm D}_a^I, {\rm D}_b^J\} F^K = & 2 i \delta^{IJ} ( \gamma^\mu)_{ab}  \partial_\mu F^K + 2 \epsilon^{IJK} (\gamma^5\gamma^\mu)_{ab} \partial_\mu {\rm d} + \cr
&+ 2 Z^{IJKM}[ -i C_{ab} \square A^M + (\gamma^5\gamma^\mu)_{ab} \partial_\mu G^M]  \cr
\{{\rm D}_a^I, {\rm D}_b^J\} G^K = & 2 i \delta^{IJ} ( \gamma^\mu)_{ab}  \partial_\mu G^K - 2 \epsilon^{IJK} (\gamma^5\gamma^\mu)_{ab} \partial^\nu F_{\mu\nu} +\cr
 &- 2 Z^{IJKM}[(\gamma^5)_{ab} \square A^M  + (\gamma^5\gamma^\mu)_{ab} \partial_\mu F^M]\cr
}\ee

\be\label{eq:IJtermsspecific2} \eqalign{
\{{\rm D}_a^I, {\rm D}_b^J\} {\rm d} =& 2 i \delta^{IJ} ( \gamma^\mu)_{ab}  \partial_\mu {\rm d} + \cr
  &+ 2 \epsilon^{IJK} ((\gamma^5)_{ab} \square A^K - i  C_{ab} \square B^K + (\gamma^5\gamma^\mu)_{ab} \partial_\mu F^K) \cr
\{{\rm D}_a^I, {\rm D}_b^J\} A_\nu = &  2 i \delta^{IJ} ( \gamma^\mu)_{ab}  F_{\mu\nu} + \cr
  &+ 2 \epsilon^{IJK} (i C_{ab} \partial_\nu A^K + (\gamma^5)_{ab} \partial_\nu B^K - (\gamma^5\gamma_\nu)_{ab} G^K) \cr
\{{\rm D}_a^I, {\rm D}_b^J\} \lambda_c = &  2 i \delta^{IJ} ( \gamma^\mu)_{ab}  \partial_\mu \lambda_c + i \epsilon^{IJK} [-C_{ab} (\gamma^\mu)_c^{~d} + (\gamma^5)_{ab}(\gamma^5\gamma^\mu?)_c^d? + \cr &~~~~~~~~~~~~~~~~~~~~~~~~~~~~~~~~~+(\gamma^5\gamma^\nu?)_ab?(\gamma^5\gamma_\nu\gamma^\mu?)_c^d?] \partial_\mu \psi_d^K \cr
\{{\rm D}_a^I, {\rm D}_b^J\} \psi_c^K = &  2 i \delta^{IJ} ( \gamma^\mu)_{ab}  \partial_\mu \psi_c^K - i \epsilon^{IJK} [-C_{ab} (\gamma^\mu)_c^{~d} + (\gamma^5)_{ab}(\gamma^5\gamma^\mu?)_c^d? + \cr &~~~~~~~~~~~~~~~~~~~~~~~~~~~~~~~~~~~~~+(\gamma^5\gamma^\nu?)_ab?(\gamma^5\gamma_\nu\gamma^\mu?)_c^d?]\partial_\mu \lambda_d  + \cr  &~~~~~~~~~~~~~~~~~~- i Z^{IJKM}[C_{ab}(\gamma^\mu?)_c^d? + (\gamma^5)_{ab}(\gamma^5\gamma^\mu?)_c^d? +  \cr &~~~~~~~~~~~~~~~~~~~~~~~~~+(\gamma^5\gamma^\nu?)_{ab}?(\gamma^5\gamma_\nu\gamma^\mu?)_c^d?] \partial_\mu \psi_d^M  
}\ee 

\noindent and for the cross terms

\be\label{eq:crosstermsspecific1}\eqalign{
\{{\rm D}_a, {\rm D}_b^I\} A^J = & 2 i \epsilon^{IJK}C_{ab} F^K  \cr
\{{\rm D}_a, {\rm D}_b^I\} B^J = & 2 i \epsilon^{IJK}C_{ab} G^K \cr
\{{\rm D}_a, {\rm D}_b^I\} F^J = & 2 i \epsilon^{IJK} C_{ab} \square A^K \cr
\{{\rm D}_a, {\rm D}_b^I\} G^J = & 2 i \epsilon^{IJK}C_{ab} \square B^K \cr
\{{\rm D}_a, {\rm D}_b^I\}\lambda_c = &0
}\ee
\be\label{eq:crosstermsspecific2}\eqalign{
\{{\rm D}_a, {\rm D}_b^I\}{\rm d} = &0 \cr
\{{\rm D}_a, {\rm D}_b^I\}A_\nu = & 2 i C_{ab} \partial_\nu A^I - 2 (\gamma^5)_{ab} \partial_\nu B^I \cr
\{{\rm D}_a, {\rm D}_b^I\} \psi_c^J = & 2i \epsilon^{IJK}C_{ab}(\gamma^\mu?)_c^d? \partial_\mu \psi_d^K 
}\ee
where
\be
   Z^{IJKM} \equiv \delta^{IM}\delta^{JK} - \delta^{IK}\delta^{JM}
\ee

\subsubsection{Central Charges and Internal Symmetries}
We will use the notation $(A^J,F^K)$ to indicate, for instance, the presence of a non-zero term involving the field $F^K$ on the right hand side of the anti-commutator $\{ {\rm D}_a^I, {\rm D}_b^J\}A^K$ and vice-versa. In this notation, we list the following fields which are coupled through a central charge or internal symmetry:
\be\label{eq:SUYMCharges}
\left.\eqalign{
&(A^J,F^K),~~~(A^J,G^K),~~~(B^J,G^K),  \cr&(A^J,{\rm d}),~~~(B^J,{\rm d}),~~~(G^J, A_\mu), \cr
&(F^J,G^K),~~~(F^J,{\rm d}), \cr
&(\psi_a^J, \lambda_b),~~~ (\psi_a^J,\psi_b^K)
}
\right\}
\begin{array}{c}
       \mbox{fields coupled by a central charge} \\
       \mbox{or internal symmetry}
\end{array}
\ee 
In addition, the algebra couples the following fields through a $U(1)$ gauge symmetry
\be
   (A_\mu, A^K),~~~(A_\mu, B^K),~~~\mbox{fields coupled through a gauge symmetry}
\ee

 In section~\ref{sec:ExtraSymmetries}, we will show how these central charges and internal symmetries can be used to uncover several first and second order Lagrangian symmetries.
We note that this algebra is absent of central charges and internal symmetries between 
\be\left.\eqalign{
&(F^J,A_\mu),~~~(A_\mu,{\rm d}),~~~(B^J,F^K), \cr
&(B^J,A^K),~~~(G^J, {\rm d})
}\right\}
\begin{array}{c}
   \mbox{fields \emph{not} coupled through} \\
   \mbox{a central charge or internal symmetry}
\end{array}
\ee  

\subsection{Reduction to \texorpdfstring{${\mathcal N} = 2$}{{\mathcal N} = 2} Systems}
Before we fully investigate the first and second order Lagrangian symmetries, we will investigate how to split the ${\mathcal N} = 4$ system into the ${\mathcal N} = 2$ FH and VM systems.  When we do this, some of the central charges and internal symmetries vanish.  In fact, in the case of the ${\mathcal N} = 2$ VM system \emph{all} of these vanish, and the algebra has \emph{no} information on first and second order Lagrangian symmetries.  This is of course because the ${\mathcal N} = 2$ VM algebra closes.

First making the following definitions
\be\label{eq:Dtilde}
   \tilde{{\rm D}}_a^1 \equiv {\rm D}_a,~~~\tilde{{\rm D}}_a^2 \equiv {\rm D}_a^1
\ee
where $i = 1,2$ labels the two supersymmetries of the embedded systems, we next make field redefinitions to manifest the embedded systems.  The embedded ${\mathcal N} = 2$ VM system is composed of half of the fields of the ${\mathcal N} = 4$ system:
\be\eqalign{
    A &\equiv A^1,~~~ B \equiv B^1,~~~ F\equiv F^1,~~~G\equiv G^1, \cr
  &A_\mu,~~~{\rm d},~~~\zeta_a^1 \equiv \psi_a^1,~~~\zeta_a^2 \equiv \lambda_a
}\ee
and the embedded ${\mathcal N} = 2$ FH system is composed of the other half
\be\label{eq:CCReduction}\eqalign{
   \tilde{A}^1 &\equiv A^2,~~~\tilde{A}^2 \equiv A^3,~~~\tilde{B}^1 \equiv B^2,~~~\tilde{B}^2 \equiv B^3, \cr
   \tilde{F}^1 &\equiv F^2,~~~\tilde{F}^2 \equiv F^3,~~~\tilde{G}^1 \equiv G^2,~~~\tilde{G}^2 \equiv G^3, \cr
   \tilde{\psi}_a^1 &\equiv \psi_a^2,~~~\tilde{\psi}_a^2 \equiv \psi_a^3
}\ee

\subsubsection{Reduction to \texorpdfstring{${\mathcal N} = 2$}{{\mathcal N} = 2} VM}
The resulting ${\mathcal N} = 2$ VM algebra is
\be\eqalign{
  \tilde{{\rm D}}_a^i A &= \zeta_a^i, \cr
  \tilde{{\rm D}}_a^i B &= i (\gamma^5)_a^{~b} \zeta_b^i, \cr
  \tilde{{\rm D}}_a^i F &= (\gamma^{\mu})_a^{~b} \partial_{\mu}\zeta_b^i,\cr
  \tilde{{\rm D}}_a^i G &= i (\sigma^3)^{ij}(\gamma^5\gamma^\mu)_a^{~b}\partial_\mu \zeta_b^j,\cr
  \tilde{{\rm D}}_a^i A_\mu &= i(\sigma^2)^{ij}(\gamma_\mu)_a^{~b}\zeta_b^j,\cr
  \tilde{{\rm D}}_a^i {\rm d} &= i (\sigma^1)^{ij}(\gamma^5 \gamma^{\mu})_a^{~b} \partial_{\mu}\zeta_b^j, \cr
  \tilde{{\rm D}}_a^i \zeta_b^j &= \delta^{ij}(i(\gamma^{\mu})_{ab}\partial_{\mu} A - (\gamma^5\gamma^{\mu})_{ab} \partial_{\mu} B - i C_{ab} F) + (\sigma^3)^{ij}(\gamma^5)_{ab} G+\cr
  &~~~- i(\sigma^2)^{ij} \frac{1}{2}(\sigma^{\mu\nu})_{ab}F_{\mu\nu} + (\sigma^1)^{ij}(\gamma^5)_{ab}{\rm d},
}\ee

\noindent where 
\be\eqalign{
              (\sigma^1)^{ij} &= \left(\begin{array}{l l}
                              0 & 1 \\
                              1 & 0 
                   \end{array}
              \right),~~~
              (\sigma^2)^{ij} = \left(\begin{array}{l l}
                              0 & -i \\
                              i & 0 
                   \end{array}
              \right),~~~
              (\sigma^3)^{ij} = \left(\begin{array}{l l}
                              1 & 0 \\
                              0 & -1 
                   \end{array}
              \right),
}\ee

\noindent

\noindent and
\be\eqalign{
 \zeta_b^1 &= \psi_b,~~~\zeta_b^2 = \lambda_b.
}\ee

The algebra reduces to
\begin{align}
   \{\tilde{{\rm D}}_a^i, \tilde{{\rm D}}_b^j\} {\mathcal V} &= 2 i \delta^{ij} (\gamma^{\mu})_{ab} \partial_\mu {\mathcal V}
\\
\label{eq:CVAvAlgebra}\{\tilde{{\rm D}}_a^i, \tilde{{\rm D}}_b^j\}A_{\nu} &= 2 i \delta^{ij}(\gamma^{\mu})_{ab} F_{\mu\nu}+ i(\sigma^2)^{ij}(2 i C_{ab} \partial_{\nu}A - 2(\gamma^5)_{ab} \partial_{\nu}B).
\end{align}

\noindent where
\be\eqalign{
   {\mathcal V} &= (A, B, F, G, {\rm d}, \psi_c, \lambda_c). 
}\ee
So this algebra closes up to gauge transformations and all the central charges and internal symmetries from the overarching ${\mathcal N} = 4$ algebra have vanished, aside from the $U(1)$ gauge symmetries.  The algebra, therefore, contains no information on extra symmetries of the Lagrangian.  

\subsubsection{Reduction to \texorpdfstring{${\mathcal N}=2$}{N=2} FH}
The transformation laws for the embedded ${\mathcal N} = 2$ FH system are
\be\eqalign{
\tilde{{\rm D}}_a^i\tilde{A}^j &= \delta^{ij}\tilde{\psi}_a^1 + i (\sigma^2)^{ij}\tilde{\psi}_a^2, \cr
\tilde{{\rm D}}_a^i\tilde{B}^j &= i (\gamma^5)_a^{~b}~[~(\sigma^3)^{ij}\tilde{\psi}_b^1 + (\sigma^1)^{ij}\tilde{\psi}_b^2~], \cr
\tilde{{\rm D}}_a^i\tilde{F}^j &=  (\gamma^\mu)_a^{~b}\partial_{\mu}[~\delta^{ij}\tilde{\psi}_b^1 + i(\sigma^2)^{ij}\tilde{\psi}_b^2~], \cr
\tilde{{\rm D}}_a^i\tilde{G}^j &= i (\gamma^5\gamma^\mu)_a^{~b}\partial_{\mu}[~(\sigma^3)^{ij}\tilde{\psi}_b^1 + (\sigma^1)^{ij}\tilde{\psi}_b^2~], \cr
\tilde{{\rm D}}_a^i\tilde{\psi}_b^1 &= i(\gamma^\mu)_{ab}\partial_\mu \tilde{A}^i - i C_{ab}\tilde{F}^i + (\sigma^3)^{ij}[(\gamma^5)_{ab}\tilde{G}^j - (\gamma^5\gamma^\mu)_{ab}\partial_\mu\tilde{B}^j], \cr
\tilde{{\rm D}}_a^i\tilde{\psi}_b^2 &= (\sigma^2)^{ij}[-(\gamma^\mu)_{ab}\partial_\mu \tilde{A}^j +  C_{ab}\tilde{F}^j] + (\sigma^1)^{ij}[(\gamma^5)_{ab}\tilde{G}^j - (\gamma^5\gamma^\mu)_{ab}\partial_\mu\tilde{B}^j]
}\ee
with algebra
\be\label{eq:CCSpecificAlgebra}\eqalign{
\{\tilde{{\rm D}}_a^i, \tilde{{\rm D}}_b^j\} \tilde{A}^k &= 2 i \delta^{ij}(\gamma^\mu)_{ab}\partial_\mu \tilde{A}^k - 2 i \tilde{Z}^{ijkm} C_{ab} \tilde{F}^m, \cr
\{\tilde{{\rm D}}_a^i, \tilde{{\rm D}}_b^j\} \tilde{B}^k &= 2 i \delta^{ij}(\gamma^\mu)_{ab}\partial_\mu \tilde{B}^k - 2 i \tilde{Z}^{ijkm} C_{ab} \tilde{G}^m, \cr
\{\tilde{{\rm D}}_a^i, \tilde{{\rm D}}_b^j\} \tilde{F}^k&= 2 i \delta^{ij}(\gamma^\mu)_{ab}\partial_\mu \tilde{F}^k - 2 i \tilde{Z}^{ijkm} C_{ab} \square \tilde{A}^m, \cr
\{\tilde{{\rm D}}_a^i, \tilde{{\rm D}}_b^j\} \tilde{G}^k&= 2 i \delta^{ij}(\gamma^\mu)_{ab}\partial_\mu \tilde{G}^k - 2 i \tilde{Z}^{ijkm} C_{ab} \square \tilde{B}^m, \cr
\{\tilde{{\rm D}}_a^i, \tilde{{\rm D}}_b^j\}\tilde{\psi}_c^k &= 2 i \delta^{ij}(\gamma^\mu)_{ab}\partial_\mu \tilde{\psi}_c^1 - 2 i \tilde{Z}^{ijkm}C_{ab}(\gamma^\mu)_c^{~d}\partial_\mu \tilde{\psi}_d^m
}\ee
where
\be
   \tilde{Z}^{ijkm} \equiv \delta^{im}\delta^{jk} - \delta^{ik}\delta^{jm},~~~i,j,k,m = 1,2.
\ee

So only the couplings $(A^J, G^K)$ and $(F^J, G^K)$ have vanished from the overarching ${\mathcal N} = 4$ theory.  Couplings still remain between $(\tilde{A}^j, \tilde{F}^k)$ and $(\tilde{B}^j, \tilde{G}^k)$ and $(\tilde{\psi}_a^i, \tilde{\psi}_b^j)$.

\section{Extra Symmetries of the Lagrangian}\label{sec:ExtraSymmetries}
Here begins the main result of the paper.   We list the first order bosonic symmetries unveiled directly by the central charges and internal symmetries.  We next calculate from these symmetries first order fermionic and second order bosonic symmetries of the Lagrangian.  We will notice that more symmetries exist which are not revealed by this algebra. We discuss the ${\mathcal N} = 4$ SUSY-YM system first and the ${\mathcal N} = 2$ FH system last.

\subsection{First Order Bosonic Symmetries}\label{sec:1stbosonsa}
Contracting the coupling from the anticommutator on $A^J$ and $F^J$ in Eq.~(\ref{eq:crosstermsspecific1}) with the Grassmann spinors $\varepsilon^a$ and $\chi^b_I$ results in the first order bosonic symmetry of the Lagrangian
\be
  \delta_{BS3a}^{(1)} \left(\begin{array}{c}
                               A^J \\  F^J 
                          \end{array}\right) \equiv \frac{\varepsilon^a \chi_I^b}{2 i} \{{\rm D}_a,{\rm D}_b^I \} \left(\begin{array}{c}
                               A^J \\  F^J 
                          \end{array}\right) = \varepsilon^a\chi^b_I\epsilon^{IJK}C_{ab}  \left(\begin{array}{c}
                    F^K \\  \square A^K   
\end{array}\right).
\ee

Interestingly, contracting the coupling from the anticommutators on $A^K$ and $F^K$ in Eq.~(\ref{eq:IJtermsspecific1}) with the Grassmann spinors $\varepsilon^a_I$ and $\chi^b_J$ results in a very similar first order bosonic symmetry of the Lagrangian
\be
  \delta_{BS3b}^{(1)} \left(\begin{array}{c}
                               A^K \\  F^K 
                          \end{array}\right) \equiv  \varepsilon^{a}_I \chi^b_J Z^{IJKM}C_{ab}\left(\begin{array}{c}
                    F^M \\  \square A^M   
\end{array}\right).
\ee

In fact, these two symmetries are identical, and we can define them succinctly as:
\be
  \delta_{BS3}^{(1)}(T) \left(\begin{array}{c}
                               A^K \\  F^K 
                          \end{array}\right) \equiv  T^{KM}\left(\begin{array}{c}
                    F^M \\  \square A^M   
\end{array}\right).
\ee
where
\be
T^{KM} \equiv \left\{\begin{array}{l}
                  \varepsilon^{a}_I \chi^b_J Z^{IJKM}C_{ab} \\
                         \mbox{or} \\
                  \varepsilon^a \chi^b_J \epsilon^{JKM} C_{ab}
             \end{array}
             \right.
\ee
The unique first order bosonic symmetries revealed by all the central charges and internal symmetries in this way are:
\be\eqalign{
  \delta_{BS1}^{(1)}(P) \left(\begin{array}{c}
                             A^K \\ {\rm d} 
                          \end{array}\right) \equiv  P^K  \left(\begin{array}{c}
                   -{\rm d}  \\ \square A^K  
\end{array}\right),~~~ \delta_{BS2}^{(1)}(Q) \left(\begin{array}{c}
                             B^K \\ {\rm d} 
                          \end{array}\right) \equiv  Q^K  \left(\begin{array}{c}
                   -{\rm d}  \\ \square B^K  
\end{array}\right)
}\ee
\be
  \delta_{BS3}^{(1)}(T) \left(\begin{array}{c}
                               A^K \\  F^K 
                          \end{array}\right) \equiv  T^{KM}  \left(\begin{array}{c}
                    F^M \\  \square A^M   
\end{array}\right)
\ee
\be
  \delta_{BS4}^{(1)}(T) \left(\begin{array}{c}
                             B^K \\  G^K  
                          \end{array}\right) \equiv  T^{KM}   \left(\begin{array}{c}
                    G^M \\ \square B^M  
\end{array}\right),~~~  \delta_{BS5}^{(1)}(W) \left(\begin{array}{c}
                           A^J   \\  G^J
                          \end{array}\right) \equiv  W^{JK} \left(\begin{array}{c}
                    G^K \\ \square A^K 
\end{array}\right)
\ee
\be
  \delta_{BS6}^{(1)}(V) \left(\begin{array}{c}
                             F^J \\  G^J
                          \end{array}\right) \equiv (V^\mu)^{JK}\partial_\mu  \left(\begin{array}{c}
                  G^K   \\  -F^K 
\end{array}\right)
\ee
\be
  \delta_{BS7}^{(1)}(U) \left(\begin{array}{c}
                             F^K \\  {\rm d}
                          \end{array}\right) \equiv  (U^\mu)^K \partial_\mu   \left(\begin{array}{c}
                    {\rm d} \\  F^K
\end{array}\right)
\ee
\be
  \delta_{BS8}^{(1)}(U) \left(\begin{array}{c}
                              G^K \\  A_\nu
                          \end{array}\right) \equiv (U^\mu)^K  \left(\begin{array}{c}
                  \partial^\nu F_{\mu\nu}   \\  \eta_{\mu\nu} G^K
\end{array}\right)
\ee
\be
  \delta_{BS9}^{(1)}(Q) \left(\begin{array}{c}
                           \lambda_c   \\  \psi_c^K
                          \end{array}\right) \equiv   Q^K (\gamma^\mu)_c^{~d}\partial_\mu \left(\begin{array}{c}
                  \psi_d^K   \\  -\lambda_d
\end{array}\right)
\ee
\be
  \delta_{BS10}^{(1)}(U) \left(\begin{array}{c}
                           \lambda_c   \\  \psi_c^K
                          \end{array}\right) \equiv   (U^\nu)^K (\gamma^5\gamma_\nu\gamma^\mu)_c^{~d}\partial_\mu \left(\begin{array}{c}
                  \psi_d^K   \\  -\lambda_d
\end{array}\right)
\ee
\be
  \delta_{BS11}^{(1)}(P) \left(\begin{array}{c}
                           \lambda_c   \\  \psi_c^K
                          \end{array}\right) \equiv   P^K (\gamma^5\gamma^\mu)_c^{~d}\partial_\mu \left(\begin{array}{c}
                  \psi_d^K   \\  -\lambda_d
\end{array}\right)
\ee
\begin{align}
  \delta_{BS12}^{(1)}(W) \psi^K_c &\equiv W^{KM}(\gamma^5\gamma^\mu)_c^{~d}\partial_\mu\psi_d^M \\
\delta_{BS13}^{(1)}(V) \psi^K_c &\equiv (V^\nu)^{KM}(\gamma^5\gamma_\nu\gamma^\mu)_c^{~d} \partial_\mu\psi_d^M\\
\delta_{BS14}^{(1)}(T) \psi^K_c &\equiv T^{KM}(\gamma^\mu)_c^{~d}\partial_\mu\psi_d^M
\end{align}

\noindent along with the $U(1)$ gauge symmetries
\be\eqalign{
  \delta_G A_\nu \equiv Q^K \partial_\nu A^K&,~~~\delta_G A_\nu \equiv P^K \partial_\nu B^K, \cr
\delta A_\nu \equiv \varepsilon^a \chi^b_I C_{ab} \partial_\nu A^I&,~~~\delta A_\nu \equiv \varepsilon^a \chi^b_I(\gamma^5)_{ab} \partial_\nu B^I
}\ee

\noindent where
\be\eqalign{
   P^K \equiv \varepsilon^{a}_I \chi^b_J \epsilon^{IJK}(\gamma^5)_{ab} 
&~~~Q^K \equiv  \varepsilon^{a}_I 
\chi^b_J \epsilon^{IJK}C_{ab}, \cr
   T^{KM} \equiv \left\{\begin{array}{l}
                  \varepsilon^{a}_I \chi^b_J Z^{IJKM}C_{ab} \\
                         \mbox{or} \\
                  \varepsilon^a \chi^b_J \epsilon^{JKM} C_{ab}
             \end{array}
             \right.,&~~~(U^\mu)^K \equiv \varepsilon^{a}_I \chi^b_J \epsilon^{IJK}(\gamma^5 \gamma^\mu)_{ab}, \cr
W^{KM} \equiv \left\{\begin{array}{l} \varepsilon^a \chi^b_J \epsilon^{JKM}(\gamma^5)_{ab} \\
                 \mbox{or} \\
                   \varepsilon^a_I \chi^b_J Z^{IJKM} (\gamma^5)_{ab}       
                 \end{array}
                \right.
,&~~~(V^{\mu})^{KM} \equiv \left\{\begin{array}{l}
                          \varepsilon^a \chi^b_J \epsilon^{JKM} (\gamma^5\gamma^\mu)_{ab}
                          \\ \mbox{or} \\
                             \varepsilon^a_I \chi^b_J Z^{IJKM} (\gamma^5 \gamma^\mu)_{ab}
                  \end{array}
                  \right.
}\ee
The following identity proves useful in directly verifying these as Lagrangian symmetries:
\be
  (\gamma^5\gamma^{(\mu}\gamma_{\alpha}\gamma^{\nu)})^{(ab)} = 0
\ee
\noindent where $(~~)$ denotes symmetrization, i.e., $(\gamma^\mu)^{(ab)} = (\gamma^\mu)^{ab} + (\gamma^\mu)^{ba}$. 

It is interesting to note here that because of the absence of $B^J$ to $F^J$ coupling in the algebra, this method fails to uncover the first order bosonic symmetry of the Lagrangian
\be
  \delta_{BS15}^{(1)}(T) \left(\begin{array}{c}
                               B^K \\  F^K 
                          \end{array}\right) \equiv  T^{KM}  \left(\begin{array}{c}
                    F^M \\  \square B^M   
\end{array}\right)
\ee
In addition, Lagrangian symmetries such as
\be
  \delta_{BS16}^{(1)}(U) \left(\begin{array}{c}
                             G^K \\  {\rm d}
                          \end{array}\right) \equiv  (U^\mu)^K \partial_\mu   \left(\begin{array}{c}
                    {\rm d} \\  G^K
\end{array}\right)
\ee
\be
  \delta_{BS17}^{(1)}(U) \left(\begin{array}{c}
                              F^K \\  A_\nu
                          \end{array}\right) \equiv (U^\mu)^K  \left(\begin{array}{c}
                  \partial^\nu F_{\mu\nu}   \\  \eta_{\mu\nu} F^K
\end{array}\right)
\ee
also are not manifest in the algebra.  We will leave all such symmetries not manifested by the algebra out of the remaining calculations of second order bosonic and first order fermionic symmetries, as we are investigating how the absence of these symmetries fails to uncover further symmetries down the line.

\subsection{Second Order Bosonic Symmetries}\label{sec:2ndbosonsa}
By taking the commutators of each of the first order bosonic symmetries with each other, we reveal second order bosonic symmetries.  This procedure will sometimes lead to redundant symmetries as in
\be\eqalign{ 
\delta^{(2)}_{BS1a}(P_1,P_2)A^K &\equiv [\delta_{BS1}^{(1)}(P_1), \delta_{BS1}^{(1)}(P_2)] A^K = \Lambda_{1,1}^{KJ}(P_1,P_2)\square A^J  \cr
\delta_{BS1b}^{(2)}(T_1,T_2)A^K &\equiv [\delta_{BS3}^{(1)}(T_1),\delta_{BS3}^{(1)}(T_2)] A^K = \Lambda_{3,3}^{JK}(T_1, T_2) \square A^J\cr
\delta_{BS1c}^{(2)}(W_1,W_2) A^J &\equiv [ \delta_{BS5}^{(1)}(W_1), \delta_{BS5}^{(1)}(W_2) ] A^J= \Lambda_{5,5}^{IJ}(W_1,W_2)\square A^I
}\ee 
where
\be\eqalign{
\Lambda_{1,1}^{KJ}(P_1, P_2) &\equiv P_{[1}^{K} P_{2]}^{J},\cr
\Lambda_{3,3}^{KJ}(T_1,T_2) &\equiv T_{[1}^{KM}T_{2]}^{MJ}, \cr
  \Lambda_{5,5}^{IJ}(W_1,W_2) &\equiv W_{[1}^{KI} W_{2]}^{JK}
}\ee
We can succinctly write these three redundant symmetries as one
\be
\delta_{BS1}^{(2)}(\Lambda_1) A^K \equiv \Lambda_1^{[KJ]} \square A^J\ee 
where $(\Lambda_1)^{KJ}$ is an arbitrary $3 \times 3$ matrix and $[~]$ denotes antisymmetrization: 
\be (\Lambda_1)^{[KJ]} = (\Lambda_1)^{KJ} - (\Lambda_1)^{JK}.\ee  In Appendix~\ref{app:SecondOrderBosonicSymmetriesExplicit}, we list all the second order bosonic symmetries which are calculated in this way, including their redundancies.  Here, we list only the unique symmetries, written in terms of the arbitrary matrices $(\Lambda_1)^{KJ}$, $(\Lambda_2^{\mu\nu})^{JK}$, $(\Lambda_3)^{IJ}$, $(\Lambda_4^\mu)^K$, $\Lambda_5^K$, and $(\Lambda_6^{\mu\nu})^{J}$:
\be\eqalign{
~~~~~~~~~~~~~~~~~~~\delta^{(2)}_{BS1}(\Lambda_1)A^K \equiv  \Lambda_1^{[KJ]}\square A^J,~~~ &\delta_{BS2}^{(2)}(\Lambda_1) B^K \equiv \Lambda_1^{[KJ]} \square B^J \cr
\delta_{BS3}^{(2)}(\Lambda_1)F^K \equiv \Lambda_1^{[KJ]} \square F^J,~~~
&\delta_{BS4}^{(2)}(\Lambda_1)G^K \equiv \Lambda_1^{[KJ]} \square G^K \cr
  \delta_{BS5}^{(2)}(\Lambda_2) F^J \equiv (\Lambda_2^{\mu\nu})^{[IJ]}\partial_\mu\partial_\nu F^I,~~~
  &\delta_{BS6}^{(2)}(\Lambda_2) G^J \equiv (\Lambda_2^{\mu\nu})^{[IJ]} \partial_\mu\partial_\nu G^I \cr
\delta_{BS7}^{(2)}(\Lambda_2) A_\nu \equiv \eta_{\nu\beta}&(\Lambda_2^{[\mu\beta]})^{JJ}\partial^\alpha F_{\mu\alpha} 
}\ee
\be
\delta_{BS8}^{(2)}(\Lambda_1) \left(\begin{array}{l}
                                 A^K \\
                                 B^K
                               \end{array}\right) \equiv \Lambda_1^{IJ} \left(\begin{array}{l}
           \delta^{IK}\square B^J \\
           -\delta^{JK}\square A^I
           \end{array}
           \right)
\ee

\be
\delta_{BS9}^{(2)}(\Lambda_3) \left(\begin{array}{l}
                                  A^K \\ F^K
                               \end{array}
                               \right) \equiv  (\Lambda_3^\mu)^{IJ} \left(\begin{array}{l}
       \delta^{IK} \partial_\mu F^J \\
   \delta^{JK}\partial_\mu\square A^I
                               \end{array}
                               \right)
\ee

\be\eqalign{
\delta_{BS10}^{(2)}(\Lambda_3) \left(\begin{array}{l}
              B^K \\ F^K \end{array}\right) 
             &\equiv  (\Lambda_{3}^\mu)^{IJ}\left(\begin{array}{l}\delta^{IK}\partial_\mu F^J \\  \delta^{KJ}\partial_{\mu} \square B^I\end{array}\right)
}\ee
\be
  \delta_{BS11}^{(2)}(\Lambda_3) \left(\begin{array}{ll}
                                 A^J   \\   G^J
                                 \end{array}\right)
     \equiv (\Lambda_{3}^\mu)^{IK}\left(\begin{array}{ll}
                      \delta^{IJ}\partial_\mu G^K              \\   \delta^{JK}\partial_\mu\square A^I
                                 \end{array}\right)
\ee
\be
\delta_{BS12}^{(2)}(\Lambda_4) \left(\begin{array}{l}
                          A^K \\{\rm d}
                        \end{array}\right)  \equiv (\Lambda_{4}^\mu)^K\left(\begin{array}{l}\partial_\mu {\rm d}  \\ \partial_\mu \square A^K
\end{array}\right)
\ee
\be
  \delta_{BS13}^{(2)}(\Lambda_4) \left(\begin{array}{ll}
                                  A^J  \\  A_\nu
                                 \end{array}\right)
            \equiv (\Lambda_{4}^\mu)^J\left(\begin{array}{ll}
                       \partial^\nu F_{\mu\nu}             \\
       \eta_{\mu\nu}\square A^J                          \end{array}\right)
\ee
\be
  \delta_{BS14}^{(2)}(\Lambda_4) \left(\begin{array}{ll}
                                  B^J  \\  A_\nu
                                 \end{array}\right)
            \equiv (\Lambda_{4}^\mu)^J\left(\begin{array}{ll}
                       \partial^\nu F_{\mu\nu}             \\
       \eta_{\mu\nu}\square B^J                          \end{array}\right)
\ee
\be
\delta_{BS15}^{(2)}(\Lambda_5) \left(\begin{array}{l}F^K \\{\rm d} \end{array}\right)
\equiv \Lambda_5^K\left(\begin{array}{l} \square  {\rm d} \\ -\square  F^K
\end{array}\right)
\ee
\be
\delta_{BS16}^{(2)}(\Lambda_5) \left(\begin{array}{l}G^K \\ {\rm d} \end{array}\right)\equiv \Lambda_{5}^K\left(\begin{array}{l} \square {\rm d} \\ - \square G^K \end{array}\right)
\ee
\be
  \delta_{BS17}^{(2)}(\Lambda_6) \left(\begin{array}{ll}
                                    G^J \\   {\rm d}
                                 \end{array}\right)
            \equiv  (\Lambda_{6}^{\mu\nu})^J\left(\begin{array}{ll}
                         \partial_\mu\partial_\nu {\rm d}         \\ -\partial_\mu\partial_\nu G^J
                                 \end{array}\right)
\ee
\be
  \delta_{BS18}^{(2)}(\Lambda_1) \left(\begin{array}{ll}
                                    F^J\\  G^J
                                 \end{array}\right)
            \equiv \Lambda_{1}^{IK}\left(\begin{array}{ll}
                       \delta^{IJ}\square G^K             \\  -\square \delta^{JK}F^I
                                 \end{array}\right)
\ee
\be
  \delta_{BS19}^{(2)}(\Lambda_6) \left(\begin{array}{ll}
                                 F^J   \\  A_\alpha
                                 \end{array}\right)
            \equiv  (\Lambda_{6}^{\mu\nu})^J(U,V)\left(\begin{array}{ll}
                           \partial_\nu\partial^\alpha F_{\mu\alpha}        \\  -\eta_{\mu\alpha}\partial_\nu F^J
                                 \end{array}\right)
\ee
and
\begin{align}
\delta_{BS20}^{(2)}(\Lambda_1)\psi_c^K \equiv & \Lambda_{1}^{[JK]} \square \psi_c^J \\
\delta_{BS21}^{(2)}(\Lambda_2)\psi_c^K \equiv & [(\Lambda_{2}^{\rho\sigma})^{KJ}-(\Lambda_{2}^{\sigma\rho})^{JK}] (\gamma_\rho\gamma^\mu\gamma_\sigma\gamma^\nu)_c^{~d} \partial_\mu\partial_\nu\psi_d^J \\
   \delta_{BS22}^{(2)}(\Lambda_2) \lambda_c \equiv &(\Lambda_{2}^{[\mu\nu]})^{KK} (\gamma_\mu\gamma^\alpha\gamma_\nu\gamma^\beta?)_c^d? \partial_\alpha\partial_\beta \lambda_d, \\
\delta_{BS23}^{(2)}(\Lambda_3)\psi_c^K \equiv  &(\Lambda_{3}^\mu)^{[JK]}(\gamma^5\gamma_\mu)_c^{~d}\square\psi_d^J \\
   \delta_{BS24}^{(2)}(\Lambda_3) \lambda_c \equiv  &(\Lambda_{3}^{\nu})^{KK}(\gamma^5\gamma^\mu?)_c^d? \partial_\mu\partial_\nu \lambda_d\\
  \delta_{BS25}^{(2)}(\Lambda_3)\psi_c^K  \equiv &(\Lambda_{3}^\mu)^{KJ} (\gamma^5\gamma^\nu?)_c^d? \partial_\mu\partial_\nu \psi_d^J\\
\delta_{BS26}^{(2)}(\Lambda_3)\psi_c^K \equiv & (\Lambda_{3}^\mu)^{[JK]}(\gamma_\mu)_c^{~d}\square \psi_d^J + 2(\Lambda_{3}^\mu)^{KJ}(\gamma^\nu)_{c}^{~d}\partial_\mu \partial_\nu \psi_d^J \\
   \delta_{BS27}^{(2)}(\Lambda_3) \lambda_c \equiv  &  (\Lambda_{3}^{\mu})^{KK} (\gamma^\nu?)_c^d? \partial_\mu\partial_\nu \lambda_d
\\
\delta_{BS28}^{(2)}(\Lambda_1)\psi_c^K \equiv & \Lambda_{1}^{KJ}(\gamma^5)_c^{~d}\square \psi_d^J    \\
   \delta_{BS29}^{(2)}(\Lambda_1) \lambda_c \equiv &  \Lambda_{1}^{KK} (\gamma^5?)_c^d? \square \lambda_d
\end{align}
\be
\delta_{BS30}^{(2)}(\Lambda_5) \left(\begin{array}{l}\lambda_c \\ \psi_c^K \end{array}\right) \equiv \Lambda_{5}^K(\gamma^5)_c^{~d}\left(\begin{array}{l}
\square \psi_d^K \\ \square \lambda_d \end{array}\right)
\ee
\be
\delta_{BS31}^{(2)}(\Lambda_5) \left(\begin{array}{l} \lambda_c \\ \psi_c^K \end{array}\right) \equiv \Lambda_{5}^{K} \left(\begin{array}{l} \square \psi^K_c \\  -\square \lambda_c \end{array}\right) 
\ee
\be\eqalign{
\delta_{BS32}^{(2)}(\Lambda_4) \left(\begin{array}{l} \lambda_c \\  \psi_c^K \end{array}\right) &\equiv (\Lambda_{4}^\alpha)^K\left(\begin{array}{l} (\gamma^5\gamma^\nu\gamma_\alpha\gamma^\nu)_c^{~d} \partial_\mu\partial_\nu \psi_d^K \\  
(\gamma^5\gamma_\alpha)_c^{~d}\square \lambda_d \end{array}\right)
}\ee
\be\eqalign{
\delta_{BS33}^{(2)} (\Lambda_4) \left( \begin{array}{l} \lambda_c \\ \psi_c^K \end{array} \right) &\equiv (\Lambda_{4}^{\mu})^K \left( \begin{array}{l} (\gamma^5\gamma_\mu?)_c^d? \square \psi_d^K \\ (\gamma^5\gamma^\alpha\gamma_\mu\gamma^\beta?)_c^d?\partial_\alpha\partial_\beta \lambda_d \end{array} \right)
}\ee
\be\eqalign{
\delta_{BS34}^{(2)} (\Lambda_4) \left( \begin{array}{l} \lambda_c \\ \psi_c^K \end{array} \right) &\equiv (\Lambda_{4}^\mu)^K \left( \begin{array}{l} (\gamma_\mu?)_c^d? \square \psi_d^K \\ (\gamma^\nu\gamma_\mu\gamma^\alpha?)_c^d?\partial_\nu\partial_\alpha \lambda_d \end{array} \right) 
}\ee
\be\eqalign{
\delta_{BS35}^{(2)} (\Lambda_4) \left( \begin{array}{l} \lambda_c \\ \psi_c^K \end{array} \right) &\equiv (\Lambda_{4}^\mu)^K \left( \begin{array}{l}  (\gamma^\alpha \gamma_\mu \gamma^\beta?)_c^d? \partial_\alpha\partial_\beta\psi_d^K \\ (\gamma_\mu?)_c^d?\square \lambda_d \end{array} \right),
}\ee
\be\eqalign{
\delta_{BS36}^{(2)} (\Lambda_6) \left( \begin{array}{l} \lambda_c \\ \psi_c^K \end{array} \right) &\equiv (\Lambda_{6}^{\mu\nu})^K \left( \begin{array}{l} (\gamma_\nu\gamma^\alpha\gamma_\mu\gamma^\beta?)_c^d? \partial_\alpha\partial_\beta \psi_d^K \\ -(\gamma_\mu\gamma^\alpha\gamma_\nu\gamma^\beta?)_c^d?\partial_\alpha\partial_\beta \lambda_d \end{array} \right),
}\ee

This analysis seems to not miss any second order bosonic symmetries which act on the fermions $\lambda_a$ and $\psi^J_a$.  However, the missing first order bosonic symmetries alluded to previously which act on the bosons clearly manifest themselves here in missing second order bosonic symmetries.  Basically, as the fields $A^J$ and $B^J$ enter the Lagrangian in the same way, they should have the same first and second order symmetries.  The same should hold for $F^J$ and $G^J$.  But clearly since, for example, the algebra is \emph{not} symmetric between exchange of $A^J \leftrightarrow B^J$ or $F^J \leftrightarrow G^J$, Lagrangian symmetries involving these field pairs will be missed when generated from the algebra in the manner presented here. 

\subsection{First Order Fermionic Symmetries}\label{sec:1stfermionsa}
Analogous to how we found the second order bosonic symmetries, we can uncover first order \emph{fermionic} symmetries through calculations such as:
\be
\delta_{FS19}^{(1)}(P) \left(
\begin{array}{c}
                            {\rm d} \\ \psi_b^J \end{array}\right) \equiv
-\varepsilon^a [ {\rm D}_a, \delta_{BS1}^{(1)}(P) ] \left(\begin{array}{c}
                            {\rm d} \\ \psi_b^J \end{array}\right) =
    \varepsilon^a P^J \left(\begin{array}{c}
                           \square\psi_a^J \\  i (\gamma^\mu)_{ab} \partial_\mu {\rm d} \end{array}\right)
\ee
All such possible calculations are listed in the Appendix~\ref{app:FirstOrderFermionicSymmetriesExplicit}, some of which are redundant as in the second order bosonic case.  Here is listed only the unique symmetries. 
\be
\delta_{FS1}^{(1)}(P) \left(\begin{array}{c}
                            A^K \\ \psi_b^J \end{array}\right) \equiv
    \varepsilon^a_J P^K \left(\begin{array}{c}         i(\gamma^5\gamma^\mu?)_a^b?\partial_\mu\psi_b^J
 \\ (\gamma^5)_{ab} \square A^K \end{array}\right)
\ee
\be
 \delta_{FS2}^{(1)}(P) \left(\begin{array}{c}
                            A^J \\ \lambda_b \end{array}\right) \equiv
    \varepsilon^a P^J \left(\begin{array}{c}
                           i(\gamma^5\gamma^\mu?)_a^b?\partial_\mu\lambda_b \\   (\gamma^5)_{ab} \square A^J \end{array}\right)
\ee
\be
 \delta_{FS3}^{(1)}(Q) \left(\begin{array}{c}
                            B^J \\ \lambda_b \end{array}\right) \equiv
    \varepsilon^a Q^J \left(\begin{array}{c}
                           i(\gamma^5\gamma^\mu?)_a^b?\partial_\mu\lambda_b \\   (\gamma^5)_{ab} \square B^J \end{array}\right)
\ee
\be
\delta_{FS4}^{(1)}(Q) \left(\begin{array}{c}
                            B^K \\ \psi_b^I \end{array}\right) \equiv
    \varepsilon^a_I Q^K \left(\begin{array}{c}         i(\gamma^5\gamma^\mu?)_a^b?\partial_\mu\psi_b^I
 \\ (\gamma^5)_{ab} \square B^K \end{array}\right)
\ee
\be
 \delta_{FS5}^{(1)}(P) \left(\begin{array}{c}
                            B^K \\ \lambda_b \end{array}\right) \equiv
    \varepsilon^a P^K \left(\begin{array}{c}                         i(\gamma^\mu?)_a^b?\partial_\mu\lambda_b \\   C_{ab} \square B^K \end{array}\right)
\ee
\be
\delta_{FS6}^{(1)}(P) \left(\begin{array}{c}
                            B^J \\ \psi_b^K \end{array}\right) \equiv
    \varepsilon^a_J P^K \left(\begin{array}{c}         i(\gamma^\mu?)_a^b?\partial_\mu\psi_b^K
 \\ C_{ab} \square B^J \end{array}\right)
\ee
\be
\delta_{FS7}^{(1)}(Q) \left(\begin{array}{c}
                            A^J \\ \lambda_b \end{array}\right) \equiv
    \varepsilon^a Q^J \left(\begin{array}{c}         (\gamma^\mu?)_a^b?\partial_\mu\lambda_b
 \\ -iC_{ab} \square A^J \end{array}\right) 
\ee
\be
\delta_{FS8}^{(1)}(Q) \left(\begin{array}{c}
                         A^I \\ \psi_b^K \end{array}\right) \equiv
    \varepsilon^a_I Q^{K} \left(\begin{array}{c} (\gamma^\mu?)_a^b?\partial_\mu\psi_b^K 
 \\ -i C_{ab} \square A^I \end{array}\right) 
\ee
and
\be
\delta_{FS9}^{(1)}(P) \left(\begin{array}{c}
                            F^K \\ \lambda_b \end{array}\right) \equiv
    \varepsilon^a P^{K}\left(\begin{array}{c} (\gamma^5?)_a^b?\square \lambda_b 
 \\ i (\gamma^5\gamma^\mu)_{ab} \partial_\mu F^K \end{array}\right)
\ee
\be
\delta_{FS10}^{(1)}(P) \left(\begin{array}{c}
                            F^J \\ \psi_b^K \end{array}\right) \equiv
    \varepsilon^a_J P^K \left(\begin{array}{c}         (\gamma^5?)_a^b?\square\psi_b^K
 \\i (\gamma^5\gamma^\mu)_{ab} \partial_\mu F^J \end{array}\right)
\ee
\be
\delta_{FS11}^{(1)}(Q) \left(\begin{array}{c}
                            G^J \\ \lambda_b \end{array}\right) \equiv
    \varepsilon^a Q^{J} \left(\begin{array}{c}
                           (\gamma^5?)_a^b?\square \lambda_b
\\ i(\gamma^5\gamma^\mu)_{ab}\partial_\mu G^J \end{array}\right)
\ee
\be
\delta_{FS12}^{(1)}(Q) \left(\begin{array}{c}
                            G^I \\ \psi_b^K \end{array}\right) \equiv
    \varepsilon^a_I Q^{K} \left(\begin{array}{c}
                           (\gamma^5?)_a^b?\square \psi_b^K
\\ i(\gamma^5\gamma^\mu)_{ab}\partial_\mu G^I \end{array}\right)
\ee
\be
 \delta_{FS13}^{(1)}(Q) \left(\begin{array}{c}
                            {\rm d} \\ \psi_b^J \end{array}\right) \equiv
    \varepsilon^a Q^J \left(\begin{array}{c}
                           i(\gamma^5)_{a}^{~b}\square\psi_b^J \\  - (\gamma^5\gamma^\mu)_{ab} \partial_\mu {\rm d} \end{array}\right)
\ee
\be
 \delta_{FS14}^{(1)}(Q) \left(\begin{array}{c}
                            {\rm d}\\ \lambda_b \end{array}\right) \equiv
    \varepsilon^a_I Q^{I} \left(\begin{array}{c}
 i(\gamma^5?)_a^b? \square \lambda_b
\\ -(\gamma^5\gamma^\mu?)_ab?\partial_\mu {\rm d}
\end{array}\right)
\ee
and
\be
\delta_{FS15}^{(1)}(Q) \left(\begin{array}{c}
                            F^I \\ \psi_b^K \end{array}\right) \equiv
    \varepsilon^a_I Q^{K} \left(\begin{array}{c} 
\square \psi_a^K
\\ i(\gamma^\mu)_{ab}\partial_\mu F^I \end{array}\right)
\ee
\be
\delta_{FS16}^{(1)}(Q) \left(\begin{array}{c}
                            F^J \\ \lambda_b \end{array}\right) \equiv
    \varepsilon^a Q^{J} \left(\begin{array}{c}  \square \lambda_a
\\ i(\gamma^\mu?)_ab?\partial_\mu F^J \end{array}\right)
\ee
\be
\delta_{FS17}^{(1)}(P) \left(\begin{array}{c}
                            G^K \\ \lambda_b \end{array}\right) \equiv
    \varepsilon^a P^{K} \left(\begin{array}{c}
                           \square \lambda_a
\\ i(\gamma^\mu)_{ab}\partial_\mu G^K \end{array}\right)
\ee
\be
\delta_{FS18}^{(1)}(P) \left(\begin{array}{c}
                            G^J \\ \psi_b^K \end{array}\right) \equiv
    \varepsilon^a_J P^K \left(\begin{array}{c}          \square\psi_a^K
 \\i(\gamma^\mu)_{ab} \partial_\mu G^J \end{array}\right)
\ee
\be
 \delta_{FS19}^{(1)}(P) \left(\begin{array}{c}
                            {\rm d} \\ \psi_b^J \end{array}\right) \equiv
    \varepsilon^a P^J \left(\begin{array}{c}
                           \square\psi_a^J \\  i (\gamma^\mu)_{ab} \partial_\mu {\rm d} \end{array}\right)
\ee
\be
 \delta_{FS20}^{(1)}(P) \left(\begin{array}{c}
                            {\rm d} \\ \lambda_b \end{array}\right) \equiv
    \varepsilon^a_I P^{I} \left(\begin{array}{c}  \square \lambda_a
\\ i(\gamma^\mu?)_ab?\partial_\mu {\rm d} \end{array}\right)
\ee
and
\be
\delta_{FS21}^{(1)}(T) \left(\begin{array}{c}
                            A^J \\ \psi_b^J \end{array}\right) \equiv
    \varepsilon^a T^{JM} \left(\begin{array}{c}
                           (\gamma^\mu)_a^{~b}\partial_\mu \psi_b^M \\  i C_{ab}\square A^M \end{array}\right)
\ee
\be
 \delta_{FS22}^{(1)}(W) \left(\begin{array}{c}
                            B^J \\ \psi_b^J \end{array}\right) \equiv
    \varepsilon^a W^{JM} \left(\begin{array}{c}
                           (\gamma^\mu?)_a^b? \partial_\mu \psi_b^M
 \\ i C_{ab}\square B^M \end{array}\right) 
\ee
\be
 \delta_{FS23}^{(1)}(T) \left(\begin{array}{c}
                            A^K \\ \lambda_b \end{array}\right) \equiv
    \varepsilon^a_I T^{IK} \left(\begin{array}{c} i(\gamma^\mu?)_a^b?\partial_\mu \lambda_b
\\ C_{ab}\square A^K \end{array}\right)
\ee
\be
\delta_{FS24}^{(1)}(T) \left(\begin{array}{c}
                            A^M \\ \psi_b^J \end{array}\right)
\equiv
    \varepsilon^a_I \epsilon^{IJK}T^{KM} \left(\begin{array}{c}         i(\gamma^\mu?)_a^b?\partial_\mu\psi_b^J
 \\ C_{ab} \square A^M \end{array}\right)
\ee
\be
\delta_{FS25}^{(1)}(T) \left(\begin{array}{c}
                            A^J \\ \psi_b^M \end{array}\right) \equiv
    \varepsilon^a_I \epsilon^{IJK}T^{KM} \left(\begin{array}{c} i(\gamma^\mu?)_a^b?\partial_\mu\psi_b^M
 \\ C_{ab} \square A^J \end{array}\right)
\ee
\be
\delta_{FS26}^{(1)}(W) \left(\begin{array}{c}
                            B^J \\ \psi_b^M \end{array}\right) \equiv
    \varepsilon^a_I \epsilon^{IJK}W^{KM} \left(\begin{array}{c}
    i(\gamma^\mu?)_a^b?\partial_\mu\psi_b^M
 \\ C_{ab}\square B^J \end{array}\right) 
\ee
and
\be
\delta_{FS27}^{(1)}(W) \left(\begin{array}{c}
                            A^J \\ \psi_b^J \end{array}\right) \equiv
    \varepsilon^a W^{JM} \left(\begin{array}{c}
         (\gamma^5\gamma^\mu?)_a^b?\partial_\mu\psi_b^M
 \\ i(\gamma^5)_{ab}\square A^M \end{array}\right)
\ee
\be
 \delta_{FS28}^{(1)}(T) \left(\begin{array}{c}
                            B^J \\ \psi_b^J \end{array}\right) \equiv
    \varepsilon^a T^{JM} \left(\begin{array}{c}
                           (\gamma^5\gamma^\mu)_a^{~b}\partial_\mu \psi_b^M \\  i (\gamma^5)_{ab}\square B^M \end{array}\right)
\ee
\be
 \delta_{FS29}^{(1)}(W) \left(\begin{array}{c}
                            A^M \\ \lambda_b \end{array}\right) \equiv
    \varepsilon^a_I W^{IM} \left(\begin{array}{c} i(\gamma^5\gamma^\mu?)_a^b?\partial_\mu \lambda_b
\\ (\gamma^5)_{ab}\square A^M \end{array}\right)
\ee
\be
\delta_{FS30}^{(1)}(W) \left(\begin{array}{c}
                            A^J \\ \psi_b^M \end{array}\right) \equiv
    \varepsilon^a_I \epsilon^{IJK}W^{KM} \left(\begin{array}{c} i(\gamma^5\gamma^\mu?)_a^b?\partial_\mu\psi_b^M 
 \\ (\gamma^5)_{ab} \square A^J \end{array}\right)
\ee
\be
\delta_{FS31}^{(1)}(W) \left(\begin{array}{c}
                            A^M \\ \psi_b^J \end{array}\right) \equiv
    \varepsilon^a_I\epsilon^{IJK} W^{KM} \left(\begin{array}{c}        i(\gamma^5\gamma^\mu?)_a^b?\partial_\mu\psi_b^J
 \\ (\gamma^5)_{ab} \square A^M \end{array}\right)
\ee
\be\eqalign{
 \delta_{FS32}^{(1)}(T) \left(\begin{array}{c}
                            B^K \\ \lambda_b \end{array}\right) &\equiv
    \varepsilon^a_I T^{IK} \left(\begin{array}{c} i(\gamma^5\gamma^\mu?)_a^b?\partial_\mu \lambda_b
\\ (\gamma^5)_{ab}\square B^K \end{array}\right)
}\ee
\be
\delta_{FS33}^{(1)}(T) \left(\begin{array}{c}
                            B^J \\ \psi_b^M \end{array}\right) \equiv
    \varepsilon^a_I \epsilon^{IJK}T^{KM} \left(\begin{array}{c}
    i(\gamma^5\gamma^\mu?)_a^b?\partial_\mu\psi_b^M
 \\ (\gamma^5)_{ab}\square B^J \end{array}\right)
\ee
\be
\delta_{FS34}^{(1)}(T)\left(\begin{array}{c}
                            B^M \\ \psi_b^J \end{array}\right) \equiv
    \varepsilon^a_I \epsilon^{IJK}T^{KM} \left(\begin{array}{c}
    i(\gamma^5\gamma^\mu?)_a^b?\partial_\mu\psi_b^J
 \\ (\gamma^5)_{ab}\square B^M \end{array}\right)
\ee
and
\be
 \delta_{FS35}^{(1)}(T) \left(\begin{array}{c}
                            G^J \\ \psi_b^J \end{array}\right) \equiv
    \varepsilon^a T^{JM} \left(\begin{array}{c}
                           i (\gamma^5)_{a}^{~b}\square \psi_b^M \\  (\gamma^5\gamma^\mu)_{ab}\partial_\mu G^M \end{array}\right)
\ee
\be
 \delta_{FS36}^{(1)}(W) \left(\begin{array}{c}
                            F^
J \\ \psi_b^J \end{array}\right) \equiv
    \varepsilon^a W^{JM} \left(\begin{array}{c}
                           i(\gamma^5?)_a^b?\square\psi_b^M
\\ (\gamma^5\gamma^\mu?)_ab?\partial_\mu F^M \end{array}\right)
\ee
\be
 \delta_{FS37}^{(1)}(W) \left(\begin{array}{c}
                            F^J \\ \psi_b^M \end{array}\right) \equiv
    \varepsilon^a_I \epsilon^{IJK}W^{KM} \left(\begin{array}{c} 
(\gamma^5?)_a^b?\square \psi_b^M
\\ i(\gamma^5\gamma^\mu)_{ab}\partial_\mu F^J \end{array}\right)
\ee
\be
\delta_{FS38}^{(1)}(T) \left(\begin{array}{c}
                            G^K \\ \lambda_b \end{array}\right) \equiv
    \varepsilon^a_I T^{IK} \left(\begin{array}{c}
                           (\gamma^5?)_a^b?\square \lambda_b
\\ i(\gamma^5\gamma^\mu)_{ab}\partial_\mu G^K \end{array}\right)
\ee
\be
 \delta_{FS39}^{(1)}(T) \left(\begin{array}{c}
                            G^K \\ \psi_b^J \end{array}\right) \equiv
    \varepsilon^a_I \epsilon^{IJK}T^{KM} \left(\begin{array}{c}
                          (\gamma^5?)_a^b?\square\psi_b^N
 \\ i(\gamma^5\gamma^\mu)_{ab}\partial_\mu G^M \end{array}\right) 
\ee
\be
\delta_{FS40}^{(1)}(T) \left(\begin{array}{c}
                            G^J \\ \psi_b^M \end{array}\right) \equiv
    \varepsilon^a_I \epsilon^{IJK}T^{KM} \left(\begin{array}{c}
(\gamma^5?)_a^b?\square\psi_b^M
\\ i(\gamma^5\gamma^\mu)_{ab}\partial_\mu G^J \end{array}\right) 
\ee
\be
 \delta_{FS41}^{(1)}(T) \left(\begin{array}{c}
                            {\rm d} \\ \psi_b^M \end{array}\right) \equiv
    \varepsilon^a_I T^{IM} \left(\begin{array}{c} 
  (\gamma^5?)_a^b?\square\psi_b^M
\\i (\gamma^5\gamma^\mu)_{ab}\partial_\mu {\rm d} \end{array}\right)
\ee
and
\be
\delta_{FS42}^{(1)}(V) \left(\begin{array}{c}
                            F^J \\ \psi_b^J \end{array}\right) \equiv
    \varepsilon^a (V^\mu)^{JM} \left(\begin{array}{c}
  i(\gamma^5\gamma^\nu)_a^{~b}\partial_\mu\partial_\nu \psi_b^M \\    (\gamma^5)_{ab}\partial_\mu F^M 
\end{array}\right)
\ee
\be
 \delta_{FS43}^{(1)}(V) \left(\begin{array}{c}
                            F^J \\ \psi_b^J \end{array}\right) \equiv
    \varepsilon^a (V^\rho)^{JM} \left(\begin{array}{c}
                           (\gamma^5\gamma^\nu\gamma_\rho\gamma^\mu?)_a^b?\partial_\mu\partial_\nu\psi_b^M
\\ i(\gamma^5\gamma_\rho\gamma^\mu?)_ba?\partial_\mu F^M \end{array}\right)
\ee
\be
 \delta_{FS44}^{(1)}(V) \left(\begin{array}{c}
                            F^M \\ \lambda_b \end{array}\right) \equiv
    \varepsilon^a_I (V^\mu)^{IM} \left(\begin{array}{c} (\gamma^5\gamma^\nu?)_a^b?\partial_\mu\partial_\nu \lambda_b
\\ i(\gamma^5)_{ab}\partial_\mu F^M \end{array}\right)
\ee
\be
\delta_{FS45}^{(1)}(V) \left(\begin{array}{c}
                            F^M \\ \psi_b^J \end{array}\right) \equiv
    \varepsilon^a_I \epsilon^{IJK}(V^\mu)^{KM} \left(\begin{array}{c}        (\gamma^5\gamma^\nu?)_a^b?\partial_\mu\partial_\nu\psi_b^J
 \\ i(\gamma^5)_{ab} \partial_\mu F^M \end{array}\right)
\ee
\be
 \delta_{FS46}^{(1)}(U) \left(\begin{array}{c}
                            F^J \\ \lambda_b \end{array}\right) \equiv
    \varepsilon^a (U^\mu)^J \left(\begin{array}{c}
                           (\gamma^5\gamma^\nu?)_a^b?\partial_\mu\partial_\nu\lambda_b \\  i(\gamma^5)_{ab} \partial_\mu F^J \end{array}\right)
\ee
\be
\delta_{FS47}^{(1)}(U) \left(\begin{array}{c}
                            F^K \\ \psi_b^I \end{array}\right) \equiv
    \varepsilon^a_I (U^\mu)^K \left(\begin{array}{c}         (\gamma^5\gamma^\nu?)_a^b?\partial_\mu\partial_\nu\psi_b^I
 \\ i(\gamma^5)_{ab} \partial_\mu F^K \end{array}\right)
\ee
\be
 \delta_{FS48}^{(1)}(V) \left(\begin{array}{c}
                            F^J \\ \psi_b^M \end{array}\right) \equiv
    \varepsilon^a_I \epsilon^{IJK}(V^\rho)^{KM} \left(\begin{array}{c} 
(\gamma^5\gamma^\mu\gamma_\rho\gamma^\nu?)_a^b?\partial_\mu\partial_\nu \psi_b^M
\\ -i(\gamma^5\gamma_\rho\gamma^\mu)_{ba}\partial_\mu F^J \end{array}\right)
\ee
\be
 \delta_{FS49}^{(1)}(U) \left(\begin{array}{c}
                            F^K \\ \lambda_b \end{array}\right) \equiv
    \varepsilon^a (U^\rho)^K \left(\begin{array}{c}
                           (\gamma^5\gamma^\mu\gamma_\rho\gamma^\nu?)_a^b?\partial_\mu\partial_\nu\lambda_b \\ - i(\gamma^5\gamma_\rho\gamma^\mu)_{ba} \partial_\mu F^K \end{array}\right)
\ee
\be
 \delta_{FS50}^{(1)}(U) \left(\begin{array}{c}
                            F^J \\ \psi_b^K \end{array}\right) \equiv
    \varepsilon^a_J (U^\rho)^K \left(\begin{array}{c}                       (\gamma^5\gamma^\mu\gamma_\rho\gamma^\nu?)_a^b? \partial_\mu\partial_\nu\psi_b^K \\ -i(\gamma^5\gamma_\rho\gamma^\mu)_{ba} \partial_\mu F^J \end{array}\right)
\ee
and
\be
\delta_{FS51}^{(1)}(U) \left(\begin{array}{c}
                            G^J \\ \lambda_b \end{array}\right) \equiv
    \varepsilon^a (U^\mu)^J \left(\begin{array}{c}
       \partial^\nu\partial_{[\mu}(\gamma_{\nu]}?)_a^b? \lambda_b \\  (?\sigma^\nu_\mu?)_{ab}\partial_\nu G^J 
\end{array}\right)
\ee
\be
\delta_{FS52}^{(1)}(U) \left(\begin{array}{c}
                            G^K \\ \psi_b^I \end{array}\right) \equiv
    \varepsilon^a_I (U^\mu)^K \left(\begin{array}{c}         \partial^\nu\partial_{[\mu}(\gamma_{\nu]}?)_a^b?\psi_b^I
 \\ (?\sigma^\nu_\mu?)_{ab} \partial_\nu G^K \end{array}\right)
\ee
\be
\delta_{FS53}^{(1)}(U) \left(\begin{array}{c}
                            G^K \\ \lambda_b \end{array}\right) \equiv
    \varepsilon^a (U^\rho)^K \left(\begin{array}{c}
       i(\gamma^\mu\gamma_\rho\gamma^\nu?)_a^b? \partial_\mu\partial_\nu \lambda_b \\  (\gamma_\rho\gamma^\mu)_{ba}\partial_\mu G^K 
\end{array}\right)
\ee
\be
 \delta_{FS54}^{(1)}(U) \left(\begin{array}{c}
                            {\rm d} \\ \psi_b^J \end{array}\right) \equiv
    \varepsilon^a (U^\mu)^J \left(\begin{array}{c}
                           (\gamma^\nu)_{a}^{~b}\partial_\mu\partial_\nu\psi_b^J \\  i C_{ab}\partial_\mu {\rm d} \end{array}\right)
\ee
\be
 \delta_{FS55}^{(1)}(U) \left(\begin{array}{c}
                            {\rm d} \\ \lambda_b \end{array}\right) \equiv
    \varepsilon^a_I (U^\mu)^{I} \left(\begin{array}{c}
 (\gamma^\nu?)_a^b? \partial_\mu\partial_\nu \lambda_b
\\ i C_{ab} \partial_\mu {\rm d} 
\end{array}\right)
\ee
\be
 \delta_{FS56}^{(1)}(V) \left(\begin{array}{c}
                            G^J \\ \psi_b^J \end{array}\right) \equiv
    \varepsilon^a (V^\rho)^{JM} \left(\begin{array}{c}
                           i (\gamma^\mu\gamma_\rho\gamma^\nu?)_a^b?\partial_\mu\partial_\nu \psi_b^M
\\ -(\gamma_\rho\gamma^\mu)_{ba}\partial_\mu G^M \end{array}\right)
\ee
\be
 \delta_{FS57}^{(1)}(V) \left(\begin{array}{c}
                            G^J \\ \psi_b^J \end{array}\right) \equiv
    \varepsilon^a (V^\mu)^{JM} \left(\begin{array}{c}                     (\gamma^\nu?)_a^b?\partial_\mu\partial_\nu\psi_b^M\\ -i C_{ab}\partial_\mu G^M \end{array}\right)
\ee
\be
\delta_{FS58}^{(1)}(U) \left(\begin{array}{c}
                            G^J \\ \psi_b^K \end{array}\right) \equiv
    \varepsilon^a_J (U^\rho)^K \left(\begin{array}{c}
       i (\gamma^\mu\gamma_\rho\gamma^\nu?)_a^b? \partial_\mu\partial_\nu \psi_b^K \\  (\gamma_\rho\gamma^\mu)_{ba}\partial_\nu G^J 
\end{array}\right)
\ee
\be
\delta_{FS59}^{(1)}(V) \left(\begin{array}{c}
                            G^M \\ \lambda_b \end{array}\right) \equiv
    \varepsilon^a_I (V^\mu)^{IM} \left(\begin{array}{c}
                           (\gamma^\nu?)_a^b?\partial_\mu\partial_\nu \lambda_b
\\ iC_{ab}\partial_\mu G^M \end{array}\right)
\ee
\be
 \delta_{FS60}^{(1)}(V) \left(\begin{array}{c}
                            G^M \\ \psi_b^J \end{array}\right) \equiv
    \varepsilon^a_I \epsilon^{IJK}(V^\mu)^{KM} \left(\begin{array}{c}
      (\gamma^\nu?)_a^b? \partial_\mu\partial_\nu\psi_a^N
 \\ i C_{ab}\partial_\mu G^M \end{array}\right) 
\ee
\be
\delta_{FS61}^{(1)}(V) \left(\begin{array}{c}
                            G^J \\ \psi_b^M \end{array}\right) \equiv
    \varepsilon^a_I \epsilon^{IJK}(V^\rho)^{KM} \left(\begin{array}{c}
i(\gamma^\mu\gamma_\rho\gamma^\nu?)_a^b?\partial_\mu\partial_\nu\psi_b^M
\\ (\gamma_\rho\gamma^\mu)_{ba}\partial_\mu G^J \end{array}\right)
\ee
\be
 \delta_{FS62}^{(1)}(U) \left(\begin{array}{c}
                            {\rm d} \\ \lambda_b \end{array}\right) \equiv
    \varepsilon^a_I (U^\rho)^{I} \left(\begin{array}{c}
 i (\gamma^\mu\gamma_\rho\gamma^\nu?)_a^b? \partial_\mu\partial_\nu \lambda_b
\\ (\gamma_\rho\gamma^\mu)_{ba} \partial_\mu {\rm d} 
\end{array}\right)
\ee
\be
 \delta_{FS63}^{(1)}(V) \left(\begin{array}{c}
                           {\rm d} \\ \psi_b^M \end{array}\right) \equiv
    \varepsilon^a_I (V^\rho)^{IM} \left(\begin{array}{c} 
 i (\gamma^\mu\gamma_\rho\gamma^\nu?)_a^b?\partial_\mu\partial_\nu\psi_b^M
\\ (\gamma_\rho\gamma^\mu)_{ba}\partial_\mu {\rm d} \end{array}\right)
\ee
and
\be\eqalign{
\delta_{FS64}^{(1)}(P) \left(\begin{array}{c}
                            A_\mu \\ \lambda_b \end{array}\right) &\equiv
    \varepsilon^a_I P^{I} \left(\begin{array}{c}                           (\gamma^5\gamma_\mu\gamma^\nu)_a^{~b}\partial_\nu \lambda_b \\   -i(\gamma^5\gamma^\beta)_{ab}\partial^\alpha F_{\alpha\beta}\end{array}\right) \cr
}\ee
\be\eqalign{
\delta_{FS65}^{(1)}(P) \left(\begin{array}{c}
                            A_\mu \\ \psi_b^K \end{array}\right) &\equiv
    \varepsilon^a P^{K} \left(\begin{array}{c}                           (\gamma^5\gamma_\mu\gamma^\nu)_a^{~b}\partial_\nu \psi_b^K \\   -i(\gamma^5\gamma^\beta)_{ab}\partial^\alpha F_{\alpha\beta}\end{array}\right) \cr
}\ee
\be\eqalign{
\delta_{FS66}^{(1)}(Q) \left(\begin{array}{c}
                            A_\mu \\ \psi_b^J \end{array}\right) &\equiv
    \varepsilon^a Q^{J} \left(\begin{array}{c}
                           -(\gamma_\mu\gamma^\nu)_a^{~b}\partial_\nu \psi_b^J \\   \frac{1}{2}(\gamma^\alpha\sigma^{\mu\nu})_{ba}\partial_\alpha F_{\mu\nu}\end{array}\right) \cr
}\ee
\be\eqalign{
 \delta_{FS67}^{(1)}(Q) \left(\begin{array}{c}
                            A_\mu \\ \lambda_b \end{array}\right) &\equiv
    \varepsilon^a_I Q^{I} \left(\begin{array}{c} 
-(\gamma_\mu\gamma^\nu?)_a^b?\partial_\nu\lambda_b
\\ \frac{1}{2}(\gamma^\alpha\sigma^{\mu\nu})_{ba}\partial_\alpha F_{\mu\nu} \end{array}\right)
 \cr
}\ee
\be
\eqalign{ 
 \delta_{FS68}^{(1)}(T) \left(\begin{array}{c}
                            A_\mu \\ \psi_b^M \end{array}\right) &\equiv
    \varepsilon^a_I T^{IM} \left(\begin{array}{c}
    (\gamma_\mu\gamma^\nu?)_a^b?\partial_\nu\psi_b^M
 \\ -\frac{1}{2}(\gamma^\alpha\sigma^{\mu\nu})_{ba}\partial_\alpha F_{\mu\nu} \end{array}\right) 
}\ee
\be
 \delta_{FS69}^{(1)}(W) \left(\begin{array}{c}
                            A_\mu \\ \psi_b^M \end{array}\right)\equiv
    \varepsilon^a_I W^{IM} \left(\begin{array}{c}
    -(\gamma^5\gamma_\mu\gamma^\nu?)_a^b?\partial_\nu\psi_b^M
 \\ -\frac{1}{2}(\gamma^5\gamma^\alpha\sigma^{\mu\nu})_{ba}\partial_\alpha F_{\mu\nu} \end{array}\right) 
\ee
\be
 \delta_{FS70}^{(1)}(U) \left(\begin{array}{c}
                            A_\mu \\ \psi_b^J \end{array}\right) \equiv
    \varepsilon^a (U_\mu)^J \left(\begin{array}{c}
  i(\gamma^5\gamma^\nu)_{a}^{~b}\partial_\nu \psi_b^J \\  - (\gamma^5)_{ab} \partial_\nu ?F^\mu\nu? \end{array}\right)
\ee
\be
 \delta_{FS71}^{(1)}(U) \left(\begin{array}{c}
                            A_\mu \\ \lambda_b \end{array}\right) \equiv   \varepsilon^a_I (U_\mu)^{I}\left(\begin{array}{c}
 -i(\gamma^5\gamma^\nu?)_a^b? \partial_\nu \lambda_b
\\ (\gamma^5)_{ab} \partial_\nu ?F^\mu\nu? 
\end{array}\right)
\ee
\be
 \delta_{FS72}^{(1)}(U) \left(\begin{array}{c}
                            A_\mu \\ \lambda_b \end{array}\right) \equiv   \varepsilon^a_I (U^\rho)^{I}\left(\begin{array}{c}
 (\gamma^5\gamma_\mu\gamma_\rho\gamma^\nu?)_a^b? \partial_\nu \lambda_b
\\ \frac{1}{2}(\gamma^5\gamma_\rho\gamma^\nu\sigma^{\alpha\beta})_{ba} \partial_\nu ?F_\alpha\beta? 
\end{array}\right)
\ee
\be
 \delta_{FS73}^{(1)}(V) \left(\begin{array}{c}
                            A_\mu \\ \psi_b^M \end{array}\right) \equiv
    \varepsilon^a_I (V^\rho)^{IM} \left(\begin{array}{c}
    (\gamma^5\gamma_\mu\gamma_\rho\gamma^\nu?)_a^b?\partial_\nu\psi_b^M
 \\ \frac{1}{2} (\gamma^5\gamma_\rho\gamma^\nu\sigma^{\alpha\beta})_{ba}\partial_\nu F_{\alpha\beta} \end{array}\right) 
\ee
and
\be
\delta_{FS74}^{(1)}(U) \left(\begin{array}{c}
                            A^J \\ \psi_b^K \end{array}\right) \equiv
    \varepsilon^a_J (U^\rho)^{K} \left(\begin{array}{c} -(\gamma^5\gamma_\rho\gamma^\mu?)_a^b?\partial_\mu\psi_b^K
 \\ i (\gamma^5\gamma_\rho)_{ab} \square A^J \end{array}\right) 
\ee
\be
\delta_{FS75}^{(1)}(V) \left(\begin{array}{c}
                            A^J \\ \psi_b^M \end{array}\right) \equiv
    \varepsilon^a_I \epsilon^{IJK}(V^\rho)^{KM} \left(\begin{array}{c} -(\gamma^5\gamma_\rho\gamma^\mu?)_a^b?\partial_\mu\psi_b^M
 \\ i (\gamma^5\gamma_\rho)_{ab} \square A^J \end{array}\right) 
\ee
\be
\delta_{FS76}^{(1)}(V) \left(\begin{array}{c}
                            A^J \\ \psi_b^J \end{array}\right) \equiv
    \varepsilon^a (V^\rho)^{JM} \left(\begin{array}{c}
         (\gamma^5\gamma_\rho\gamma^\mu?)_a^b?\partial_\mu\psi_b^M
 \\ i(\gamma^5\gamma_\rho)_{ab} \square A^M \end{array}\right)
\ee
\be
\delta_{FS77}^{(1)}(U) \left(\begin{array}{c}
                            A^K \\ \lambda_b \end{array}\right) \equiv
    \varepsilon^a (U^\rho)^{K} \left(\begin{array}{c} (\gamma^5\gamma_\rho\gamma^\mu?)_a^b?\partial_\mu\lambda_b
 \\ -i (\gamma^5\gamma_\rho)_{ab} \square A^K \end{array}\right) 
\ee
\be
\delta_{FS78}^{(1)}(U) \left(\begin{array}{c}
                            B^K \\ \lambda_b \end{array}\right) \equiv
    \varepsilon^a (U^\rho)^{K} \left(\begin{array}{c}
    i(\gamma_\rho\gamma^\mu?)_a^b?\partial_\mu\lambda_b
 \\ (\gamma_\rho)_{ab}\square B^K \end{array}\right) 
\ee
\be
\delta_{FS79}^{(1)}(V) \left(\begin{array}{c}
                            B^J \\ \psi_b^M \end{array}\right) \equiv
    \varepsilon^a_I \epsilon^{IJK}(V^\rho)^{KM} \left(\begin{array}{c}
    i(\gamma_\rho\gamma^\mu?)_a^b?\partial_\mu\psi_b^M
 \\ (\gamma_\rho)_{ab}\square B^J \end{array}\right) 
\ee
\be
\delta_{FS80}^{(1)}(U) \left(\begin{array}{c}
                            B^J \\ \psi_b^K \end{array}\right) \equiv
    \varepsilon^a_J (U^\rho)^{K} \left(\begin{array}{c}
    i(\gamma_\rho\gamma^\mu?)_a^b?\partial_\mu\psi_b^K
 \\ (\gamma_\rho)_{ab}\square B^J \end{array}\right) 
\ee
\be
 \delta_{FS81}^{(1)}(V) \left(\begin{array}{c}
                            B^J \\ \psi_b^J \end{array}\right) \equiv
    \varepsilon^a (V^\rho)^{JM} \left(\begin{array}{c}
                           -i (\gamma_\rho\gamma^\mu?)_a^b? \partial_\mu \psi_b^M
 \\ (\gamma_\rho)_{ab}\square B^M \end{array}\right)
\ee
and
\be
 \delta_{FS82}^{(1)}(T) \left(\begin{array}{c}
                            F^J \\ \psi_b^J \end{array}\right) \equiv
    \varepsilon^a T^{JM} \left(\begin{array}{c}
                           i \square \psi_a^M \\   (\gamma^\mu)_{ab}\partial_\mu F^M \end{array}\right)
\ee
\be
\delta_{FS83}^{(1)}(W) \left(\begin{array}{c}
                            G^J \\ \psi_b^J \end{array}\right) \equiv
    \varepsilon^a W^{JM} \left(\begin{array}{c}
                           i \square \psi_a^M
\\ (\gamma^\mu)_{ab}\partial_\mu G^M \end{array}\right)
\ee
\be
 \delta_{FS84}^{(1)}(T) \left(\begin{array}{c}
                            F^K \\ \lambda_b \end{array}\right) \equiv
    \varepsilon^a_I T^{IK} \left(\begin{array}{c}
                           \square \lambda_a
\\ i(\gamma^\mu)_{ab}\partial_\mu F^K \end{array}\right)
\ee
\be
 \delta_{FS85}^{(1)}(T) \left(\begin{array}{c}
                            F^J \\ \psi_b^M \end{array}\right) \equiv
    \varepsilon^a_I \epsilon^{IJK}T^{KM} \left(\begin{array}{c} 
\square \psi_a^M
\\ i(\gamma^\mu)_{ab}\partial_\mu F^J \end{array}\right)
\ee
\be
 \delta_{FS86}^{(1)}(T) \left(\begin{array}{c}
                            F^M \\ \psi_b^J \end{array}\right) \equiv
    \varepsilon^a_I \epsilon^{IJK}T^{KM} \left(\begin{array}{c}
\square\psi_a^J
 \\ i (\gamma^\mu)_{ab}\partial_\mu F^M \end{array}\right) 
\ee
\be
 \delta_{FS87}^{(1)}(W) \left(\begin{array}{c}
                            G^M \\ \lambda_b \end{array}\right) \equiv
    \varepsilon^a_I W^{IM} \left(\begin{array}{c}
                           \square \lambda_a
\\ i(\gamma^\mu)_{ab}\partial_\mu G^M \end{array}\right)
\ee
\be
\delta_{FS88}^{(1)}(W) \left(\begin{array}{c}
                            G^J \\ \psi_b^M \end{array}\right) \equiv
    \varepsilon^a_I \epsilon^{IJK}W^{KM} \left(\begin{array}{c}
                           \square \psi_a^M
\\ i(\gamma^\mu)_{ab}\partial_\mu G^J \end{array}\right)
\ee
\be
 \delta_{FS89}^{(1)}(W) \left(\begin{array}{c}
                            G^M \\ \psi_b^J \end{array}\right) \equiv
    \varepsilon^a_I\epsilon^{IJK} W^{KM} \left(\begin{array}{c}
                           \square\psi_a^J
 \\ i (\gamma^\mu)_{ab}\partial_\mu G^M \end{array}\right) 
\ee
\be
 \delta_{FS90}^{(1)}(W) \left(\begin{array}{c}
                            {\rm d} \\ \psi_b^M \end{array}\right) \equiv
    \varepsilon^a_I W^{IM} \left(\begin{array}{c} 
\square \psi_a^M
\\ i(\gamma^\mu)_{ab}\partial_\mu {\rm d} \end{array}\right)
\ee

\subsection{Symmetries of the \texorpdfstring{${\mathcal N} = 2$}{{\mathcal N} = 2} FH Lagrangian}
The symmetries of the ${\mathcal N} = 2$ FH system follow analogously from the ${\mathcal N} = 4$ calculations.  The first order bosonic symmetries of the ${\mathcal N} = 2$ FH system calculated from the central charges and internal symmetries are
\be
  \tilde{\delta}_{BS1}^{(1)}(\tilde{T}) \left(\begin{array}{c}
                               \tilde{A}^k \\  \tilde{F}^k 
                          \end{array}\right) \equiv  \tilde{T}^{km}  \left(\begin{array}{c}
                    \tilde{F}^m \\  \square \tilde{A}^m   
\end{array}\right)
\ee
\be
  \tilde{\delta}_{BS2}^{(1)}(\tilde{T}) \left(\begin{array}{c}
                             \tilde{B}^k \\  \tilde{G}^k  
                          \end{array}\right) \equiv  \tilde{T}^{km}   \left(\begin{array}{c}
                    \tilde{G}^m \\ \square \tilde{B}^m  
\end{array}\right)
\ee
\be
\tilde{\delta}_{BS3}^{(1)}(\tilde{T}) \tilde{\psi}^k_c \equiv \tilde{T}^{km}(\gamma^\mu)_c^{~d}\partial_\mu\tilde{\psi}_d^m
\ee
with
\be
   \tilde{T}^{km} \equiv \tilde{R}^{ijkm}C_{ab}\varepsilon^{a}_i\chi^b_j
\ee
where $i,j,k,m=1,2$, and $\varepsilon^a_i$ and $\chi^b_j$ are once again infinitesimal Grassmann spinors.   Here, we clearly notice the absence of symmetries between $A^J \leftrightarrow B^J$, $A^J \leftrightarrow G^J$, $B^J \leftrightarrow F^J$, and $G^J \leftrightarrow F^J$.  As in the ${\mathcal N} = 4$ case, this is a direct result of the absence of coupling terms between these fields in the algebra.

Interestingly, we find that the second order bosonic symmetries calculated from these first order symmetries all vanish identically
\begin{align}
\tilde{\delta}_{BS1}^{(2)}(\tilde{T}_1,\tilde{T}_2)\tilde{A}^k \equiv & [\tilde{\delta}_{BS1}^{(1)}(\tilde{T}_1), \tilde{\delta}_{BS1}^{(1)}(\tilde{T}_2)] \tilde{A}^k = \tilde{\Lambda}_{1}^{jk}(\tilde{T}_1, \tilde{T}_2) \square \tilde{A}^j = 0 \\
\tilde{\delta}_{BS2}^{(2)}(\tilde{T}_1,\tilde{T}_2) \tilde{B}^k \equiv &
[\tilde{\delta}_{BS2}^{(1)}(\tilde{T}_1), \tilde{\delta}_{BS2}^{(1)}(\tilde{T}_2)] \tilde{B}^k = \tilde{\Lambda}_{1}^{jk}(\tilde{T}_1, \tilde{T}_2)\square \tilde{B}^j = 0 \\
\tilde{\delta}_{BS3}^{(2)}(\tilde{T}_1,\tilde{T}_2)\tilde{F}^k \equiv & [\tilde{\delta}_{BS1}^{(1)}(\tilde{T}_1), \tilde{\delta}_{BS1}^{(1)}(\tilde{T}_2)] \tilde{F}^k = \tilde{\Lambda}_{1}^{jk}(\tilde{T}_1, \tilde{T}_2)\square \tilde{F}^j = 0 \\
\tilde{\delta}_{BS4}^{(2)}(\tilde{T}_1,\tilde{T}_2)\tilde{G}^k \equiv & [\tilde{\delta}_{BS2}^{(1)}(T_1), \tilde{\delta}_{BS2}^{(1)}(\tilde{T}_2) ] \tilde{G}^k = \tilde{\Lambda}_{1}^{jk}(\tilde{T}_1, \tilde{T}_2)\square \tilde{G}^j = 0 \\
\tilde{\delta}_{BS5}^{(2)}(\tilde{T}_1, \tilde{T}_2)\tilde{\psi}_c^k \equiv & [ \tilde{\delta}_{BS3}^{(1)}(\tilde{T}_1), \tilde{\delta}_{BS3}^{(1)}(\tilde{T}_2) ] \tilde{\psi}_c^k = \tilde{\Lambda}_{1}^{jk}(\tilde{T}_1, \tilde{T}_2)\square \tilde{\psi}_c^J   = 0
\end{align}
as
\be
    \tilde{\Lambda}_{1,1}^{jk}(\tilde{T_1},\tilde{T_2}) \equiv \tilde{T}_{[1}^{jm}\tilde{T}_{2]}^{mk} = 0,~~~j,k,m = 1,2
\ee
even though for a general matrix $\tilde{\Lambda}^{jk}$,
\be\eqalign{
   \tilde{\delta}_{BS}^{(2)} \tilde{\chi}_C^j &\equiv \tilde{\Lambda}^{[jk]} \square \tilde{\chi}_C^k, \cr
   \tilde{\chi} _C^j &\equiv (\tilde{A}^j, \tilde{B}^j, \tilde{F}^j, \tilde{G}^j, \tilde{\psi}_c^j)
}\ee
is still a symmetry of the ${\mathcal N}=2$ FH Lagrangian.

On the other hand, several first order fermionic symmetries still remain after reduction to the ${\mathcal N} = 2$ FH system:
\be\eqalign{
\tilde{\delta}_{FS1}^{(1)}(\tilde{T}) \left(\begin{array}{c}
                            \tilde{A}^k \\ \tilde{\psi}_b^1 \end{array}\right) &\equiv
    \varepsilon^a_i \tilde{T}^{ik} \left(\begin{array}{c} -(\gamma^\mu?)_a^b?\partial_\mu\tilde{\psi}_b^1
 \\ i C_{ab} \square \tilde{A}^k \end{array}\right) \cr
 \tilde{\delta}_{FS2}^{(1)}(\tilde{T}) \left(\begin{array}{c}
                            \tilde{F}^k \\ \tilde{\psi}_b^1 \end{array}\right) &\equiv
    \varepsilon^a_i \tilde{T}^{ik} \left(\begin{array}{c}
                           \square\tilde{\psi}_a^1
 \\ i(\gamma^\mu)_{ab}\partial_\mu \tilde{F}^k \end{array}\right) 
 \cr
 \tilde{\delta}_{FS3}^{(1)}(\tilde{T}) \left(\begin{array}{c}
                            \tilde{A}^k \\ \tilde{\psi}_b^2 \end{array}\right) &\equiv
    \varepsilon^a_i (\sigma^2)^{ij} \tilde{T}^{jk} \left(\begin{array}{c} i(\gamma^\mu?)_a^b?\partial_\mu \tilde{\psi}_b^2
\\ C_{ab}\square \tilde{A}^k \end{array}\right)
 \cr
\tilde{\delta}_{FS4}^{(1)}(\tilde{T}) \left(\begin{array}{c}
                            \tilde{F}^k \\ \tilde{\psi}_b^2 \end{array}\right) &\equiv
    \varepsilon^a_i (\sigma^2)^{ij}T^{jk} \left(\begin{array}{c}
                           -i \square \tilde{\psi}_a^2
\\ (\gamma^\mu)_{ab}\partial_\mu \tilde{F}^k \end{array}\right)
}\ee
\be\eqalign{
\tilde{\delta}_{FS5}^{(1)}(\tilde{T}) \left(\begin{array}{c}
                            \tilde{B}^k \\ \tilde{\psi}_b^1 \end{array}\right) &\equiv
    \varepsilon^a_i (\sigma^3)^{ij}\tilde{T}^{jk} \left(\begin{array}{c}        i(\gamma^5\gamma^\mu?)_a^b?\partial_\mu\tilde{\psi}_b^1
 \\ (\gamma^5)_{ab} \square \tilde{B}^k \end{array}\right) \cr
 \tilde{\delta}_{FS6}^{(1)}(\tilde{T}) \left(\begin{array}{c}
                            \tilde{G}^k \\ \tilde{\psi}_b^1 \end{array}\right) &\equiv
    \varepsilon^a_i (\sigma^3)^{ij}\tilde{T}^{jk} \left(\begin{array}{c}
                 -i(\gamma^5?)_a^b?\square\tilde{\psi}_b^1
 \\ (\gamma^5\gamma^\mu)_{ab}\partial_\mu \tilde{G}^k \end{array}\right) 
 \cr
\tilde{\delta}_{FS7}^{(1)}(\tilde{T}) \left(\begin{array}{c}
                            \tilde{B}^k \\ \tilde{\psi}_b^2 \end{array}\right) &\equiv
    \varepsilon^a_i (\sigma^1)^{ij}\tilde{T}^{jk} \left(\begin{array}{c}        i(\gamma^5\gamma^\mu?)_a^b?\partial_\mu\tilde{\psi}_b^2
 \\ (\gamma^5)_{ab} \square \tilde{B}^k \end{array}\right)
 \cr
 \tilde{\delta}_{FS8}^{(1)}(\tilde{T}) \left(\begin{array}{c}
                            \tilde{G}^k \\ \tilde{\psi}_b^2 \end{array}\right) &\equiv
    \varepsilon^a_i (\sigma^1)^{ij}\tilde{T}^{jk} \left(\begin{array}{c}
                 -i(\gamma^5?)_a^b?\square\tilde{\psi}_b^2
 \\ (\gamma^5\gamma^\mu)_{ab}\partial_\mu \tilde{G}^k \end{array}\right) 
}\ee

These are only the unique symmetries uncovered via this method, the redundant calculations being shown once again in Appendix~\ref{app:FirstOrderFermionicSymmetriesCCExplicit}.  Here we notice as in the bosonic case, that these fermionic symmetries are not themselves symmetric with respect to $A^J \leftrightarrow B^J$ and $F^J \leftrightarrow G^J$.  Again, this is a direct result of the absence of the corresponding central charge or internal symmetry in the algebra.

\section{Conclusion}

The $d=4$, ${\mathcal N} = 4$ SUSY-YM system is important to many theoretical models in physics today.  As it is a conformal field theory, it's possible that its study can lead to further understanding of `walking' theories such as technicolor. In string theory, the AdS/CFT correspondence relates calculations of $d=4$, ${\mathcal N} = 4$ SUSY-YM to classical supergravity calculations on $AdS_5 \times S^5$, where the correspondence is weak to strong and vice versa. In an effort to more accurately describe the standard model, this has been taken further to include correspondences to gauge theories with running couplings.  Even so, the problem of how to augment the dynamical theory of $d=4$, ${\mathcal N} = 4$ SUSY-YM with a \emph{finite} number of auxiliary fields such that the algebra closes has been unsolved for quite some time.  A solution to this problem would be helpful to more fully understand these aforementioned theories relating to conformal field theories.  

In this paper, we chose a particular set of auxiliary fields for $d=4$, ${\mathcal N} = 4$ SUSY-YM and catalogued the Lagrangian symmetries manifest in the central charges and internal symmetries of the resulting algebra.  It was noted how not all possible Lagrangian symmetries can be uncovered this way, as certain central charges and internal symmetries are missing from the algebra.  We reinforce here that all results presented are from straightforward, actual calculations with no assumptions of centrality.  For instance, we have directly calculated that the SUSY-YM Lagrangian in Eq.~(\ref{eq:SUSYYMLagrangian}) is invariant with respect to the transformation laws in Eqs.~(\ref{eq:chi0specific}), (\ref{eq:V0specific}), (\ref{eq:chiIspecific}), and (\ref{eq:VIspecific}).  We have directly calculated that these transformation laws satisfy the anti-commutation relations in Eqs.~(\ref{eq:00termsspecific1}), (\ref{eq:IJtermsspecific1}), (\ref{eq:IJtermsspecific2}), (\ref{eq:crosstermsspecific1}), and (\ref{eq:crosstermsspecific2}).  The main result of this paper is how these transformation laws and anti-commutators lead by direct calculation to the first and second order Lagrangian symmetries presented in section~\ref{sec:ExtraSymmetries}.

Furthermore, reduction of this particular ${\mathcal N} = 4$ system to the ${\mathcal N} = 2$ Fayet hypermultiplet and ${\mathcal N} = 2$  vector multiplet was shown to follow from our direct calculations.  Here it was noticed how in this reduction, central charges and internal symmetries are lost from the algebra.  In the case of the vector multiplet, all charges and internal symmetries are lost as the algebra closes.  In the case of the Fayet hypermultiplet, some central charges and internal symmetries remain, as this algebra does not close.

Finally, we make a note on quantization of non-closed systems such as the ${\mathcal N} = 4$ SUSY-YM system investigated in detail in this paper.  In general, non-closure of an algebra leads to an added difficulty in the quantization procedure.  Perhaps the most ubiquitous example is the criticality of string theory.  For quantum non-critical strings, one must solve the Liouville theory.  This is not necessary in the case of critical strings~\cite{Polyakov:1981rd,Polyakov:1981re}.  In the case of our results of the ${\mathcal N} = 4$ SUSY-YM system, we have laid out our results in the hopes of eventually obtaining a closed system, in the sense of Eq.~(\ref{eq:ClosedAlgebra}), without an infinite number of auxiliary fields.  For instead quantization of the non-closed system presented, the specific forms of the non-closure terms we calculated are important in the same vein as the Liouville theory for non-critical strings.  We leave this quantization as a future project.

\vspace{.05in}
 \begin{center}
 \parbox{4in}{{\it ``It is while you are patiently toiling at the little tasks of life that the meaning and shape of the great whole of life dawn on you.''}\,\,-\,\, Phillips Brooks}
 \end{center}

\section*{Acknowledgements}
This research was supported in part by the endowment of the John S.~Toll Professorship, the University of Maryland 
Center for String \& Particle Theory, National Science Foundation Grant PHY-0354401.  SJG \& KS offer additional gratitude to the M.\ L.\ K. Visiting Professorship program and to the 
M.\ I.\ T.\ Center for Theoretical Physics for support and hospitality extended during the undertaking of this work.  We would also like to thank Michael Faux for pointing out an incorrect wording with respect to the definition of central charges.

\appendix
\section*{Appendix}
\section{Explicit Calculation of First Order Fermionic Symmetries}

In this appendix, we explain in more detail the procedure which led us to the symmetries presented in the body of the paper.  Many symmetries found in this manner are redundant, and those presented in the paper are the unique symmetries found through this procedure.
\subsection{\texorpdfstring{${\mathcal N} = 4$}{{\mathcal N} = 4} SUSY-YM}
\subsubsection{Second Order Bosonic Symmetries}\label{app:SecondOrderBosonicSymmetriesExplicit}
In this section of the appendix, we explicitly show how the second order bosonic symmetries are discovered through the ${\mathcal N} = 4$ algebra.  Several are redundant, and in the body of the paper, only the unique symmetries were listed.
\begin{align}
\delta^{(2)}_{BS1}(P_1,P_2)A^K &\equiv [\delta_{BS1}^{(1)}(P_1), \delta_{BS1}^{(1)}(P_2)] A^K = \Lambda_{1,1}^{KJ}(P_1,P_2)\square A^J \\
\delta_{BS1}^{(2)}(T_1,T_2)A^K & \equiv  [\delta_{BS3}^{(1)}(T_1),\delta_{BS3}^{(1)}(T_2)] A^K = \Lambda_{3,3}^{JK}(T_1, T_2) \square A^J \\
\delta_{BS2}^{(2)}(Q_1,Q_2) B^K & \equiv  [ \delta_{BS2}^{(1)}(Q_1),\delta_{BS2}^{(1)}(Q_2) ] B^K = \Lambda_{2,2}^{KJ}(Q_1, Q_2) \square B^J \\
\delta_{BS2}^{(2)}(T_1,T_2) B^K &\equiv
[\delta_{BS4}^{(1)}(T_1),\delta_{BS4}^{(1)}(T_2)] B^K = \Lambda_{4,4}^{JK}(T_1, T_2)\square B^J 
\end{align}
\begin{align}
\delta_{BS3}^{(2)}(T_1,T_2)F^K &\equiv [\delta_{BS3}^{(1)}(T_1),\delta_{BS3}^{(1)}(T_2)] F^K = \Lambda_{3,3}^{JK}(T_1, T_2)\square F^J \\
\delta_{BS4}^{(2)}(T_1,T_2)G^K &\equiv [\delta_{BS4}^{(1)}(T_1),\delta_{BS4}^{(1)}(T_2) ] G^K = \Lambda_{4,4}^{JK}(T_1, T_2)\square G^J \\
\delta_{BS5}^{(2)}(U_1,U_2) F^K &\equiv [\delta_{BS7}^{(1)}(U_1),\delta_{BS7}^{(1)}(U_2)] F^K = (\Lambda_{7,7}^{\mu\nu})^{JK}(U_1, U_2)\partial_\mu\partial_\nu F^J \\
\delta_{BS6}^{(2)}(U_1,U_2) G^K &\equiv [\delta_{BS8}^{(1)}(U_1),\delta_{BS8}^{(1)}(U_2)] G^K +  \eta_{\mu\nu}  (\Lambda^{\mu\nu}_{8,8})^{JK} (U_1, U_2)\square G^K \nonumber\\
  &= (\Lambda^{\mu\nu}_{8,8})^{JK} (U_1, U_2)\partial_\mu\partial_\nu G^J \\
\delta_{BS7}^{(2)}(U_1,U_2) A_\nu &\equiv [\delta_{BS8}^{(1)}(U_1),\delta_{BS8}^{(1)}(U_2)] A_\nu = \eta_{\nu\beta}(\Lambda^{\mu\beta}_{8,8})^{JJ}(U_1, U_2)\partial^\alpha F_{\mu\alpha}
\end{align}
with

\be\eqalign{
    \Lambda_{1,1}^{KJ}(P_1, P_2) &\equiv P_{[1}^{K} P_{2]}^{J},\cr
\Lambda_{3,3}^{KJ}(T_1,T_2) &= \Lambda_{4,4}^{KJ}(T_1,T_2)\equiv T_{[1}^{KM}T_{2]}^{MJ}, \cr
\Lambda_{2,2}^{KJ}(Q_1,Q_2) &\equiv Q_{[1}^K Q_{2]}^J, \cr
   (\Lambda^{\mu\nu}_{7,7})^{JK}(U_1,U_2) &= (\Lambda^{\mu\nu}_{8,8})^{JK}(U_1,U_2) \equiv (U_{[1}^{\mu})^{J}(U_{2]}^\nu)^{K}, \cr 
}\ee

\noindent and
\be\eqalign{
\delta_{BS8}^{(2)}(P,Q) \left(\begin{array}{l}
                                 A^K \\
                                 B^K
                               \end{array}\right) \equiv & [ \delta_{BS1}^{(1)}(P), \delta_{BS2}^{(1)}(Q) ] \left(\begin{array}{l}
                                 A^K \\
                                 B^K
                               \end{array}\right) \cr
                               = &\Lambda_{1,2}^{IJ}(P,Q)\left(\begin{array}{l}
           - \delta^{IK}\square B^J \\
           \delta^{JK}\square A^I
           \end{array}
           \right)
}\ee
\be\eqalign{
\delta_{BS9}^{(2)}(P,U) \left(\begin{array}{l}
                                  A^K \\ F^K
                               \end{array}
                               \right) \equiv & [ \delta_{BS1}^{(1)}(P),\delta_{BS7}^{(1)}(U) ] \left(\begin{array}{l} A^K \\ F^K
                               \end{array}
                               \right) \cr
                               = & -(\Lambda_{1,7}^\mu)^{IJ}(P,U) \left(\begin{array}{l}
       \delta^{IK} \partial_\mu F^J \\
   \delta^{JK}\partial_\mu\square A^I
                               \end{array}
                               \right)
}\ee
\be\eqalign{
\delta_{BS12}^{(2)}(T,U) \left(\begin{array}{l}
                          A^K \\{\rm d}
                        \end{array}\right) \equiv & [\delta_{BS3}^{(1)}(T), \delta_{BS8}^{(1)}(U)] 
\left(\begin{array}{l}
                          A^K \\{\rm d}
                        \end{array}\right) \cr = & -(\Lambda_{3,8}^\mu)^K(T,U)\left(\begin{array}{l}\partial_\mu {\rm d}  \\ \partial_\mu \square A^K
\end{array}\right) 
}\ee
\be\eqalign{
\delta_{BS14}^{(2)}(T,U)\left( \begin{array}{ll}
                                  B^K \\ A_\nu \end{array} \right) \equiv & [ \delta_{BS4}^{(1)}(T), \delta_{BS8}^{(1)}(U) ] \left( \begin{array}{ll}
                                  B^K \\ A_\nu \end{array} \right)
\cr = &  -(\Lambda_{4,8}^\mu)^K(T,U) \left(\begin{array}{ll}
          \partial^\nu F_{\mu\nu} \\ \eta_{\mu\nu}\square B^K
             \end{array} \right) 
}\ee
\be\eqalign{
\delta_{BS10}^{(2)}(Q,U) \left(\begin{array}{l}
              B^K \\ F^K \end{array}\right) 
             \equiv & [ \delta_{BS2}^{(1)}(Q), \delta_{BS7}^{(1)}(U) ] \left(\begin{array}{l}
              B^K \\ F^K \end{array}\right) \cr = & (\Lambda_{2,7}^\mu)^{IJ}(Q,U)\left(\begin{array}{l}\delta^{IK}\partial_\mu F^J \\  \delta^{KJ}\partial_{\mu} \square B^I
 \end{array}\right) 
}\ee
\be\eqalign{
\delta_{BS15}^{(2)}(P,T) \left(\begin{array}{l}F^K \\{\rm d} \end{array}\right)
 \equiv  & [ \delta_{BS1}^{(1)}(P), \delta_{BS3}^{(1)}(T) ] \left(\begin{array}{l} F^K \\{\rm d} \end{array}\right) \cr = & \Lambda_{1,3}^K(P,T)\left(\begin{array}{l} -\square  {\rm d}\\ \square  F^K
\end{array}\right)
}\ee
\be\eqalign{
\delta_{BS16}^{(2)}(Q,T) \left(\begin{array}{l}G^K \\ {\rm d} \end{array}\right) \equiv & [\delta_{BS2}^{(1)}(Q), \delta_{BS4}^{(1)}(T) ] \left(\begin{array}{l} G^K \\ {\rm d} \end{array}\right) \cr = &  \Lambda_{2,4}^K(Q,T)\left(\begin{array}{l} \square {\rm d} \\ - \square G^K \end{array}\right) 
}\ee

\noindent with
\be\eqalign{
\Lambda_{1,2}^{JK}(P,Q) &= - \Lambda_{2,1}^{KJ}(Q,P) \equiv  P^J Q^K, \cr
(\Lambda^\mu_{1,7})^{JK}(P,U) &= -(\Lambda^\mu_{7,1})^{KJ}(U,P) \equiv P^J (U^{\mu})^K, \cr
(\Lambda^\mu_{3,7})^K(T,U) &= -(\Lambda^\mu_{5,2})^K(U,T)  \cr
&=(\Lambda^\mu_{4,12})^K(T,U)= -(\Lambda^\mu_{8,4})^K(U,T) \equiv T^{KM} (U^\mu)^M,\cr
(\Lambda_{2,7}^\mu)^{JK}(Q,U) &= -(\Lambda_{7,2}^\mu)^{KJ}(U,Q) \equiv Q^J(U^\mu)^K, \cr
\Lambda_{1,3}^K(P,T) &= -\Lambda_{3,1}^K(T,P) \equiv P^M T^{MK}, \cr
\Lambda_{2,4}^K(Q,T) &= - \Lambda_{4,2}^{K}(T,Q) \cr
 &= \Lambda_{11,14}^{K}(Q,T) = -\Lambda_{14,11}^{K}(T,Q) \equiv Q^M T^{MK},
}\ee

\noindent and
\noindent
\be\eqalign{
\delta_{BS30}^{(2)}(Q,W) \left(\begin{array}{l}\lambda_c \\ \psi_c^K \end{array}\right) \equiv & [ \delta_{BS9}^{(1)}(Q), \delta_{BS12}^{(1)}(W) ] \left(\begin{array}{l}\lambda_c \\ \psi_c^K \end{array}\right) \cr = &  \Lambda_{9,12}^K(Q,W)(\gamma^5)_c^{~d}\left(\begin{array}{l}
\square \psi_c^K \\ \square \lambda_d \end{array}\right) 
}\ee
\be\eqalign{
\delta_{BS32}^{(2)}(Q,V) \left(\begin{array}{l} \lambda_c \\  \psi_c^K \end{array}\right) \equiv & [ \delta_{BS9}^{(1)}(Q), \delta_{BS13}^{(1)}(V) ] \left(\begin{array}{l} \lambda_c \\  \psi_c^K \end{array}\right)
   \cr = &(\Lambda_{9,13}^\alpha)^K(Q,V)\left(\begin{array}{l} (\gamma^5\gamma^\mu\gamma_\alpha\gamma^\nu)_c^{~d} \\ \partial_\mu\partial_\nu \psi_d^K 
(\gamma^5\gamma_\alpha)_c^{~d}\square \lambda_d \end{array}\right)
}\ee
\be\eqalign{
\delta_{BS31}^{(2)}(Q,T) \left(\begin{array}{l} \lambda_c \\ \psi_c^K \end{array}\right) \equiv & [ \delta_{BS9}^{(1)}(Q), \delta_{BS14}^{(1)}(T) ] \left(\begin{array}{l} \lambda_c \\ \psi_c^K \end{array}\right) \cr = &   \Lambda_{9,14}^{K}(Q,T) \left(\begin{array}{l} -\square \psi^K_c \\  \square \lambda_c \end{array}\right)  
}\ee
\begin{align}
\delta_{BS26}^{(2)}(W,V)\psi_c^K &\equiv [ \delta_{BS12}^{(1)}(W), \delta_{BS13}^{(1)}(V) ] \psi_c^K \nonumber\\
= & (\Lambda_{12,13}^\mu)^{[JK]}(\gamma_\mu)_c^{~d}\square \psi_d^J + 2(\Lambda_{12,13}^\mu)^{KJ}(\gamma^\nu)_{c}^{~d}\partial_\mu \partial_\nu \psi_d^J \\
\delta_{BS28}^{(2)}(W,T)\psi_c^K \equiv & [ \delta_{BS12}^{(1)}(W), \delta_{BS14}^{(1)}(T) ] \psi_c^K = -\Lambda_{12,14}^{KJ}(W,T)(\gamma^5)_c^{~d}\square \psi_d^J \\
\delta_{BS23}^{(2)}(V,T)\psi_c^K \equiv & [ \delta_{BS13}^{(1)}(V), \delta_{BS14}^{(1)}(T) ] \psi_c^K + \delta^{(2)}_{BS25}(V,T)\psi_c^K \nonumber\\ = & (\Lambda_{13,14}^\alpha)^{[JK]}(V,T)(\gamma^5\gamma_\alpha)_c^{~d}\square\psi_d^J \\
\delta_{BS20}^{(2)}(Q_1,Q_2)\psi_c^K \equiv & [ \delta_{BS9}^{(1)}(Q_1), \delta_{BS9}^{(1)}(Q_2) ] \psi_c^K = - \Lambda_{11,11}^{JK}(Q_1, Q_2) \square \psi_c^J \\
\delta_{BS20}^{(2)}(W_1,W_2)\psi_c^K \equiv &[ \delta_{BS12}^{(1)}(W_1), \delta_{BS12}^{(1)}(W_2) ] \psi_c^K = \Lambda_{12,12}^{KJ}(W_1,W_2)\square \psi_c^J \\
\delta_{BS21}^{(2)}(V_1,V_2)\psi_c^K \equiv & [ \delta_{BS13}^{(1)}(V_1), \delta_{BS13}^{(1)}(V_2) ] \psi_c^K \nonumber\\ = &  -(\Lambda_{13,13}^{\rho\sigma})^{KJ}(V_1,V_2) (\gamma_\rho\gamma^\mu\gamma_\sigma\gamma^\nu)_c^{~d} \partial_\mu\partial_\nu\psi_d^J \\
\delta_{BS20}^{(2)}(T_1,T_2)\psi_c^K \equiv &[ \delta_{BS10}^{(1)}(T_1), \delta_{BS10}^{(1)}(T_2) ] \psi_c^K = -\Lambda_{14,14}^{KJ}(T_1, T_2)\square \psi_c^J   
\end{align}
\noindent with
\be\eqalign{
\Lambda_{9,9}^{JK}(Q_1,Q_2) &\equiv Q_{[1}^J Q_{2]}^K, \cr
\Lambda_{9,12}^{K}(Q,W) &= -\Lambda_{12,11}^{K}(W,Q) \equiv Q^M W^{MK}, \cr
(\Lambda^\mu_{9,13})^{K}(Q,V) &= -\Lambda_{13,11}^{K}(V,Q) \equiv Q^M(V^\mu)^{MK}, \cr
\Lambda_{12,12}^{JK}(W_1,W_2) &\equiv W_{[1}^{JM} W_{2]}^{MK},\cr
(\Lambda_{12,13}^\mu)^{JK}(W,V) &= -(\Lambda_{13,12}^\mu)^{KJ}(V,W) \equiv W^{JM}(V^\mu)^{MK}, \cr
\Lambda_{12,14}^{JK}(W,T) &= - \Lambda_{14,12}^{KJ}(T,W) \equiv W^{M(J}T^{K)M}, 
}\ee
\be\eqalign{
(\Lambda_{13,13}^{\rho\sigma})^{KJ}(V_1,V_2) &\equiv (V_{[1}^\rho)^{KM}(V_{2]}^\sigma)^{MJ} ,\cr
(\Lambda_{13,14}^W)^{JK}(V,T) &= -(\Lambda_{14,13}^W)^{KJ}(T,V) \equiv (V^W)^{JM}T^{MK} ,\cr
\Lambda_{14,14}^{KJ}(T_1,T_2) &\equiv T_{[1}^{KM}T_{2]}^{MJ}
}\ee

\noindent and

\begin{align}
  \delta_{BS1}^{(2)}(W_1,W_2) A^J \equiv &[ \delta_{BS5}^{(1)}(W_1), \delta_{BS5}^{(1)}(W_2) ] A^J= \Lambda_{5,5}^{IJ}(W_1,W_2)\square A^I \\
  \delta_{BS4}^{(2)}(W_1,W_2) G^J \equiv & [ \delta_{BS5}^{(1)}(W_1), \delta_{BS5}^{(1)}(W_2) ] G^J= \Lambda_{5,5}^{IJ}(W_1,W_2)\square G^I \\
  \delta_{BS5}^{(2)}(V_1,V_2) F^J \equiv & [ \delta_{BS6}^{(1)}(V_1), \delta_{BS6}^{(1)}(V_2) ]F^J= - (\Lambda_{6,6}^{\mu\nu})^{IJ}(V_1,V_2) \partial_\mu\partial_\nu F^I \\
  \delta_{BS6}^{(2)}(V_1,V_2) G^J \equiv & [ \delta_{BS6}^{(1)}(V_1), \delta_{BS6}^{(1)}(V_2) ] G^J = -(\Lambda_{6,6}^{\mu\nu})^{IJ}(V_1,V_2)\partial_\mu\partial_\nu G^I
\end{align}

\be\eqalign{
  \delta_{BS16}^{(2)}(P,W) \left(\begin{array}{ll}
                            G^J        \\   {\rm d}
                                 \end{array}\right)
             \equiv & [ \delta_{BS1}^{(1)}(P), \delta_{BS5}^{(1)}(W) ] \left(\begin{array}{ll}
                            G^J        \\   {\rm d}
                                 \end{array}\right) \cr = &  \Lambda_{1,5}^J(P,W)\left(\begin{array}{ll}
                         -\square {\rm d}           \\     \square G^J
                                 \end{array}\right)
}\ee
\be\eqalign{
  \delta_{BS18}^{(2)}(T,W) \left(\begin{array}{ll}
                                    F^J\\  G^J
                                 \end{array}\right)
             \equiv & [ \delta_{BS3}^{(1)}(T), \delta_{BS5}^{(1)}(W) ] \left(\begin{array}{ll}
                                    F^J\\  G^J
                                 \end{array}\right)\cr = &  \Lambda_{3,5}^{IK}(T,W)\left(\begin{array}{ll}
                       -\delta^{IJ}\square G^K             \\  \square \delta^{JK}F^I
                                 \end{array}\right)
}\ee
\be\eqalign{
  \delta_{BS8}^{(2)}(T,W) \left(\begin{array}{ll}
                                    A^J\\ B^J
                                 \end{array}\right)
             \equiv & [ \delta_{BS4}^{(1)}(T), \delta_{BS5}^{(1)}(W) ] \left(\begin{array}{ll}
                                    A^J\\ B^J
                                 \end{array}\right)\cr = &  \Lambda_{4,5}^{IK}(T,W)\left(\begin{array}{ll}
                          \delta^{JK}\square B^I          \\   -\delta^{IJ}\square A^K
                                 \end{array}\right)
}\ee
\be\eqalign{
  \delta_{BS13}^{(2)}(U,W) \left(\begin{array}{ll}
                                  A^J  \\  A_\nu
                                 \end{array}\right)
             \equiv & [ \delta_{BS8}^{(1)}(U), \delta_{BS5}^{(1)}(W) ] \left(\begin{array}{ll}
                                  A^J  \\  A_\nu
                                 \end{array}\right) \cr = & -(\Lambda_{8,5}^\mu)^J(U,W)\left(\begin{array}{ll}
                       \partial^\nu F_{\mu\nu}             \\
       \eta_{\mu\nu}\square A^J                          \end{array}\right)
}\ee
\be\eqalign{
  \delta_{BS9}^{(2)}(W,V) \left(\begin{array}{ll}
                                 A^J   \\   F^J
                                 \end{array}\right)
             \equiv & -[ \delta_{BS5}^{(1)}(W), \delta_{BS6}^{(1)}(V) ] \left(\begin{array}{ll}
                                 A^J   \\   F^J
                                 \end{array}\right)\cr = &  -(\Lambda_{5,6}^\mu)^{IK}(W,V)\left(\begin{array}{ll}
                         \delta^{IJ}\partial_\mu F^K           \\    \delta^{JK}\partial_\mu\square A^I
                                 \end{array}\right)
}\ee
\be\eqalign{
  \delta_{BS11}^{(2)}(T,V) \left(\begin{array}{ll}
                                 A^J   \\   G^J
                                 \end{array}\right)
             \equiv & -[ \delta_{BS3}^{(1)}(T), \delta_{BS6}^{(1)}(V) ] \left(\begin{array}{ll}
                                 A^J   \\   G^J
                                 \end{array}\right)\cr = &  (\Lambda_{3,6}^\mu)^{IK}(T,V)\left(\begin{array}{ll}
                      \delta^{IJ}\partial_\mu G^K              \\   \delta^{JK}\partial_\mu\square A^I
                                 \end{array}\right)
}\ee
\be\eqalign{
  \delta_{BS10}^{(2)}(T,V) \left(\begin{array}{ll}
                                 B^J   \\  F^J
                                 \end{array}\right)
             \equiv & -[ \delta_{BS4}^{(1)}(T), \delta_{BS6}^{(1)}(V) ] \left(\begin{array}{ll}
                                 B^J   \\  F^J
                                 \end{array}\right) \cr = &  -\Lambda_{4,6}^\mu)^{IK}(T,V)\left(\begin{array}{ll}
                           \delta^{IJ}(\partial_\mu F^K         \\  \delta^{JK}\partial_\mu \square B^I
                                 \end{array}\right)
}\ee
\be\eqalign{
  \delta_{BS17}^{(2)}(U,V) \left(\begin{array}{ll}
                                    G^J \\   {\rm d}
                                 \end{array}\right)
             \equiv & -[ \delta_{BS7}^{(1)}(U), \delta_{BS6}^{(1)}(V) ] \left(\begin{array}{ll}
                                    G^J \\   {\rm d}
                                 \end{array}\right) \cr = &  (\Lambda_{7,6}^{\mu\nu})^J(U,V)\left(\begin{array}{ll}
                         -\partial_\mu\partial_\nu {\rm d}           \\\partial_\mu\partial_\nu G^J
                                 \end{array}\right)
}\ee
\be\eqalign{
  \delta_{BS19}^{(2)}(U,V) \left(\begin{array}{ll}
                                 F^J   \\  A_\alpha
                                 \end{array}\right)
             \equiv & -[ \delta_{BS8}^{(1)}(U), \delta_{BS6}^{(1)}(V) ] \left(\begin{array}{ll}
                                 F^J   \\  A_\alpha
                                 \end{array}\right) \cr = &  (\Lambda_{8,6}^{\mu\nu})^J(U,V)\left(\begin{array}{ll}
                           \partial_\nu\partial^\alpha F_{\mu\alpha}        \\  -\eta_{\mu\alpha}\partial_\nu F^J
                                 \end{array}\right)
}\ee

\noindent with
\be\eqalign{
  \Lambda_{5,5}^{IJ}(W_1,W_2) &\equiv W_{[1}^{KI} W_{2]}^{JK}, \cr
  \Lambda_{6,6}^{IJ}(V_1,V_2) &\equiv (V_{[1}^\mu)^{KI}(V_{2]}^{\nu})^{JK}, \cr
  \Lambda_{1,5}^J(P,W) &\equiv P^K W^{KJ}, \cr
  \Lambda_{3,5}^{IJ}(T,W) &= \Lambda_{4,5}^{IJ}(T,W) \equiv  T^{IK}W^{KJ}, \cr
  (\Lambda_{8,5}^\mu)^J(U,W) &\equiv (U^\mu)^K W^{KJ}, \cr
  (\Lambda_{5,6}^\mu)^{JI}(W,V) &\equiv  W^{JK}(V^\mu)^{KI}, \cr
  (\Lambda_{3,6}^\mu)^{JI}(T,V) &= (\Lambda_{4,6}^\mu)^{JI}(T,V) \equiv  T^{JK}(V^\mu)^{KI}, \cr
 (\Lambda_{7,6}^{\mu\nu})^J(U,V) &=  (\Lambda_{8,6}^{\mu\nu})^J(U,V) \equiv (U^\mu)^K(V^\nu)^{KJ}
}\ee
and
\be\eqalign{
   \delta_{BS24}^{(2)}(Q,U) \lambda_c \equiv &[ \delta_{BS9}^{(1)}(Q), \delta_{BS10}^{(1)}(U) ] \lambda_c \cr = &-2 (\Lambda_{9,10}^{\mu})^{KK}(Q,U)(\gamma^5\gamma^\nu?)_c^d? \partial_\mu\partial_\nu \lambda_d,
}\ee
\be\eqalign{
  \delta_{BS25}^{(2)}(Q,U)\psi_c^K  \equiv & [ \delta_{BS9}^{(1)}(Q), \delta_{BS10}^{(1)}(U) ] \psi_c^K \cr = &- 2 (\Lambda_{9,10}^\mu)^{KJ}(Q,U) (\gamma^5\gamma^\nu?)_c^d? \partial_\mu\partial_\nu \psi_d^J,
}\ee
\be\eqalign{
\delta_{BS34}^{(2)} (W,U) \left( \begin{array}{l} \lambda_c \\ \psi_c^K \end{array} \right) &\equiv  [ \delta_{BS12}^{(1)}(W), \delta_{BS10}^{(1)}(U) ]\left( \begin{array}{l} \lambda_c \\ \psi_c^K \end{array} \right)  \cr 
&= - (\Lambda_{12,10}^\mu)^K(W,U) \left( \begin{array}{l} (\gamma_\mu?)_c^d? \square \psi_d^K \\ (\gamma^\nu\gamma_\mu\gamma^\alpha?)_c^d?\partial_\nu\partial_\alpha \lambda_d \end{array} \right) 
}\ee
\be\eqalign{
\delta_{BS36}^{(2)} (V,U) \left( \begin{array}{l} \lambda_c \\ \psi_c^K \end{array} \right) &\equiv  [ \delta_{BS13}^{(1)}(V), \delta_{BS10}^{(1)}(U) ]\left( \begin{array}{l} \lambda_c \\ \psi_c^K \end{array} \right)  \cr 
&= (\Lambda_{13,10}^{\mu\nu})^K(V,U) \left( \begin{array}{l} -(\gamma_\nu\gamma^\alpha\gamma_\mu\gamma^\beta?)_c^d? \partial_\alpha\partial_\beta \psi_d^K \\ (\gamma_\mu\gamma^\alpha\gamma_\nu\gamma^\beta?)_c^d?\partial_\alpha\partial_\beta \lambda_d \end{array} \right)
}\ee
\be\eqalign{
\delta_{BS33}^{(2)} (T,U) \left( \begin{array}{l} \lambda_c \\ \psi_c^K \end{array} \right) &\equiv  [ \delta_{BS14}^{(1)}(T), \delta_{BS10}^{(1)}(U) ]\left( \begin{array}{l} \lambda_c \\ \psi_c^K \end{array} \right)  \cr 
&= -(\Lambda_{14,10}^{\mu})^K(T,U) \left( \begin{array}{l} (\gamma^5\gamma_\mu?)_c^d? \square \psi_d^K \\ (\gamma^5\gamma^\alpha\gamma_\mu\gamma^\beta?)_c^d?\partial_\alpha\partial_\beta \lambda_d \end{array} \right)
}\ee
\be\eqalign{
   \delta_{BS22}^{(2)}(U_1,U_2) \lambda_c \equiv & [ \delta_{BS10}^{(1)}(U_1), \delta_{BS10}^{(1)}(U_2) ] \lambda_c \cr = & (\Lambda_{10,10}^{\mu\nu})^{KK}(U_1,U_2) (\gamma_\mu\gamma^\alpha\gamma_\nu\gamma^\beta?)_c^d? \partial_\alpha\partial_\beta \lambda_d
}\ee
\be\eqalign{
   \delta_{BS21}^{(2)}(U_1,U_2) \psi_c^K \equiv  &[ \delta_{BS10}^{(1)}(U_1), \delta_{BS10}^{(1)}(U_2) ] \psi_c^K \cr = & (\Lambda_{10,10}^{\mu\nu})^{KJ}(U_1,U_2) (\gamma_\mu\gamma^\alpha\gamma_\nu\gamma^\beta?)_c^d? \partial_\alpha\partial_\beta \psi_d^J
}\ee
\begin{equation}
   \delta_{BS27}^{(2)}(U,P) \lambda_c \equiv [ \delta_{BS10}^{(1)}(U), \delta_{BS11}^{(1)}(P) ] \lambda_c =  2(\Lambda_{10,11}^{\mu})^{KK}(U,P) (\gamma^\nu?)_c^d? \partial_\mu\partial_\nu \lambda_d
\end{equation}
\be\eqalign{
   \delta_{BS26}^{(2)}(U,P) \psi_c^K \equiv & [ \delta_{BS10}^{(1)}(U), \delta_{BS11}^{(1)}(P) ] \psi_c^K   \cr  = & (\Lambda_{10,11}^{\mu})^{JK}(U,P)(\gamma^\alpha\gamma_\mu\gamma^\beta?)_c^d? \partial_\alpha\partial_\beta\psi_d^J \cr
     &+ (\Lambda_{10,11}^\mu)^{KJ}(U,P)(\gamma_\mu?)_c^d?\square\psi_d^J
}\ee
\begin{align}
   \delta_{BS29}^{(2)}(Q,P) \lambda_c &\equiv [ \delta_{BS9}^{(1)}(Q), \delta_{BS11}^{(1)}(P) ] \lambda_c =  -\Lambda_{9,11}^{KK}(Q,P) (\gamma^5?)_c^d? \square \lambda_d \\
   \delta_{BS28}^{(2)}(Q,P) \psi_c^K &\equiv [ \delta_{BS9}^{(1)}(Q), \delta_{BS11}^{(1)}(P) ] \psi_c^K  = -\Lambda_{9,11}^{KJ}(Q,P)(\gamma^5?)_c^d? \square\psi_d^J
\end{align}
\be\eqalign{
\delta_{BS31}^{(2)} (W,P) \left( \begin{array}{l} \lambda_c \\ \psi_c^K \end{array} \right) \equiv & [ \delta_{BS12}^{(1)}(W), \delta_{BS11}^{(1)}(P) ]\left( \begin{array}{l} \lambda_c \\ \psi_c^K \end{array} \right) \cr = &\Lambda_{9,11}^K(W,P) \left( \begin{array}{l}  \square \psi_c^K \\ -\square \lambda_d \end{array} \right)
}\ee
\be\eqalign{
\delta_{BS35}^{(2)} (V,P) \left( \begin{array}{l} \lambda_c \\ \psi_c^K \end{array} \right) &\equiv  [ \delta_{BS13}^{(1)}(V), \delta_{BS11}^{(1)}(P) ]\left( \begin{array}{l} \lambda_c \\ \psi_c^K \end{array} \right)  \cr
&= (\Lambda_{13,11}^\mu)^K(V,P) \left( \begin{array}{l}  (\gamma^\alpha \gamma_\mu \gamma^\beta?)_c^d? \partial_\alpha\partial_\beta\psi_d^K \\ (\gamma_\mu?)_c^d?\square \lambda_d \end{array} \right)
}\ee
\be\eqalign{
\delta_{BS30}^{(2)} (T,P) \left( \begin{array}{l} \lambda_c \\ \psi_c^K \end{array} \right) &\equiv  [ \delta_{BS14}^{(1)}(T), \delta_{BS11}^{(1)}(P) ]\left( \begin{array}{l} \lambda_c \\ \psi_c^K \end{array} \right)  \cr &= -(\Lambda_{14,11}^\mu)^K(T,P) \left( \begin{array}{l}  (\gamma^5?)_c^d? \square \psi_d^K \\ (\gamma^5?)_c^d?\square \lambda_d \end{array} \right)
}\ee
\be\eqalign{
   \delta_{BS20}^{(2)}(P_1,P_2) \psi_c^K &\equiv [ \delta_{BS11}^{(1)}(P_1), \delta_{BS11}^{(1)}(P_2) ] \psi_c^K  = -\Lambda_{11,11}^{KJ}(P_1,P_2)\square\psi_c^J,
}\ee
with
\be\eqalign{
(\Lambda_{9,10}^\mu)^{JK}(Q,U) &\equiv Q^J(U^\mu)^K,~~~(\Lambda_{12,10}^\mu)^K(W,U) = W^{KM}(U^\mu)^M, \cr
(\Lambda_{13,10}^{\mu\nu})^{K}(V,U) &\equiv (V^\mu)^{KM}(U^\nu)^M,~~~ (\Lambda_{14,10}^\mu)^K(T,U) = T^{KM}(U^\mu)^M, \cr
(\Lambda_{10,10}^{\mu\nu})^{KJ}(U_1,U_2) &\equiv (U_{[1}^\mu)^{K}(U_{2]}^\nu)^J,~~~ (\Lambda_{10,11}^\mu)^{KM}(U,P) \equiv (U^\mu)^{K}P^M, \cr
\Lambda_{9,11}^{KJ}(Q,P) &\equiv Q^{(K}P^{J)},~~~ \Lambda_{12,11}^K(W,P) \equiv W^{KM}P^M, \cr
(\Lambda_{13,11}^\mu)^{K}(V,P) &\equiv  (V^\mu)^{KM}P^{M)},~~~ \Lambda_{14,11}^K(T,P) \equiv T^{KM}P^M, \cr
\Lambda_{11,11}^{KM}(P_1,P_2) &\equiv P_{[1}^K P_{2]}^M
}\ee

\noindent where $[~~]$ denotes antisymmetry, i.e. $U_{[1}^{J} U_{2]}^{K} = U_1^J U_2^K - U_2^J U_1^K$.

\subsubsection{Fermionic Symmetries}\label{app:FirstOrderFermionicSymmetriesExplicit}
Taking the commutators or ${\rm D}_a$ and ${\rm D}_a^I$ with the first order bosonic symmetries for the ${\mathcal N} = 4$ SUSY-YM system, we find several first order \emph{fermionic} symmetries, some of which are redundant.  The symmetries calculated below which involve $\varepsilon^a_I P^K \epsilon^{IJK}$, $\varepsilon^a_I Q^K \epsilon^{IJK}$, and $\varepsilon^a_I (U^\rho)^K \epsilon^{IJK}$ are redefined through
\be\eqalign{
\varepsilon^a_I P^K \epsilon^{IJK} \to \varepsilon^a P^J \cr 
\varepsilon^a_I Q^K \epsilon^{IJK} \to \varepsilon^a Q^J\cr 
\varepsilon^a_I (U^\rho)^K \epsilon^{IJK} \to \varepsilon^a (U^\rho)^J 
}\ee
as symmetries defined either way are equivalent for the Lagrangian.  In section~\ref{sec:1stfermionsa}, all symmetries are listed using this redefinition where applicable.
\begin{align}
 \delta_{FS19}^{(1)}(P) \left(
\begin{array}{c}
                           {\rm d} \\ \psi_b^J \end{array}\right) \equiv &
    \varepsilon^a P^J \left(\begin{array}{c}
                           \square\psi_a^J \\  i (\gamma^\mu)_{ab} \partial_\mu {\rm d} \end{array}\right)= -\varepsilon^a [ {\rm D}_a, \delta_{BS1}^{(1)}(P) ]\left(\begin{array}{c}
                            {\rm d} \\ \psi_b^J \end{array}\right) \\
 \delta_{FS13}^{(1)}(Q) \left(\begin{array}{c}
                            {\rm d} \\ \psi_b^J \end{array}\right) \equiv &
    \varepsilon^a Q^J \left(\begin{array}{c}
                           i(\gamma^5)_{a}^{~b}\square\psi_b^J \\  - (\gamma^5\gamma^\mu)_{ab} \partial_\mu {\rm d} \end{array}\right)= \varepsilon^a [ {\rm D}_a, \delta_{BS2}^{(1)}(Q) ]\left(\begin{array}{c}
                            {\rm d} \\ \psi_b^J \end{array}\right) \\
 \delta_{FS54}^{(1)}(U) \left(\begin{array}{c}
                            {\rm d} \\ \psi_b^J \end{array}\right) \equiv &
    \varepsilon^a (U^\mu)^J \partial_\mu \left(\begin{array}{c}
                           (\gamma^\nu)_{a}^{~b}\partial_\nu\psi_b^J \\  i C_{ab} {\rm d} \end{array}\right)= \varepsilon^a [ {\rm D}_a, \delta_{BS7}^{(1)}(U) ]\left(\begin{array}{c}
                            {\rm d} \\ \psi_b^J \end{array}\right) \\
\delta_{FS70}^{(1)}(U) \left(\begin{array}{c}
                            A_\mu \\ \psi_b^J \end{array}\right) \equiv &
    \varepsilon^a (U_\mu)^J \partial^\nu \left(\begin{array}{c}
  i(\gamma^5\gamma_\nu)_{a}^{~b}\psi_b^J \\  - (\gamma^5)_{ab} ?F^\mu_\nu? \end{array}\right)= \varepsilon^a [ {\rm D}_a, \delta_{BS8}^{(1)}(U) ]\left(\begin{array}{c}
                            A_\mu \\ \psi_b^J \end{array}\right)\\
 \delta_{FS2}^{(1)}(P) \left(\begin{array}{c}
                            A^J \\ \lambda_b \end{array}\right) \equiv &
    \varepsilon^a P^J \left(\begin{array}{c}
                           i(\gamma^5\gamma^\mu?)_a^b?\partial_\mu\lambda_b \\   (\gamma^5)_{ab} \square A^J \end{array}\right)= \varepsilon^a [ {\rm D}_a, \delta_{BS1}^{(1)}(P) ]\left(\begin{array}{c}
                            A^J \\ \lambda_b \end{array}\right)
\end{align}
\begin{align}                                      
 \delta_{FS3}^{(1)}(Q) \left(\begin{array}{c}
                            B^J \\ \lambda_b \end{array}\right) \equiv &
    \varepsilon^a Q^J \left(\begin{array}{c}
                           i(\gamma^5\gamma^\mu?)_a^b?\partial_\mu\lambda_b \\   (\gamma^5)_{ab} \square B^J \end{array}\right)= -\varepsilon^a [ {\rm D}_a, \delta_{BS2}^{(1)}(Q) ]\left(\begin{array}{c}
                            B^J \\ \lambda_b \end{array}\right) \\  
 \delta_{FS51}^{(1)}(U) \left(\begin{array}{c}
                            G^J \\ \lambda_b \end{array}\right) \equiv &
    \varepsilon^a (U^\mu)^J \left(\begin{array}{c}
       \partial^\nu\partial_{[\mu}(\gamma_{\nu]}?)_a^b? \lambda_b \\  (?\sigma^\nu_\mu?)_{ab}\partial_\nu G^J 
\end{array}\right)= \varepsilon^a [ {\rm D}_a, \delta_{BS8}^{(1)}(U) ]\left(\begin{array}{c}
                           G^J \\ \lambda_b \end{array}\right) \\ 
\delta_{FS46}^{(1)}(U) \left(\begin{array}{c}
                            F^J \\ \lambda_b \end{array}\right) \equiv &
    \varepsilon^a (U^\mu)^J \left(\begin{array}{c}
                           (\gamma^5\gamma^\nu?)_a^b?\partial_\mu\partial_\nu\lambda_b \\  i(\gamma^5)_{ab} \partial_\mu F^J \end{array}\right) \cr = &-i\varepsilon^a [ {\rm D}_a, \delta_{BS7}^{(1)}(U) ] \left(\begin{array}{c}
                            F^J \\ \lambda_b \end{array}\right)
\end{align}
and from $[ {\rm D}_a, \delta_{BS3}^{(1)}(T) ]$  we have
\be\eqalign{
 \delta_{FS21}^{(1)}(T) \left(\begin{array}{c}
                            A^J \\ \psi_b^J \end{array}\right) &\equiv
    \varepsilon^a T^{JM} \left(\begin{array}{c}
                           (\gamma^\mu)_a^{~b}\partial_\mu \psi_b^M \\  i C_{ab}\square A^M \end{array}\right) \cr
 \delta_{FS82}^{(1)}(T) \left(\begin{array}{c}
                            F^J \\ \psi_b^J \end{array}\right) &\equiv
    \varepsilon^a T^{JM} \left(\begin{array}{c}
                           \square \psi_a^M \\  -i (\gamma^\mu)_{ab}\partial_\mu F^M \end{array}\right) 
}\ee
and from $[ {\rm D}_a, \delta_{BS4}^{(1)}(T) ]$
\be\eqalign{
 \delta_{FS28}^{(1)}(T) \left(\begin{array}{c}
                            B^J \\ \psi_b^J \end{array}\right) &\equiv
    \varepsilon^a T^{JM} \left(\begin{array}{c}
                           (\gamma^5\gamma^\mu)_a^{~b}\partial_\mu \psi_b^M \\  i(\gamma^5)_{ab}\square B^M \end{array}\right) \cr
 \delta_{FS35}^{(1)}(T) \left(\begin{array}{c}
                            G^J \\ \psi_b^J \end{array}\right) &\equiv
    \varepsilon^a T^{JM} \left(\begin{array}{c}
                           i (\gamma^5)_{a}^{~b}\square \psi_b^M \\  (\gamma^5\gamma^\mu)_{ab}\partial_\mu G^M \end{array}\right) 
}\ee
and from $[ {\rm D}_a, \delta_{BS9}^{(1)}(Q) ]$
\be\eqalign{
\delta_{FS66}^{(1)}(Q) \left(\begin{array}{c}
                            A_\mu \\ \psi_b^J \end{array}\right) &\equiv
    \varepsilon^a Q^{J} \left(\begin{array}{c}
                           -(\gamma_\mu\gamma^\nu)_a^{~b}\partial_\nu \psi_b^J \\   \frac{1}{2}(\gamma^\alpha\sigma^{\mu\nu})_{ba}\partial_\alpha F_{\mu\nu}\end{array}\right) \cr
 \delta_{FS13}^{(1)}(Q) \left(\begin{array}{c}
                            d \\ \psi_b^J \end{array}\right) &\equiv
    \varepsilon^a Q^{J} \left(\begin{array}{c}
                           i (\gamma^5?)_a^b?\square\psi_b^J\\ -(\gamma^5\gamma^\mu?)_ab?\partial_\mu d \end{array}\right) 
}\ee
and from $[ {\rm D}_a, \delta_{BS12}^{(1)}(W) ]$
\be\eqalign{
\delta_{FS27}^{(1)}(W) \left(\begin{array}{c}
                            A^J \\ \psi_b^J \end{array}\right) &\equiv
    \varepsilon^a W^{JM} \left(\begin{array}{c}
         (\gamma^5\gamma^\mu?)_a^b?\partial_\mu\psi_b^M
 \\ i(\gamma^5)_{ab}\square A^M \end{array}\right) \cr
 \delta_{FS22}^{(1)}(W) \left(\begin{array}{c}
                            B^J \\ \psi_b^J \end{array}\right) &\equiv
    \varepsilon^a W^{JM} \left(\begin{array}{c}
                           (\gamma^\mu?)_a^b? \partial_\mu \psi_b^M
 \\ C_{ab}\square B^M \end{array}\right) 
 \cr
 \delta_{FS36}^{(1)}(W) \left(\begin{array}{c}
                            F^
J \\ \psi_b^J \end{array}\right) &\equiv
    \varepsilon^a W^{JM} \left(\begin{array}{c}
                           i(\gamma^5?)_a^b?\square\psi_b^M
\\ (\gamma^5\gamma^\mu?)_ab?\partial_\mu F^M \end{array}\right)
 \cr
 \delta_{FS83}^{(1)}(W) \left(\begin{array}{c}
                            G^J \\ \psi_b^J \end{array}\right) &\equiv
    \varepsilon^a W^{JM} \left(\begin{array}{c}
                           i\square \psi_a^M
\\ (\gamma^\mu)_{ab}\partial_\mu G^M \end{array}\right)
}
\ee
and from $[ {\rm D}_a, \delta_{BS13}^{(1)}(V) ]$
\be\eqalign{
\delta_{FS76}^{(1)}(V) \left(\begin{array}{c}
                            A^J \\ \psi_b^J \end{array}\right) &\equiv
    \varepsilon^a (V^\rho)^{JM} \left(\begin{array}{c}
         (\gamma^5\gamma_\rho\gamma^\mu?)_a^b?\partial_\mu\psi_b^M
 \\ i(\gamma^5\gamma_\rho)_{ab} \square A^M \end{array}\right) \cr
 \delta_{FS81}^{(1)}(V) \left(\begin{array}{c}
                            B^J \\ \psi_b^J \end{array}\right) &\equiv
    \varepsilon^a (V^\rho)^{JM} \left(\begin{array}{c}
                           -i (\gamma_\rho\gamma^\mu?)_a^b? \partial_\mu \psi_b^M
 \\ (\gamma_\rho)_{ab}\square B^M \end{array}\right) 
 \cr
 \delta_{FS43}^{(1)}(V) \left(\begin{array}{c}
                            F^J \\ \psi_b^J \end{array}\right) &\equiv
    \varepsilon^a (V^\rho)^{JM} \left(\begin{array}{c}
                           (\gamma^5\gamma^\nu\gamma_\rho\gamma^\mu?)_a^b?\partial_\mu\partial_\nu\psi_b^M
\\ i(\gamma^5\gamma_\rho\gamma^\mu?)_ba?\partial_\mu F^M \end{array}\right)
 \cr
 \delta_{FS56}^{(1)}(V) \left(\begin{array}{c}
                            G^J \\ \psi_b^J \end{array}\right) &\equiv
    \varepsilon^a (V^\rho)^{JM} \left(\begin{array}{c}
                           i (\gamma^\mu\gamma_\rho\gamma^\mu?)_a^b?\partial_\mu\partial_\nu \psi_b^M
\\ -(\gamma_\rho\gamma^\mu)_{ba}\partial_\mu G^M \end{array}\right)
}
\ee
and from $[ {\rm D}_a, \delta_{BS14}^{(1)}(T) ]$
\be\eqalign{
\delta_{FS21}^{(1)}(T) \left(\begin{array}{c}
                            A^J \\ \psi_b^J \end{array}\right) &\equiv
    \varepsilon^a T^{JM} \left(\begin{array}{c}
         (\gamma^\mu?)_a^b?\partial_\mu\psi_b^M
 \\ iC_{ab} \square A^M \end{array}\right) \cr
 \delta_{FS28}^{(1)}(T) \left(\begin{array}{c}
                            B^J \\ \psi_b^J \end{array}\right) &\equiv
    \varepsilon^a T^{JM} \left(\begin{array}{c}
                           (\gamma^5\gamma^\mu?)_a^b? \partial_\mu \psi_b^M
 \\  i(\gamma^5)_{ab}\square B^M \end{array}\right) 
}\ee
\be\eqalign{
 \delta_{FS82}^{(1)}(T) \left(\begin{array}{c}
                            F^J \\ \psi_b^J \end{array}\right) &\equiv
    \varepsilon^a T^{JM} \left(\begin{array}{c}
                          i \square\psi_a^M
\\ (\gamma^\mu?)_ab?\partial_\mu F^M \end{array}\right)
 \cr
 \delta_{FS35}^{(1)}(T) \left(\begin{array}{c}
                            G^J \\ \psi_b^J \end{array}\right) &\equiv
    \varepsilon^a T^{JM} \left(\begin{array}{c}
                           i (\gamma^5?)_a^b?\square \psi_b^M
\\ (\gamma^5\gamma^\mu)_{ab}\partial_\mu G^M \end{array}\right)
}\ee
and from  $[ {\rm D}_a, \delta_{BS5}^{(1)}(W) ]$
\be\eqalign{
\delta_{FS27}^{(1)}(W) \left(\begin{array}{c}
                            A^J \\ \psi_b^J \end{array}\right) &\equiv
    \varepsilon^a W^{JM} \left(\begin{array}{c}
                           (\gamma^5\gamma^\mu)_a^{~b}\partial_\mu \psi_b^M \\   i(\gamma^5)_{ab}\square A^M\end{array}\right) \cr
 \delta_{FS83}^{(1)}(W) \left(\begin{array}{c}
                            G^J \\ \psi_b^J \end{array}\right) &\equiv
    \varepsilon^a W^{JM} \left(\begin{array}{c}
                          i \square\psi_a^M\\ (\gamma^\mu)_{ab}\partial_\mu G^M \end{array}\right) 
}\ee
and from $[ {\rm D}_a, \delta_{BS6}^{(1)}(V) ]$
\be\eqalign{
\delta_{FS42}^{(1)}(V) \left(\begin{array}{c}
                            F^J \\ \psi_b^J \end{array}\right) &\equiv
    \varepsilon^a (V^\mu)^{JM} \left(\begin{array}{c}
  i(\gamma^5\gamma^\nu)_a^{~b}\partial_\mu\partial_\nu \psi_b^M \\   (\gamma^5)_{ab}\partial_\mu F^M 
\end{array}\right) \cr
 \delta_{FS57}^{(1)}(V) \left(\begin{array}{c}
                            G^J \\ \psi_b^J \end{array}\right) &\equiv
    \varepsilon^a (V^\mu)^{JM} \left(\begin{array}{c}                     (\gamma^\nu?)_a^b?\partial_\mu\partial_\nu\psi_b^M\\ -i C_{ab}\partial_\mu G^M \end{array}\right) 
}
\ee

and from $[ {\rm D}_a, \delta_{BS9}^{(1)}(Q) ]$
\be\eqalign{
\delta_{FS7}^{(1)}(Q) \left(\begin{array}{c}
                            A^J \\ \lambda_b \end{array}\right) &\equiv
    \varepsilon^a Q^J \left(\begin{array}{c}         (\gamma^\mu?)_a^b?\partial_\mu\lambda_b
 \\ -iC_{ab} \square A^J \end{array}\right) \cr
 \delta_{FS3}^{(1)}(Q) \left(\begin{array}{c}
                            B^J \\ \lambda_b \end{array}\right) &\equiv
    \varepsilon^a Q^{J} \left(\begin{array}{c}
                           i (\gamma^5\gamma^\mu?)_a^b? \partial_\mu \lambda_b
 \\ (\gamma^5)_{ab}\square B^J \end{array}\right) 
 \cr
 \delta_{FS16}^{(1)}(Q) \left(\begin{array}{c}
                            F^J \\ \lambda_b \end{array}\right) &\equiv
    \varepsilon^a Q^{J} \left(\begin{array}{c}  \square \lambda_a
\\ i(\gamma^\mu?)_ab?\partial_\mu F^J \end{array}\right)
 \cr
 \delta_{FS11}^{(1)}(Q) \left(\begin{array}{c}
                            G^J \\ \lambda_b \end{array}\right) &\equiv
    \varepsilon^a Q^{J} \left(\begin{array}{c}
                           i (\gamma^5?)_a^b?\square \lambda_b
\\ -(\gamma^5\gamma^\mu)_{ab}\partial_\mu G^J \end{array}\right)
}\ee
\noindent and
and from $[ {\rm D}_a^I, \delta_{BS1}^{(1)}(P) ]$
\be\eqalign{
\delta_{FS1}^{(1)}(P) \left(\begin{array}{c}
                            A^K \\ \psi_b^J \end{array}\right) &\equiv
    \varepsilon^a_I P^K \left(\begin{array}{c}         i(\gamma^5\gamma^\mu?)_a^b?\partial_\mu\psi_b^I
 \\ \delta^{IJ}(\gamma^5)_{ab} \square A^K \end{array}\right) \cr
 \delta_{FS20}^{(1)}(P) \left(\begin{array}{c}
                            {\rm d} \\ \lambda_b \end{array}\right) &\equiv
    \varepsilon^a_I P^{I} \left(\begin{array}{c}  \square \lambda_a
\\ i(\gamma^\mu?)_ab?\partial_\mu {\rm d} \end{array}\right)
}\ee
\be\eqalign{
 \delta_{FS19}^{(1)}(P) \left(\begin{array}{c}
                            {\rm d} \\ \psi_b^J \end{array}\right) &\equiv
    \varepsilon^a_I P^{K} \epsilon^{IJK} \left(\begin{array}{c}
                            \square \psi_b^J
 \\ i(\gamma^\mu)_{ab}\partial_\mu {\rm d} \end{array}\right) 
 \cr
 &\to \varepsilon^a P^{J}\left(\begin{array}{c}
                            \square \psi_b^J
 \\ i(\gamma^\mu)_{ab}\partial_\mu {\rm d} \end{array}\right)
 }\ee
and from $[ {\rm D}_a^I, \delta_{BS3}^{(1)}(T) ]$
\be\eqalign{
\delta_{FS24}^{(1)}(T) \left(\begin{array}{c}
                            A^M \\ \psi_b^J \end{array}\right) &\equiv
    \varepsilon^a_I \epsilon^{IJK}T^{KM} \left(\begin{array}{c} i(\gamma^\mu?)_a^b?\partial_\mu\psi_b^J
 \\ C_{ab} \square A^M \end{array}\right) \cr
 \delta_{FS86}^{(1)}(T) \left(\begin{array}{c}
                            F^M \\ \psi_b^J \end{array}\right) &\equiv
    \varepsilon^a_I \epsilon^{IJK}T^{KM} \left(\begin{array}{c}
\square\psi_a^J
 \\ i (\gamma^\mu)_{ab}\partial_\mu F^M \end{array}%
\right)
}
\ee


\subsection{\texorpdfstring{${\mathcal N} = 2$}{{\mathcal N} = 2} FH}\label{app:FirstOrderFermionicSymmetriesCCExplicit}
In this section, we list all of the ${\mathcal N} = 2$ FH fermionic first order symmetries uncovered via our method, including the redundant ones.  Only the unique symmetries were listed in the body of the paper. From from $[ \tilde{{\rm D}}_a^i, \tilde{\delta}_{BS1}^{(1)}(\tilde{T}) ]$ we find the symmetries
\be\eqalign{
\tilde{\delta}_{FS1}^{(1)}(\tilde{T}) \left(\begin{array}{c}
                            \tilde{A}^k \\ \tilde{\psi}_b^1 \end{array}\right) &\equiv
    \varepsilon^a_i \tilde{T}^{ik} \left(\begin{array}{c} -(\gamma^\mu?)_a^b?\partial_\mu\tilde{\psi}_b^1
 \\ i C_{ab} \square \tilde{A}^k \end{array}\right) \cr
 \tilde{\delta}_{FS2}^{(1)}(\tilde{T}) \left(\begin{array}{c}
                            \tilde{F}^k \\ \tilde{\psi}_b^1 \end{array}\right) &\equiv
    \varepsilon^a_i \tilde{T}^{ik} \left(\begin{array}{c}
                           \square\tilde{\psi}_a^1
 \\ i(\gamma^\mu)_{ab}\partial_\mu \tilde{F}^k \end{array}\right) 
 \cr
 \tilde{\delta}_{FS3}^{(1)}(\tilde{T}) \left(\begin{array}{c}
                            \tilde{A}^k \\ \tilde{\psi}_b^2 \end{array}\right) &\equiv
    \varepsilon^a_i (\sigma^2)^{ij} \tilde{T}^{jk} \left(\begin{array}{c} i(\gamma^\mu?)_a^b?\partial_\mu \tilde{\psi}_b^2
\\ C_{ab}\square \tilde{A}^k \end{array}\right)
 \cr
\tilde{\delta}_{FS4}^{(1)}(\tilde{T}) \left(\begin{array}{c}
                            \tilde{F}^k \\ \tilde{\psi}_b^2 \end{array}\right) &\equiv
    \varepsilon^a_i (\sigma^2)^{ij}T^{jk} \left(\begin{array}{c}
                           -i \square \tilde{\psi}_a^2
\\ (\gamma^\mu)_{ab}\partial_\mu \tilde{F}^k \end{array}\right)
}\ee
and from $[ \tilde{{\rm D}}_a^i, \tilde{\delta}_{BS2}^{(1)}(\tilde{T}) ]$
\be\eqalign{
\tilde{\delta}_{FS5}^{(1)}(\tilde{T}) \left(\begin{array}{c}
                            \tilde{B}^k \\ \tilde{\psi}_b^1 \end{array}\right) &\equiv
    \varepsilon^a_i (\sigma^3)^{ij}\tilde{T}^{jk} \left(\begin{array}{c}        i(\gamma^5\gamma^\mu?)_a^b?\partial_\mu\tilde{\psi}_b^1
 \\ (\gamma^5)_{ab} \square \tilde{B}^k \end{array}\right) \cr
 \tilde{\delta}_{FS6}^{(1)}(\tilde{T}) \left(\begin{array}{c}
                            \tilde{G}^k \\ \tilde{\psi}_b^1 \end{array}\right) &\equiv
    \varepsilon^a_i (\sigma^3)^{ij}\tilde{T}^{jk} \left(\begin{array}{c}
                 -i(\gamma^5?)_a^b?\square\tilde{\psi}_b^1
 \\ (\gamma^5\gamma^\mu)_{ab}\partial_\mu \tilde{G}^k \end{array}\right) 
 }\ee
\be\eqalign{
\tilde{\delta}_{FS7}^{(1)}(\tilde{T}) \left(\begin{array}{c}
                            \tilde{B}^k \\ \tilde{\psi}_b^2 \end{array}\right) &\equiv
    \varepsilon^a_i (\sigma^1)^{ij}\tilde{T}^{jk} \left(\begin{array}{c}        i(\gamma^5\gamma^\mu?)_a^b?\partial_\mu\tilde{\psi}_b^2
 \\ (\gamma^5)_{ab} \square \tilde{B}^k \end{array}\right)
 \cr
 \tilde{\delta}_{FS8}^{(1)}(\tilde{T}) \left(\begin{array}{c}
                            \tilde{G}^k \\ \tilde{\psi}_b^2 \end{array}\right) &\equiv
    \varepsilon^a_i (\sigma^1)^{ij}\tilde{T}^{jk} \left(\begin{array}{c}
                 -i(\gamma^5?)_a^b?\square\tilde{\psi}_b^2
 \\ (\gamma^5\gamma^\mu)_{ab}\partial_\mu \tilde{G}^k \end{array}\right) 
}\ee
Calculation of $[ \tilde{{\rm D}}_a^i, \tilde{\delta}_{BS3}^{(1)}(\tilde{T}) ]$ uncovers no new symmetries, just these same eight again:
\be\eqalign{
\tilde{\delta}_{FS1}^{(1)}(\tilde{T}) \left(\begin{array}{c}
                            \tilde{A}^k \\ \tilde{\psi}_b^1 \end{array}\right) &\equiv
    \varepsilon^a_i (\sigma^2)^{ik}\tilde{T}^{12} \left(\begin{array}{c} i(\gamma^\mu?)_a^b?\partial_\mu\tilde{\psi}_b^1
 \\ C_{ab} \square \tilde{A}^k \end{array}\right) \cr
 \tilde{\delta}_{FS2}^{(1)}(\tilde{T}) \left(\begin{array}{c}
                            \tilde{F}^k \\ \tilde{\psi}_b^1 \end{array}\right) &\equiv
    \varepsilon^a_i (\sigma^2)^{ik}\tilde{T}^{12} \left(\begin{array}{c}
                           i \square\tilde{\psi}_a^1
 \\ -(\gamma^\mu)_{ab}\partial_\mu \tilde{F}^k \end{array}\right) 
 \cr
 \tilde{\delta}_{FS3}^{(1)}(\tilde{T}) \left(\begin{array}{c}
                            \tilde{A}^k \\ \tilde{\psi}_b^2 \end{array}\right) &\equiv
    \varepsilon^a_k \tilde{T}^{12} \left(\begin{array}{c} -(\gamma^\mu?)_a^b?\partial_\mu \tilde{\psi}_b^2
\\ i C_{ab}\square \tilde{A}^k \end{array}\right)
 \cr
\tilde{\delta}_{FS4}^{(1)}(\tilde{T}) \left(\begin{array}{c}
                            \tilde{F}^k \\ \tilde{\psi}_b^2 \end{array}\right) &\equiv
    \varepsilon^a_k \tilde{T}^{12} \left(\begin{array}{c}
                           \square \tilde{\psi}_a^2
\\ i(\gamma^\mu)_{ab}\partial_\mu \tilde{F}^k \end{array}\right) \cr
\tilde{\delta}_{FS5}^{(1)}(\tilde{T}) \left(\begin{array}{c}
                            \tilde{B}^k \\ \tilde{\psi}_b^1 \end{array}\right) &\equiv
    \varepsilon^a_i (\sigma^1)^{ik}\tilde{T}^{12} \left(\begin{array}{c}        i(\gamma^5\gamma^\mu?)_a^b?\partial_\mu\tilde{\psi}_b^1
 \\ (\gamma^5)_{ab} \square \tilde{B}^k \end{array}\right) \cr
 \tilde{\delta}_{FS6}^{(1)}(\tilde{T}) \left(\begin{array}{c}
                            \tilde{G}^k \\ \tilde{\psi}_b^1 \end{array}\right) &\equiv
    \varepsilon^a_i (\sigma^1)^{ik}\tilde{T}^{12} \left(\begin{array}{c}
                 i(\gamma^5?)_a^b?\square\tilde{\psi}_b^1
 \\ -(\gamma^5\gamma^\mu)_{ab}\partial_\mu \tilde{G}^k \end{array}\right) 
 \cr
\tilde{\delta}_{FS7}^{(1)}(\tilde{T}) \left(\begin{array}{c}
                            \tilde{B}^k \\ \tilde{\psi}_b^2 \end{array}\right) &\equiv
    \varepsilon^a_i (\sigma^3)^{ik}\tilde{T}^{12} \left(\begin{array}{c}        i(\gamma^5\gamma^\mu?)_a^b?\partial_\mu\tilde{\psi}_b^2
 \\ (\gamma^5)_{ab} \square \tilde{B}^k \end{array}\right)
 \cr
 \tilde{\delta}_{FS8}^{(1)}(\tilde{T}) \left(\begin{array}{c}
                            \tilde{G}^k \\ \tilde{\psi}_b^2 \end{array}\right) &\equiv
    \varepsilon^a_i (\sigma^3)^{ik}\tilde{T}^{12} \left(\begin{array}{c}
                 -i(\gamma^5?)_a^b?\square\tilde{\psi}_b^2
 \\ (\gamma^5\gamma^\mu)_{ab}\partial_\mu \tilde{G}^k \end{array}\right)
}
\ee
under redefinitions of $\tilde{T}$.

\bibliographystyle{utphys}
\bibliography{Bibliography}

\providecommand{\href}[2]{#2}\begingroup\raggedright\begin{thebibliography}{10}

\bibitem{Maldacena:1997re}
J.~M. Maldacena, ``{The large N limit of superconformal field theories and
  supergravity},'' {\em Adv. Theor. Math. Phys.} {\bf 2} (1998)  231--252,
\href{http://arxiv.org/abs/hep-th/9711200}{{\tt arXiv:hep-th/9711200}}.

\bibitem{Klebanov:2000hb}
I.~R. Klebanov and M.~J. Strassler, ``{Supergravity and a confining gauge
  theory: Duality cascades and chiSB-resolution of naked singularities},'' {\em
  JHEP} {\bf 08} (2000)  052,
\href{http://arxiv.org/abs/hep-th/0007191}{{\tt arXiv:hep-th/0007191}}.

\bibitem{Maldacena:2000yy}
J.~M. Maldacena and C.~Nunez, ``{Towards the large N limit of pure N = 1 super
  Yang Mills},'' \href{http://dx.doi.org/10.1103/PhysRevLett.86.588}{{\em Phys.
  Rev. Lett.} {\bf 86} (2001)  588--591},
\href{http://arxiv.org/abs/hep-th/0008001}{{\tt arXiv:hep-th/0008001}}.

\bibitem{Cvetic:2001ma}
M.~Cvetic, G.~W. Gibbons, H.~Lu, and C.~N. Pope, ``{Supersymmetric non-singular
  fractional D2-branes and NS-NS 2-branes},''
  \href{http://dx.doi.org/10.1016/S0550-3213(01)00236-X}{{\em Nucl. Phys.} {\bf
  B606} (2001)  18--44},
\href{http://arxiv.org/abs/hep-th/0101096}{{\tt arXiv:hep-th/0101096}}.

\bibitem{Maldacena:2001pb}
J.~M. Maldacena and H.~S. Nastase, ``{The supergravity dual of a theory with
  dynamical supersymmetry breaking},'' {\em JHEP} {\bf 09} (2001)  024,
\href{http://arxiv.org/abs/hep-th/0105049}{{\tt arXiv:hep-th/0105049}}.

\bibitem{Canoura:2005uz}
F.~Canoura, J.~D. Edelstein, L.~A.~P. Zayas, A.~V. Ramallo, and D.~Vaman,
  ``{Supersymmetric branes on AdS(5) x Y**(p,q) and their field theory
  duals},'' {\em JHEP} {\bf 03} (2006)  101,
\href{http://arxiv.org/abs/hep-th/0512087}{{\tt arXiv:hep-th/0512087}}.

\bibitem{Herzog:2000rz}
C.~P. Herzog and I.~R. Klebanov, ``{Gravity duals of fractional branes in
  various dimensions},''
  \href{http://dx.doi.org/10.1103/PhysRevD.63.126005}{{\em Phys. Rev.} {\bf
  D63} (2001)  126005},
\href{http://arxiv.org/abs/hep-th/0101020}{{\tt arXiv:hep-th/0101020}}.

\bibitem{Herzog:2002ss}
C.~P. Herzog, ``{String tensions and three dimensional confining gauge
  theories},'' \href{http://dx.doi.org/10.1103/PhysRevD.66.065009}{{\em Phys.
  Rev.} {\bf D66} (2002)  065009},
\href{http://arxiv.org/abs/hep-th/0205064}{{\tt arXiv:hep-th/0205064}}.

\bibitem{PandoZayas:2003yb}
L.~A. Pando~Zayas, J.~Sonnenschein, and D.~Vaman, ``{Regge trajectories
  revisited in the gauge / string correspondence},''
  \href{http://dx.doi.org/10.1016/j.nuclphysb.2003.12.006}{{\em Nucl. Phys.}
  {\bf B682} (2004)  3--44},
\href{http://arxiv.org/abs/hep-th/0311190}{{\tt arXiv:hep-th/0311190}}.

\bibitem{PandoZayas:2008hw}
L.~A. Pando~Zayas, V.~G.~J. Rodgers, and K.~Stiffler, ``{Luscher term for
  k-string potential from holographic one loop corrections},''
  \href{http://dx.doi.org/10.1088/1126-6708/2008/12/036}{{\em JHEP} {\bf 12}
  (2008)  036},
\href{http://arxiv.org/abs/0809.4119}{{\tt arXiv:0809.4119 [hep-th]}}.

\bibitem{Doran:2009pp}
C.~A. Doran, L.~A. Pando~Zayas, V.~G.~J. Rodgers, and K.~Stiffler, ``{Tensions
  and Luscher terms for (2+1)-dimensional k-strings from holographic models},''
  \href{http://dx.doi.org/10.1088/1126-6708/2009/11/064}{{\em JHEP} {\bf 11}
  (2009)  064},
\href{http://arxiv.org/abs/0907.1331}{{\tt arXiv:0907.1331 [hep-th]}}.

\bibitem{Stiffler:2009ma}
K.~Stiffler, ``{Mesons From string theory},''
\href{http://arxiv.org/abs/0909.5681}{{\tt arXiv:0909.5681 [hep-th]}}.

\bibitem{Stiffler:2010pz}
K.~Stiffler, ``{A walk through superstring theory with an application to
  Yang-Mills theory: k-strings and D-branes as gauge/gravity dual objects},''
  {\em University of Iowa Ph.D. Thesis, published through Proquest} (2010)
  \href{http://gradworks.umi.com/34/22/3422196.html}{3422196},
\href{http://arxiv.org/abs/1012.0021}{{\tt arXiv:1012.0021 [hep-th]}}.

\bibitem{Siegel:1981dx}
W.~Siegel and M.~Rocek, ``{On off-shell supermultiplets},''
\href{http://dx.doi.org/10.1016/0370-2693(81)90887-X}{{\em Phys. Lett.} {\bf
  B105} (1981)  275}.

\bibitem{ABJM}
O.~Aharony, O.~Bergman, D.~L. Jafferis, and J.~Maldacena, ``{N=6 superconformal
  Chern-Simons-matter theories, M2-branes and their gravity duals},''
  \href{http://dx.doi.org/10.1088/1126-6708/2008/10/091}{{\em JHEP} {\bf 0810}
  (2008)  091}, \href{http://arxiv.org/abs/0806.1218}{{\tt arXiv:0806.1218
  [hep-th]}}.

\bibitem{3DSUSY1}
S.~J. Gates, Jr. and H.~Nishino, ``{Remarks on the N=2 supersymmetric
  Chern-Simons theories},''
  \href{http://dx.doi.org/10.1016/0370-2693(92)90277-B}{{\em Phys.Lett.} {\bf
  B281} (1992)  72--80}.

\bibitem{3DSUSY2}
H.~Nishino and S.~J. Gates, Jr., ``{Chern-Simons theories with supersymmetries
  in three-dimensions},''
  \href{http://dx.doi.org/10.1142/S0217751X93001363}{{\em Int.J.Mod.Phys.} {\bf
  A8} (1993)  3371--3422}.

\bibitem{3DSUSY3}
R.~Brooks and S.~J. Gates, Jr., ``{Extended supersymmetry and superBF gauge
  theories},'' \href{http://dx.doi.org/10.1016/0550-3213(94)90600-9}{{\em
  Nucl.Phys.} {\bf B432} (1994)  205--224},
  \href{http://arxiv.org/abs/hep-th/9407147}{{\tt arXiv:hep-th/9407147
  [hep-th]}}.

\bibitem{3DSUSY4}
S.~J. Gates, Jr. and L.~Rana, ``{A Theory of spinning particles for large N
  extended supersymmetry},''
  \href{http://dx.doi.org/10.1016/0370-2693(95)00474-Y}{{\em Phys.Lett.} {\bf
  B352} (1995)  50--58}, \href{http://arxiv.org/abs/hep-th/9504025}{{\tt
  arXiv:hep-th/9504025 [hep-th]}}.

\bibitem{Galperin1984}
A.~Galperin, E.~Ivanov, S.~Kalitsyn, V.~Ogievetsky, and E.~Sokatchev {\em
  Class. Quant.Grav.} {\bf 1} (1984)  469.

\bibitem{Galperin1985}
A.~Galperin, E.~Ivanov, V.~Ogievetsky, and E.~Sokatchev {\em Class.
  Quant.Grav.} {\bf 2} (1985)  647.

\bibitem{Gates:1984nk}
S.~J. Gates, Jr., C.~Hull, and M.~Rocek, ``{Twisted Multiplets and New
  Supersymmetric Nonlinear Sigma Models},''
  \href{http://dx.doi.org/10.1016/0550-3213(84)90592-3}{{\em Nucl.Phys.} {\bf
  B248} (1984)  157}.

\bibitem{Karlhede:1984vr}
A.~Karlhede, U.~Lindstrom, and M.~Rocek, ``{Selfinteracting tensor multiplets
  in N=2 superspace},''
  \href{http://dx.doi.org/10.1016/0370-2693(84)90120-5}{{\em Phys.Lett.} {\bf
  B147} (1984)  297}.

\bibitem{Fayet:1975yi}
P.~Fayet, ``{Fermi-Bose hypersymmetry},''
\href{http://dx.doi.org/10.1016/0550-3213(76)90458-2}{{\em Nucl. Phys.} {\bf
  B113} (1976)  135}.

\bibitem{Gates:2009me}
S.~J. Gates, Jr., J.~Gonzales, B.~MacGregor, J.~Parker, R.~Polo-Sherk, V.~G.~J.
  Rodgers, and L.~Wassink, ``{4D, N = 1 supersymmetry genomics (I)},''
  \href{http://dx.doi.org/10.1088/1126-6708/2009/12/008}{{\em JHEP} {\bf 12}
  (2009)  008},
\href{http://arxiv.org/abs/0902.3830}{{\tt arXiv:0902.3830 [hep-th]}}.

\bibitem{Polyakov:1981rd}
A.~M. Polyakov, ``{Quantum geometry of bosonic strings},''
\href{http://dx.doi.org/10.1016/0370-2693(81)90743-7}{{\em Phys. Lett.} {\bf
  B103} (1981)  207--210}.

\bibitem{Polyakov:1981re}
A.~M. Polyakov, ``{Quantum geometry of fermionic strings},''
\href{http://dx.doi.org/10.1016/0370-2693(81)90744-9}{{\em Phys. Lett.} {\bf
  B103} (1981)  211--213}.

\end{thebibliography}\endgroup


\end{document}